\documentclass{mn}
\usepackage{amsfonts}
\usepackage{amsmath}
\usepackage{graphicx}

\def\gsim{\;\lower4pt\hbox{${\buildrel\displaystyle >\over\sim}$}\;}
\def\lsim{\;\lower4pt\hbox{${\buildrel\displaystyle <\over\sim}$}\;}
\def\grls{\;\lower4pt\hbox{${\buildrel\displaystyle >\over <}$}\;}
\begin{document}

\title[Configurations of composite MSID]{Stationary
perturbation configurations in a composite system of
stellar and coplanarly magnetized gaseous singular
isothermal discs}
\author[Y.-Q. Lou and Y. Zou]{Yu-Qing Lou$^{1,2,3}$
and Yue Zou$^1$\\$^1$Physics Department, The Tsinghua Center
for Astrophysics, Tsinghua University, Beijing 100084, China
\\$^2$Department of Astronomy and Astrophysics, The University
of Chicago, 5640 Ellis Ave, Chicago, IL 60637, USA
\\$^3$National Astronomical Observatories, Chinese Academy of
Science, A20, Datun Road, Beijing 100012, China
\\}
\date{Accepted 2004... Received 2003...; in
original form 2003}\maketitle

\begin{abstract}
We construct aligned and unaligned stationary perturbation
configurations in a composite system of stellar and coplanarly
magnetized gaseous singular isothermal discs (SIDs) coupled by 
gravity. This study extends recent analyses on (magnetized)
SIDs of Shu et al., Lou and Lou \& Shen. By this model, we intend to
provide a conceptual framework to gain insights for multi-wavelength
large-scale structural observations of disc galaxies. Both SIDs are
approximated to be razor-thin and are in a self-consistent
axisymmetric background equilibrium with power-law surface mass
densities and flat rotation curves. The gaseous SID is embedded with
a coplanar azimuthal magnetic field $B_{\theta}(r)$ of a radial
scaling $r^{-1/2}$ that is not force-free. In comparison with SID
problems studied earlier, there exist three possible classes of
stationary solutions allowed by more dynamic freedoms. To identify
physical solutions, we explore parameter space involving three
dimensionless parameters: ratio $\lambda$ of Alfv\'en speed to sound
speed in the magnetized gaseous SID, ratio $\beta$ for the square of
the stellar velocity dispersion to the gas sound speed and ratio
$\delta$ of the surface mass densities of the two SIDs. For both
aligned and unaligned spiral cases with azimuthal periodicities
$|m|\geq 2$, one of the three solution branches is always physical,
while the other two branches might become invalid when $\beta$
exceeds certain critical values. For the onset criteria from an 
axisymmetric equilibrium to aligned secular bar-like instabilities,
the corresponding $\mathcal T/|\mathcal W-\mathcal M|$ ratio, which
varies with $\lambda$, $\beta$ and $\delta$, may be considerably
lower than the oft-quoted value of $\mathcal T/|\mathcal W|\sim 0.14$,
where $\cal{T}$ is the total kinetic energy, $\cal{W}$ is the total
gravitational potential energy and $\cal{M}$ is the total magnetic
energy. For unaligned spiral cases, we examine marginal instabilities
for axisymmetric ($|m|=0$) and non-axisymmetric ($|m|>0$) disturbances.
The resulting marginal stability curves differ from the previous ones. The
case of a composite partial MSID system is also investigated to include
the gravitational effect of an axisymmetric dark matter halo on the SID
equilibrium. We further examine the phase relationship among the mass
densities of the two SIDs and azimuthal magnetic field perturbation. Our 
exact global perturbation solutions and critical points are valuable for 
testing numerical magnetohydrodynamic codes. For galactic applications, 
our model analysis contains more realistic elements and offer useful
insights for structures and dynamics of disc galaxies consisting of
stars and magnetized gas.
\end{abstract}


\section{INTRODUCTION}

In galactic contexts, we venture to formulate a theoretical
magnetohydrodynamic (MHD) disc problem to explore possible large-scale
structures and dynamics of stationary MHD density waves in a composite
system of stellar and magnetized interstellar medium (ISM) gas discs.
The two gravitationally coupled discs are treated as `fluid' and
`magnetofluid' respectively and are both expediently approximated as
razor-thin singular isothermal discs (SIDs) with the gaseous SID being
embedded with a coplanar azimuthal magnetic field. For the gravitational
effect of a massive axisymmetric dark matter halo, we prescribe a
background composite system of two coupled partial SIDs (Syer \&
Tremaine 1996; Shu et al. 2000; Lou 2002; Lou \& Shen 2003; Shen \&
Lou 2003). In our model analysis, we construct stationary aligned
and unaligned logarithmic spiral MHD perturbation configurations
in a composite system of two SIDs with flat rotation curves, and
attempt to relate various morphologies of disc galaxies,
including barred and lopsided, barred and normal spiral structures.
For possible observational diagnostics, we derive phase relationships
among perturbation patterns of the stellar surface mass density, the
gas surface mass density and the azimuthal magnetic field.

This introduction serves two purposes. The first one is to provide
a general background information relevant to the problem at hand
and the second one is to give reasons of pursuing this MHD disc 
problem.

In a pioneering study of a composite system of stellar and gas
discs coupled by gravity, Lin \& Shu (1966, 1968) used a stellar 
distribution function and a gas fluid disc description to derive 
and analyze the local dispersion relation of galactic spiral
density waves. Since then, there have been extensive theoretical
studies on perturbation configurations and stability properties of
a composite disc system, mainly in galactic contexts. Kato (1972)
investigated oscillations and overstabilities of density waves 
using a
formalism similar to that of Lin \& Shu (1966, 1968). In a two-fluid
formalism, Jog \& Solomon (1984a, b) examined the growth of local
axisymmetric perturbations in a composite disc system. Bertin \&
Romeo (1988) studied the influence of a gas disc on spiral modes
in a two-fluid model framework. Vandervoort (1991a, b) studied the
influence of interstellar gas on oscillations and stabilities of
spheroidal galaxies. The two-fluid approach was also adopted in a
stability study of a two-component disc system with finite disc
thickness by Romeo (1992). The analysis for morphologies of disc
galaxies was performed by Lowe et al. (1994). For the stability of
a composite disc system, different effective $Q_{eff}$ parameters
(Safronov 1960; Toomre 1964) have been suggested using a two-fluid
formalism by Elmegreen (1995) and Jog (1996). Recently, Lou \& Fan
(1998b) used the two-fluid formalism to study properties of open
and tight-winding spiral density-wave modes in a composite disc
system. Lou \& Shen (2003) studied stationary global perturbation
structures in a two-fluid system of SIDs and, instead of a
redefinition of a different $Q_{eff}$ parameter, Shen \& Lou (2003)
offered a more practical $D-$criterion for the axisymmetric
instability in a composite SID system.

A rich class of disc problems involves stability properties of SIDs.
There have been numerous studies on this subject since the pioneering
work of Mestel (1963) (e.g. Zang 1976; Toomre 1977; Lemos, Kalnajs \&
Lynden-Bell 1991; Lynden-Bell \& Lemos 1999; Goodman \& Evans 1999;
Charkrabarti, Laughlin \& Shu 2003). Specifically, Syer \& Tremaine
(1996) made an important breakthrough to derive semi-analytic solutions
for stationary perturbation configurations in a class of SIDs.
Shu et al. (2000) obtained stationary solutions for perturbation
configurations in an isopedically magnetized SID with a flat
rotation curve. Through numerical explorations, they interpreted
these stationary aligned and unaligned logarithmic spiral
configurations as onsets of bar-type and barred-spiral instabilities
(see also Galli et al. 2001). Different from yet complementary to
the analysis of Shu et al. (2000), Lou (2002) performed a coplanar
MHD perturbation analysis in a single background SID embedded with
an azimuthal magnetic field,
from the perspective of stationary fast and slow MHD density waves
(FMDWs and SMDWs; Fan \& Lou 1996; Lou \& Fan 1998a). Lou (2002)
also derived a form of magnetic virial theorem for an MSID and
suggested the ratio of rotation energy to the sum of gravitational
and magnetic energies to be crucial for the onset of bar-like
instability in an MSID system.
In galactic contexts, it would be more realistic to consider
large-scale structures and dynamics in a composite system of
stellar and magnetized ISM discs.
As a first step, Lou \& Shen (2003) made a foray on this model
problem, constructed stationary aligned and unaligned logarithmic
spiral configurations in such a composite SID system and further
examined axisymmetric instability properties (Shen \& Lou 2003).

In disc galaxies, the ISM disc is magnetized with the magnetic
energy density being comparable to the energy densities of
thermal gas and of relativistic cosmic-ray gas (e.g. Lou \&
Fan 2003). Information of galactic magnetic fields can be
estimated by synchrotron radio emissions from spiral galaxies.
For such a magnetized composite system, MHD will play an 
indispensable role and reveal more realistic aspects of
dynamic and diagnostic information. These important problems
(Shu et al. 2000; Lou 2002; Lou \& Shen 2003) are not only
interesting by themselves, but also serve as necessary steps
for establishing an even more realistic model. Motivated by 
this prospect (Lou \& Fan 1998b; Lou 2002; Lou \& Shen 2003), 
we construct here stationary perturbation configurations for 
aligned and unaligned logarithmic spiral cases in a composite 
system of a stellar SID and a coplanarly magnetized gaseous 
MSID, and discuss their stability properties.

We adopt a relatively simple formalism for a composite system
involving fluid and magnetofluid discs coupled by gravity. We 
provide in Section 2 an MHD description for the coplanarly 
magnetized gaseous MSID, obtain conditions for background 
axisymmetric equilibrium state for both stellar SID and gaseous 
MSID, and derive linearized equations for coplanar perturbations. 
There exist aligned and unaligned classes of global MHD 
perturbation solutions; they are analyzed in Section 3 and 
Section 4, respectively. The exact solutions of stationary 
perturbations, their stability properties and their 
corresponding phase relationships among perturbation variables 
are examined and summarized in Section 5. Details are included 
in Appendices A$-$E.

\section{FLUID-MAGNETOFLUID FORMALISM}

It would be physically more precise to adopt a distribution function
formalism in dealing with a stellar disc especially in terms of
singularities and resonances (e.g. Lin \& Shu 1966, 1968; Binney \&
Tremaine 1987). For the present purpose of modelling large-scale
stationary perturbation structures and for mathematical simplicity
(Lou \& Shen 2003), it suffices to start with the fluid-magnetofluid 
formalism, including an MHD treatment for the gaseous MSID (Lou 2002; 
Lou \& Fan 2003). In this section, we present the basic equations for 
the fluid-magnetofluid system consisting of a stellar SID and a gaseous 
MSID. For flat rotation curves, conditions on the background rotational 
equilibrium with axisymmetry can be derived. We then obtain the 
linearized equations for coplanar MHD perturbations in the composite 
MSID system.

\subsection{Basic Nonlinear MHD Equations}

The two SIDs, located at $z=0$, are both approximated as 
infinitesimally thin. In our fluid-magnetofluid treatment, 
the two SIDs are coupled through mutual gravitational
interaction. For large-scale stationary perturbations, diffusive
processes such as viscosity, ambipolar diffusion and thermal diffusion
etc. are ignored. For physical variables under consideration, we shall
use superscript or subscript $s$ to indicate an association with the
stellar SID and superscript or subscript $g$ to indicate an association
with the gaseous MSID. In cylindrical coordinates ($r$, $\theta$, $z$),
the basic fluid-magnetofluid equations for a composite MSID system can
be readily written out.

In the fluid approximation for a stellar SID, the mass
conservation, the radial component of the momentum equation
and the azimuthal component of the momentum equation are
given below in order, namely
\begin{equation}\label{1}
\frac{\partial\Sigma^s}{\partial t}+\frac{1}{r}\frac{\partial (r
\Sigma^s u^s)}{\partial r}
+\frac1{r^2}\frac{\partial(\Sigma^sj^s)}{\partial \theta}=0\ ,
\end{equation}
\begin{equation}\label{2}
\frac{\partial u^s}{\partial t}+u^s\frac{\partial u^s}{\partial
r}+\frac{j^s}{r^2}\frac{\partial
u^s}{\partial\theta}-\frac{j^{s2}}{r^3}
=-\frac{1}{\Sigma^{s}}\frac{\partial\Pi^s}{\partial r}
-\frac{\partial\phi}{\partial r}\ ,
\end{equation}
\begin{equation}\label{3}
\frac{\partial j^s}{\partial t}+u^s\frac{\partial j^s}{\partial
r}+\frac{j^s}{r^2}\frac{\partial
j^s}{\partial\theta}=-\frac{1}{\Sigma^{s}}\frac{\partial
\Pi^s}{\partial\theta}-\frac{\partial\phi}{\partial\theta}\ ,
\end{equation}
where $u^s$ is the radial component of the stellar bulk velocity,
$j^s\equiv rv^s$ is the stellar specific angular momentum along
the $\hat{z}$ direction; and $v^s$ is the azimuthal component
of the stellar bulk velocity; $\phi$ is the total gravitational
potential, $\Pi^s$ is the vertically integrated (effective)
pressure (sometimes referred to as the two-dimensional pressure),
and $\Sigma^s$ is the vertically integrated stellar mass density
(i.e. stellar surface mass density).

In the magnetofluid approximation for the gaseous MSID, the mass
conservation, the radial component of the momentum equation and
the azimuthal component of the momentum equation are given below
in order, namely
\begin{equation}\label{4}
\frac{\partial\Sigma^g}{\partial t}+\frac{1}{r}
\frac{\partial (r\Sigma^g u^g)}{\partial r}
+\frac1{r^2}\frac{\partial(\Sigma^gj^g)}{\partial\theta}=0\ ,
\end{equation}
\begin{equation}\label{5}
\begin{split}
\frac{\partial u^g}{\partial t}+u^g\frac{\partial u^g}{\partial r}
+\frac{j^g}{r^2}\frac{\partial u^g}{\partial\theta}
-\frac{j^{g2}}{r^3}=-\frac1{\Sigma^g}\frac{\partial\Pi^g}{\partial r}
-\frac{\partial\phi}{\partial r} \qquad   \\
\quad\qquad -\frac1{\Sigma^g}\int \frac{dzB_\theta}{4\pi r}
\bigg[\frac{\partial(rB_\theta)}{\partial r}
-\frac{\partial B_r}{\partial\theta}\bigg]\ ,\quad
\end{split}
\end{equation}
\begin{equation}\label{6}
\begin{split}
\frac{\partial j^g}{\partial t}+u^g\frac{\partial j^g}{\partial r}
+\frac{j^g}{r^2}\frac{\partial j^g}{\partial\theta}
=-\frac1{\Sigma^g}\frac{\partial\Pi^g}{\partial\theta}
-\frac{\partial\phi}{\partial\theta}\qquad\qquad \\
\qquad +\frac1{\Sigma^g}\int
\frac{dzB_r}{4\pi}\bigg[\frac{\partial(rB_\theta)}{\partial r}
-\frac{\partial B_r}{\partial\theta}\bigg]\ ,
\end{split}
\end{equation}
where $u^g$ is the radial component of the gas bulk velocity,
$j^g\equiv rv^g$ is the gas specific angular momentum along the
$\hat{z}$ direction, $v^g$ is the azimuthal component of the gas 
bulk velocity, $\Pi^g$ is the vertically integrated gas pressure 
(sometimes referred to as the two-dimensional gas pressure), 
$\Sigma^g$ is the vertically integrated gas mass density (i.e. 
gas surface mass density) and $B_r$ and $B_\theta$ are the radial 
and azimuthal components of magnetic field $\mathbf B$.
The last two terms on the right-hand sides of equations (\ref{5})
and (\ref{6}) are the radial and azimuthal components of the
Lorentz force due to the coplanar magnetic field. The coupling of
the two sets of fluid and magnetofluid equations $(1)-(3)$ and
$(4)-(6)$ is effected by the total gravitational potential $\phi$
through the Poisson integral, namely
\begin{equation}\label{7}
F\phi(r,\theta,t)=\oint d\psi{\int_0}^\infty
\frac{-G(\Sigma^g+\Sigma^s)\zeta d\zeta}
{[\zeta^2+r^2-2\zeta r\cos(\psi-\theta)]^{1/2}}\ ,
\end{equation}
where $F$ is a constant ratio with $0< F\leq 1$; $F\phi$ is the 
gravitational potential from the stellar SID and gaseous MSID 
together, and the fraction $(1-F)\phi$ is attributed to an 
axisymmetric dark matter halo that is unresponsive to coplanar 
MHD perturbations in the composite MSID system (e.g. Shu et al. 
2000).

The divergence-free condition for the coplanar
magnetic field ${\bf{B}}=(B_r, B_{\theta},\ 0)$ is
\begin{equation}\label{8}
\frac{\partial(rB_r)}{\partial r}
+\frac{\partial B_\theta}{\partial\theta }=0\ ,
\end{equation}
and the radial and azimuthal components
of magnetic induction equation are
\begin{equation}\label{9}
\frac{\partial B_r}{\partial t}
=\frac{1}{r}\frac{\partial}{\partial\theta}(u^gB_\theta-v^gB_r)\ ,
\end{equation}
\begin{equation}\label{10}
\frac{\partial B_\theta}{\partial t}
=-\frac{\partial}{\partial r}(u^gB_\theta-v^gB_r)\ .
\end{equation}
Equations $(1)-(10)$ form the basis of our theoretical analysis.

\subsection{Rotational MSID Equilibrium}
For a flat rotation curve in a stellar SID, we write
the corresponding angular rotation rate in the form of
\begin{equation}\label{11}
\Omega_s(r)=a_s\frac{D_s}r\ ,
\end{equation}
where $a_s$, mimicking an `effective isothermal sound speed',
represents the velocity dispersion of the stellar SID and
$D_s$ is a dimensionless parameter for stellar SID rotation.
We invoke the expedient polytropic approximation
\begin{equation}\label{12}
\Pi^s_0=a^2_s\Sigma^s_0\
\end{equation}
to relate the two-dimensional pressure and the surface mass
density. The epicyclic frequency $\kappa_s$ of the stellar
disk is defined by
\begin{equation}\label{13}
\kappa^2_s\equiv
\frac{2\Omega_s}r\frac{d}{dr}(r^2\Omega_s)=2\Omega^2_s\ .
\end{equation}
In parallel, we have the angular
rotation rate of the gaseous MSID as
\begin{equation}\label{15}
\Omega_g(r)=a_g\frac{D_g}r\ ,
\end{equation}
where $a_g$ is the isothermal sound speed of the gaseous MSID and
$D_g$ is a dimensionless parameter for MSID rotation. The polytropic
relation and the definition of the epicyclic frequency $\kappa_g$
for the gaseous MSID are simply
\begin{equation}\label{16}
\Pi^g_0=a^2_g\Sigma^g_0\ ,
\end{equation}
\begin{equation}\label{17}
\kappa^2_g=2\Omega^2_g\ .
\end{equation}
To avoid the magnetic field winding dilemma in a rotating disc
(e.g. Lou \& Fan 1998a), the background coplanar magnetic field,
which is not force-free, is taken to be purely azimuthal about
the symmetry $\hat z-$axis:
\begin{equation}\label{18}
B_\theta(r)={\cal{F}}r^{-\frac{1}2}\ ,
\end{equation}
where ${\cal{F}}$ is a constant (Lou 2002) proportional
to the encircled magnetic flux within $r$, and
\begin{equation}\label{19}
B_r=B_z=0\ .
\end{equation}
In galactic model applications, one needs to invoke a central bulge
or other processes to avoid the divergence of $B_{\theta}$ as
$r\rightarrow 0$. From the radial momemtrum equations (\ref{2})
and (\ref{5}) in a rotational equilibrium and the fact that
$F\partial \phi/\partial r=2\pi G(\Sigma_0^s+\Sigma_0^g)$ by
Poisson integral (\ref{7}), one derives the following expressions
for the background surface mass densities, namely
\begin{equation}\label{20}
\Sigma_0^s=F\frac{a_s^2(1+D_s^2)}{2\pi Gr}\frac{1}{(1+\delta)}\ ,
\end{equation}
\begin{equation}\label{21}
\Sigma_0^g=F\frac{a_g^2(1+D_g^2)-C_A^2/2}{2\pi
Gr}\frac{\delta}{(1+\delta)}\ ,
\end{equation}
where $\delta\equiv{\Sigma^g_0}/{\Sigma^s_0}$ is the surface mass
density ratio of the two coupled background SIDs
and $C_A$ is the Alfv\'en wave speed in the MSID defined by
\begin{equation}\label{22}
C_A^2\equiv\int dzB_\theta^2/(4\pi\Sigma_0)\ .
\end{equation}
From equations (\ref{20}) and (\ref{21}), it then follows that
\begin{equation}\label{86}
a_s^2(1+D_s^2)=a_g^2(1+D_g^2)-C_A^2/2\ .
\end{equation}
Physically, condition (\ref{86}) results from the basic fact that
the same total gravitational force $\partial\phi/\partial r$,
including the contribution from the dark matter halo, acts on both
the stellar SID and magnetized gaseous SID, and is very useful in
our analysis below. It should be noted that the rotation rates of
the two SIDs are different in general (Lou \& Shen 2003; Shen \&
Lou 2003). In dimensionless form, condition (\ref{86}) can be
written in the form of either
\begin{equation}\label{87}
D_g^2=\beta(1+D_s^2)-1+\lambda^2/2
\end{equation}
or
\begin{equation}\label{90}
D_s^2=\frac{1}{\beta}(1+D_g^2-\lambda^2/2) -1\ ,
\end{equation}
where parameter $\beta\equiv a_s^2/a_g^2$ stands for the square of
the ratio of the stellar velocity dispersion to the sound speed of
the MSID, and parameter $\lambda^2\equiv C_A^2/a_g^2$ stands for
the square of the ratio of the Alfv\'en speed  to the sound speed
in the MSID. In disc galaxies, the stellar velocity dispersion
$a_s$ is usually higher than the sound speed $a_g$, we naturally
focus on the case of $\beta\geq1$ (e.g. Jog \& Solomon 1984a, b;
Bertin \& Romeo 1988; Jog 1996; Elmegreen 1995; Lou \& Fan 1998b;
Lou \& Shen 2003; Shen \& Lou 2003). In this $\beta\ge 1$ regime,
it follows from condition (\ref{90}) that
\begin{equation}\label{92}
1+D_s^2\ \leq\  1+D_g^2-\lambda^2/2\ <\ 1+D_g^2\ ,
\end{equation}
implying that for $\beta\geq1$,
\begin{equation}\label{93}
D_s^2<D_g^2\ .
\end{equation}
Inequality (\ref{93}) is very important to identify physically valid
mathematical solutions of $D_s^2$ for stationary MHD perturbations. 
For specified parameters $\beta$ and $\lambda^2$, $D_s^2$ and $D_g^2$ 
are related to each other linearly by condition (\ref{87}). We 
emphasize that mathematical solutions of $D_g^2$ and $D_s^2$ become 
unphysical for either $D_g^2<0$ or $D_s^2<0$ or both. The key here 
is that by inequality (\ref{93}), we only need to consider $D_s^2>0$ 
because $D_g^2$ must also be positive. In our analysis, we mainly use 
equation (\ref{87}) to derive a cubic algebraic equation in terms of 
$D_s^2$ and examine solution properties.

\subsection{Perturbations in a Composite MSID System}
For small coplanar MHD perturbations in a composite MSID system,
basic nonlinear equations (\ref{1})$-$(\ref{10}) can be linearized
in a straightforward manner, namely
\begin{equation}\label{23}
\frac{\partial\Sigma^s_1}{\partial t}+\frac{1}{r}
\frac{\partial (r\Sigma^s_0 u^s_1)}{\partial r}
+\Omega_s\frac{\partial\Sigma^s_1}{\partial\theta}
+\frac{\Sigma^s_0}{r^2}\frac{\partial j^s_1}{\partial\theta}=0\ ,
\end{equation}
\begin{equation}\label{24}
\frac{\partial u_1^s}{\partial t}
+\Omega_s\frac{\partial u_1^s}{\partial\theta}
-2\frac{\Omega_sj_1^s}{r}=-\frac{\partial}{\partial
r}\bigg(a^2_s\frac{\Sigma^s_1}{\Sigma^s_0}+\phi_1\bigg)\ ,
\end{equation}
\begin{equation}\label{25}
\frac{\partial j^s_1}{\partial
t}+\frac{r\kappa^2_s}{2\Omega_s}u_1^s+\Omega_s\frac{\partial
j^s_1}{\partial\theta}=-\frac{\partial}{\partial
\theta}\bigg(a^2_s\frac{\Sigma^s_1}{\Sigma^s_0}+\phi_1\bigg)\
\end{equation}
for coplanar hydrodynamic perturbations in a stellar SID, and
\begin{equation}\label{26}
\frac{\partial\Sigma^g_1}{\partial t}
+\frac{1}{r}\frac{\partial (r\Sigma^g_0 u_1^g)}
{\partial r}+\Omega_g\frac{\partial\Sigma^g_1}
{\partial\theta}+\frac{\Sigma^g_0}{r^2}
\frac{\partial j^g_1}{\partial\theta}=0\ ,
\end{equation}
\begin{equation}\label{27}
\begin{split}
\frac{\partial u_1^g}{\partial t}+\Omega_g\frac{\partial u_1^g}
{\partial\theta}-2\frac{\Omega_g j_1^g}{r}
=-\frac{\partial}{\partial r}
\bigg(a^2_g\frac{\Sigma^g_1}{\Sigma^g_0}+\phi_1\bigg)
\\
-\frac1{\Sigma^g_0}\int\frac{dzB_\theta}{4\pi r}
\bigg[\frac{\partial(rb_\theta)}{\partial r}
-\frac{\partial b_r}{\partial\theta}\bigg]
\\
+\frac{C_A^2\Sigma^g_1}{2\Sigma^g_0r}
-\frac1{\Sigma^g_0}\int\frac{dzb_\theta}
{4\pi r}\frac{\partial (rB_\theta)}{\partial r}\ ,
\end{split}
\end{equation}
\begin{equation}\label{28}
\begin{split}
\frac{\partial j^g_1}{\partial
t}+\frac{r\kappa^2_g}{2\Omega_g}u_1^g
+\Omega_g\frac{\partial j^g_1}{\partial\theta}
=-\frac{\partial}{\partial\theta}
\bigg(a^2_g\frac{\Sigma^g_1}{\Sigma^g_0}+\phi_1\bigg)
\\
+\frac1{\Sigma^g_0}\int\frac{dzb_r}{4\pi }
\frac{\partial (rB_\theta)}{\partial r}
\end{split}
\end{equation}
for coplanar MHD perturbations in a gaseous MSID,
\begin{equation}\label{29}
F\phi_1=-G\oint d\psi{\int_0}^\infty\frac{(\Sigma^g_1+\Sigma^s_1)
\zeta d\zeta}{[\zeta^2+r^2-2\zeta r\cos(\psi-\theta)]^{1/2}}
\end{equation}
for the linearized Poisson integral, and
\begin{equation}\label{30}
\frac{\partial(rb_r)}{\partial r}+\frac{\partial
b_\theta}{\partial \theta }=0\ ,
\end{equation}
\begin{equation}\label{31}
\frac{\partial b_r}{\partial t}=
\frac{1}{r}\frac{\partial}
{\partial\theta}(u_1^gB_\theta-r\Omega_g b_r)\ ,
\end{equation}
\begin{equation}\label{32}
\frac{\partial b_\theta}{\partial t}
=-\frac{\partial}{\partial r}
(u_1^gB_\theta-r\Omega_g b_r)
\end{equation}
for the linearized divergence-free condition and
the linearized magnetic induction equation.

We do not consider vertical variations along $z$ direction
across the
composite MSID system. With a harmonic $\exp(i\omega t-im\theta)$
dependence for all perturbation variables, we introduce
complex radial variations $\mu_s(r)$, $\mu_g(r)$, $U_s(r)$,
$U_g(r)$, $J_s(r)$, $J_g(r)$, $V(r)$, $R(r)$ and $Z(r)$ for $
\Sigma_1^s$, $\Sigma_1^g$, $u_1^s$, $u_1^g$, $j_1^s$, $j_1^g$,
$\phi_1$, $ b_r$ and $b_\theta$, respectively. Thus, hydrodynamic
equations (\ref{23})$-$(\ref{25}) can be reduced to the form of
\begin{equation}\label{33}
i(\omega-m\Omega_s)\mu_s+\frac{1}{r}\frac{\partial}{\partial r}
(r\Sigma_0^sU_s)-\frac{im\Sigma_0^s}{r^2}J_s=0\ ,
\end{equation}
\begin{equation}\label{34}
i(\omega-m\Omega_s)U_s-\frac{2\Omega_sJ_s}{r}
=-\frac{\partial\Phi_s}{\partial r}\ ,
\end{equation}
\begin{equation}\label{35}
i(\omega-m\Omega_s)J_s+\frac{r\kappa_s^2}{2\Omega_s}U_s=im\Phi_s\
\end{equation}
for the stellar SID, where $\Phi_s\equiv a_s^2\mu_s/\Sigma^s_0+V$.
Similarly, MHD equations (\ref{26})$-$(\ref{32}) can be cast into
the form of
\begin{equation}\label{36}
i(\omega-m\Omega_g)\mu_g+\frac{1}{r}\frac{\partial}
{\partial r}(r\Sigma_0^gU_g)-\frac{im\Sigma_0^g}{r^2}J_g=0\ ,
\end{equation}
\begin{equation}\label{37}
\begin{split}
i(\omega-m\Omega_g)U_g-\frac{2\Omega_gJ_g}{r}
=-\frac{\partial\Phi_g}{\partial r}
+\frac{C_A^2\mu_g}{2\Sigma_0^gr} \qquad\qquad\ \
\\
-\frac1{\Sigma_0^g}\int\frac{dzZ}{4\pi r}
\frac{\partial (rB_\theta)}{\partial r}
-\frac1{\Sigma_0^g}\int\frac{dzB_\theta}{4\pi r}
\bigg[\frac{\partial(rZ)}{\partial r}+imR\bigg]\ ,
\end{split}
\end{equation}
\begin{equation}\label{38}
i(\omega-m\Omega_g)J_g+\frac{r\kappa_g^2}{2\Omega_g}U_g
=im\Phi_g+\frac1{\Sigma^g_0}\int\frac{dzR}{4\pi}
\frac{\partial(rB_{\theta})}{\partial r}\ ,
\end{equation}
where $\Phi_g\equiv a_g^2\mu_g/\Sigma^g_0+V$,
\begin{equation}\label{39}
\frac{\partial (rR)}{\partial r}-imZ=0\ ,
\end{equation}
\begin{equation}\label{40}
i(\omega-m\Omega_g)R+\frac{imB_\theta}rU_g=0\ ,
\end{equation}
\begin{equation}\label{41}
i\omega Z=\frac{\partial}{\partial r}(r\Omega_g
R)-\frac{\partial}{\partial r}(B_\theta U_g)
\end{equation}
for coplanar MHD perturbations in the gaseous MSID. By setting
angular frequency $\omega=0$ in coplanar MHD perturbation 
equations (\ref{29}) and (\ref{33})$-$(\ref{41}), we can
construct global stationary
MHD perturbation configurations in a composite system of MSIDs
without invoking the WKBJ or tight-winding approximation and
analyze their properties.

\section{ALIGNED MHD CONFIGURATIONS}

Coplanar perturbations in a composite MSID system can be classified
as `aligned' and `unaligned' solutions (e.g. Kalnajs 1973; Shu et al. 
2000; Lou 2002; Lou \& Shen 2003). For aligned configurations, all 
streamlines and magnetic field lines are aligned in a composite MSID 
system. For unaligned spiral configurations, neighbouring streamlines 
shift relative to each other in a systematic manner (Kalnajs 1973); 
the same physical scenario holds true for neighbouring magnetic field
lines (Lou \& Fan 1998a).

In this section, we obtain the stationary dispersion relation for
aligned coplanar MHD perturbations (both full and partial MSIDs) by
perturbation equations (\ref{29}) and (\ref{33})$-$(\ref{41}) in the
preceding section. The solution behaviours and the corresponding
phase relationships are analyzed, mainly in the context of a full
SID system (i.e. $F=1$). At the end of this section, we derive the
MHD virial theorem for a composite MSID system and suggest the
onset criterion for secular bar-like instabilities in a composite
MSID system (Ostriker \& Peebles 1973; Binney \& Tremaine 1987;
Shu et al. 2000; Lou 2002).

\subsection{Dispersion Relation for Aligned Perturbations }

To construct stationary perturbation configurations of MSID that
are aligned, we set $\omega=0$ in equations (\ref{33})$-$(\ref{41}).
Let us first set $\partial/\partial t=0$ or $\omega=0$ in equations
(\ref{36})$-$(\ref{41}) for coplanar MHD perturbations in the MSID
to obtain
\begin{equation}\label{63}
m\Omega_g\mu_g+\frac1r\frac{\partial}{\partial r}
(r\Sigma_0^giU_g)+\frac{m\Sigma_0^g}{r^2}J_g=0\ ,
\end{equation}
\begin{equation}\label{64}
\begin{split}
m\Omega_giU_g+\frac{2\Omega_gJ_g}r=\frac{\partial\Phi_g}{\partial r}
-\frac{C_A^2\mu_g}{2\Sigma_0^gr}
+\frac{C_A^2miU_g}{\Omega_gr^2}\qquad\qquad
\\
-\frac{C_A^2}{2r^{1/2}}\frac{\partial}{\partial r}
\bigg(\frac {iU_g}{m\Omega_gr^{1/2}}\bigg)
-\frac{C_A^2}{r^{1/2}}\frac{\partial}{\partial r}
\bigg[r\frac{\partial}{\partial r}
\bigg(\frac{iU_g}{m\Omega_gr^{1/2}}\bigg)\bigg],\quad
\end{split}
\end{equation}
\begin{equation}\label{65}
m\Omega_gJ_g+\frac{r\kappa_g^2}{2\Omega_g}iU_g
=-m\Phi_g+\frac{C_A^2iU_g}{2\Omega_gr}\ ,
\end{equation}
\begin{equation}\label{66}
iR=\frac{B_\theta iU_g}{\Omega_gr}\ ,
\end{equation}
\begin{equation}\label{67}
Z=-\frac im\frac{\partial(rR)}{\partial r}\ .
\end{equation}
For aligned perturbations, we take the following
potential-density pair (Shu et al. 2000; Lou 2002; 
Lou \& Shen 2003)
\begin{equation}\label{47}
\mu_s\propto 1/r\ ,
\end{equation}
\begin{equation}\label{48}
\mu_g\propto1/r\ ,
\end{equation}
\begin{equation}\label{49}
V=-\frac{2\pi Gr}{|m|}(\mu_s+\mu_g)\ ,
\end{equation}
such that
\begin{equation}\label{50}
\Phi_g\equiv a_g^2\mu_g/\Sigma^g_0+V=\hbox{constant}.
\end{equation}
For a constant $iU_g$, as will be shown presently,
combinations of equations (\ref{63}) and (\ref{65})
and expressions (\ref{47})$-$(\ref{49}) give
\begin{equation}\label{68}
J_g=-\frac{\Omega_gr^2\mu_g}{\Sigma_0^g}\ ,
\end{equation}
\begin{equation}\label{69}
iU_g=m\frac{\Phi_g-\Omega_g^2r^2\mu_g/\Sigma_0^g}
{C_A^2/(2\Omega_gr)-\Omega_gr}\ .
\end{equation}
As $\Phi_g$ is constant by equation (\ref{50}), it is clear that
$iU_g$ is another constant. Consequently, equations (\ref{68})
and (\ref{64}) give
\begin{equation}\label{70}
\bigg[m\Omega_gr-\frac{C_A^2(m^2-1/2)}{m\Omega_gr}\bigg]iU_g
-\frac{2\Omega_g^2r^2\mu_g}{\Sigma_0^g}
+\frac{C_A^2\mu_g}{2\Sigma_0^g}=0\ .
\end{equation}
Substitutions of expressions (\ref{49}), (\ref{68}) and
(\ref{69}) into equation (\ref{70}) give a relation
between $\mu_g$ and $\mu_s$, namely
\begin{equation}\label{43}
\bigg(\frac{a_g^2-\Omega_g^2r^2}{\Sigma_0^g r}
-\frac{2\pi G}{|m|}+\frac {\mathcal K}{\mathcal A}\bigg)\mu_g
=\frac{2\pi G}{|m|}\mu_s\ ,
\end{equation}
where two coefficients $\mathcal K$
and $\mathcal A$ are defined by
\begin{equation}\label{44}
\mathcal K\equiv\frac{C_A^2/2-2\Omega_g^2r^2}{\Sigma_0^gr}
\end{equation}
and
\begin{equation}\label{45}
\mathcal A\equiv
\frac{m^2\Omega_g^2r^2-C_A^2(m^2-1/2)}{C_A^2/2-\Omega_g^2r^2}\ ,
\end{equation}
respectively. In parallel, we set $\omega=0$ in equations
(\ref{33})$-$(\ref{35}) for coplanar perturbations in the
stellar disc and use relations (\ref{47})$-$(\ref{49}) to
obtain\footnote{In view of the exchange symmetry between
the stellar and gas SIDs, relation (\ref{42}) can also be
obtained by simply setting $C_A=0$ in equation (\ref{43})
and switching subscripts $g$ and $s$.}
another relation between $\mu_s$ and $\mu_g$, namely
\begin{equation}\label{42}
\bigg(m^2\frac{a_s^2-\Omega_s^2r^2}{\Sigma^s_0r}-2\pi
G|m|+\frac{2\Omega_s^2r^2}{\Sigma^s_0r}\bigg)\mu_s
=2\pi G|m|\mu_g\ .
\end{equation}
Combining equations (\ref{42}) and (\ref{43}),
we derive the stationary dispersion relation
\begin{equation}\label{46}
\begin{split}
\frac{r^2}{m^2}\times\frac1{m^2\Omega_g^2r^2-C_A^2(m^2-1/2)}
\qquad\qquad\qquad\qquad\qquad\\
\times\bigg[(m^2-2)\Omega_s^2r^2
+2\pi G|m|\Sigma_0^sr-m^2a_s^2\bigg]
\qquad\qquad\qquad\\
\times\bigg\{m^4\Omega_g^4-\bigg[2\Omega_g^2
+\bigg(\frac{C_A^2+a_g^2}{r^2}
-\frac{2\pi G\Sigma^g_0}{|m|r}\bigg)m^2
-\frac{2C_A^2}{r^2}\bigg]
\qquad\\
\times m^2\Omega_g^2
+\frac{m^2C_A^2}{r^2}\bigg[\bigg(\frac{a_g^2}{r^2}
-\frac{2\pi G\Sigma^g_0}{|m|r}
\bigg)\bigg(m^2-\frac12\bigg)-\frac{C_A^2}{4r^2}\bigg]\bigg\}
\qquad\\
=4\pi^2G^2\Sigma_0^s\Sigma_0^g\qquad\qquad
\end{split}
\end{equation}
for aligned coplanar MHD perturbations in a composite MSID system. In
the absence of the mutual gravitational coupling between the MSID and
the stellar SID, represented by the term on the right-hand side of
equation (\ref{46}), the left-hand side of equation (\ref{46}) would
give rise to two separate dispersion relations, one for the stellar
SID and one for the gaseous MSID. The first one would be
\begin{equation}\label{71}
(m^2-2)\Omega_s^2r^2+2\pi G|m|\Sigma_0^sr-m^2a_s^2=0\ ,
\end{equation}
which gives the stationary dispersion relation for aligned coplanar
perturbations in a stellar SID alone, that is, a single SID without
magnetic field. Substituting expressions
(\ref{11}) and (\ref{20}) of $\Omega_s$ and $\Sigma_0^s$ (with $F=1$)
into equation (\ref{71}), we obtain
\begin{equation}\label{72}
(|m|-1)[D_s^2(|m|+2)-|m|]=0\ ,
\end{equation}
which is simply equation (26) of Shu et al. (2000).

The second factor in the curly braces on the
left-hand side of equation (\ref{46}) is
\begin{equation}\label{73}
\begin{split}
m^4\Omega_g^4-\bigg[2\Omega_g^2
+\bigg(\frac{C_A^2+a_g^2}{r^2}-\frac{2\pi G\Sigma^g_0}{|m|r}
\bigg)m^2-\frac{2C_A^2}{r^2}\bigg]m^2\Omega_g^2
\\
+\frac{m^2C_A^2}{r^2}\bigg[\bigg(\frac{a_g^2}{r^2}-\frac{2\pi
G\Sigma^g_0}{|m|r} \bigg)(m^2-\frac12)-\frac{C_A^2}{4r^2}\bigg]=0\ ,
\ \qquad
\end{split}
\end{equation}
which is the stationary dispersion relation for aligned coplanar
MHD perturbations in the gaseous MSID alone, that is, a single
MSID with a coplanar magnetic field. Equation (\ref{73}) is 
simply equation (3.2.6) of Lou (2002).

By the above results, it is clear that equation (\ref{46})
represents the dispersion relation for stationary MHD density
waves in a composite MSID system. As both SID and MSID rotate,
the stationarity of MHD density wave patterns in an inertial
frame of reference imposes conditions on dimensionless
rotation parameter $D_s^2$ (or equivalently, $D_g^2$). For the
following analysis, we substitute expressions (\ref{11}),
(\ref{15}), (\ref{20}) and (\ref{21}) into equation (\ref{46})
to yield another form of stationary dispersion relation in a
composite full or partial MSID system, namely
\begin{equation}\label{89}
\begin{split}
\frac{1}{m^2}\times\frac1{m^2D_g^2-\lambda^2(m^2-1/2)}
\qquad\qquad\qquad\qquad\\
\times\bigg[(m^2-2)D_s^2 +F|m|\frac{1+D_s^2}{1+\delta}-m^2\bigg]
\times\bigg\{m^4D_g^4-\bigg[2D_g^2+
\\
\bigg(\lambda^2+1-F\frac{1+D_g^2-\lambda^2/2}{|m|}\frac{\delta}
{1+\delta}\bigg)m^2-2\lambda^2\bigg]\times m^2D_g^2
\qquad\\
+\bigg[\bigg(1-F\frac{1+D_g^2-\lambda^2/2}{|m|}\frac{\delta}
{1+\delta}\bigg)\bigg(m^2-\frac12\bigg)-\frac{\lambda^2}{4}\bigg]
\qquad\\
\times
m^2\lambda^2\bigg\}-F^2\frac{(1+D_s^2)(1+D_g^2-\lambda^2/2)\delta}
{(1+\delta)^2}=0\ .\qquad\qquad
\end{split}
\end{equation}
A substitution of expression (\ref{87}) for $D_g^2$ into (\ref{89})
gives a cubic algebraic equation in terms of $D_s^2$. Or equivalently,
a substitution of expression (\ref{90}) for $D_s^2$ into (\ref{89})
would yield a cubic algebraic equation in terms of $D_g^2$. As
noted earlier, we focus on the cubic equation of $D_s^2$ to identify
physical solutions of $D_s^2\ge 0$, because $D_s^2<D_g^2$ as a result
of $a_s^2>a_g^2$ in typical disc galaxies.

For a later examination of spatial phase relationship between
azimuthal magnetic field perturbation $b_{\theta}$ and the
surface mass density perturbation $\Sigma_1^g$ of the gaseous
MSID, we combine equations (\ref{66}) and (\ref{67}) to obtain
\begin{equation}\label{99}
Z=-\frac{iU_gB_\theta}{2m\Omega_gr}\ ,
\end{equation}
relating the $\theta$-component of the magnetic field
perturbation $Z$ and the radial gas flow speed perturbation
$iU_g$. Using equations (\ref{70}) and (\ref{99}), one can
eliminate $iU_g$ to obtain
\begin{equation}\label{100}
Z=-\frac{B_{\theta}\mu_g}{2\Sigma_0^g}
\frac{(2\Omega_g^2r^2-C_A^2/2)}
{[m^2\Omega_g^2r^2-C_A^2(m^2-1/2)]}\ ,
\end{equation}
that can be further reduced to
\begin{equation}\label{101}
\frac{\mu_g}{Z}=-\frac{2\Sigma_0^g}{B_{\theta}}
\frac{[m^2D_g^2-\lambda^2(m^2-1/2)]}{2D_g^2-\lambda^2/2}\ .
\end{equation}
For real $\mu_g/Z$, the sign of the right-hand side of the
equation (\ref{101}) will determine the phase relationship
between the azimuthal magnetic field perturbation $b_{\theta}$
and the gas surface mass density perturbation $\Sigma_1^g$ in
the gaseous MSID. That is, $b_{\theta}$ and $\Sigma_1^g$ are
in and out of phase for a positive and negative right-hand
side of the equation (\ref{101}), respectively.

\subsection{Axisymmetric Disturbances with $|m|=0$}

For aligned axisymmetric disturbances with $|m|=0$, it would be
inappropriate to directly use relation (\ref{46}) or (\ref{89}).
One should carefully examine equations (\ref{33})$-$(\ref{41})
with $\omega=m=0$. With $U_g=U_s=R=0$, equations (\ref{33}),
(\ref{35}), (\ref{36}), (\ref{38}), (\ref{39}) and (\ref{41})
can be satisfied, and equation (\ref{40}) is identically zero.
By choosing $Z\propto r^{-1/2}$, $\mu_g\propto r^{-1}$,
$\mu_s\propto r^{-1}$, $J_g\propto r$, $J_s\propto r$,
$V\propto\ln r$, $\Phi_g\propto\ln r+$ constant and
$\Phi_s\propto\ln r+$ constant, the two remaining equations
(\ref{34}) and (\ref{37}) are consistent with a rescaling of the
axisymmetric background. In the present context, this rescaling 
is somewhat trivial. We turn to cases of $|m|\geq 1$ below.

\subsection{Nonaxisymmetric Disturbances with $|m|\geq1$ }

\subsubsection{Behaviours of $D_s^2$ solutions }

We first come to the aligned case of $|m|=1$. For a full composite
MSID system with $F=1$, it is easy to verify that equation (\ref{89})
can be satisfied for arbitrary $D_s^2$. This is quite similar to cases
of a single SID studied by Shu et al. (2000), of a single MSID studied
by Lou (2002) and a composite SID system studied by Lou \& Shen (2003).
However, we note that for a single
partial MSID (Lou 2002) and for a composite system of two coupled
partial SIDs (Lou \& Shen 2003), such stationary aligned eccentric
$|m|=1$ perturbations are not allowed for arbitrary $D_s^2$. Likewise,
in our case of a composite partial MSID system with $F<1$, it is easy
to see that equation (\ref{89}) can no longer be satisfied for
arbitrary $D_s^2$ by simply setting $|m|=1$.

In the following, we analyze cases of $|m|\geq 2$ and focus on the
special case of $F=1$ for a full composite MSID system. As mentioned
earlier, we work in the parameter regime of $\beta\geq 1$ with a
typical disc galaxy in mind. By an extensive numerical exploration
of aligned cases from $|m|=2$ to $|m|=5$, we note empirically that
cases of $|m|>2$ are fairly similar to the case of $|m|=2$. For
this reason, we shall mainly consider the case of $|m|=2$. As
noted earlier, we can substitute expression (\ref{87}) for $D_g^2$
into equation (\ref{89}) to derive a cubic algebraic equation of
$y\equiv D_s^2$ as
\begin{eqnarray}\label{91}
\mathcal A+\mathcal B y+\mathcal C {y}^{2}+\mathcal D{y}^{3}=0\ ,
\end{eqnarray}
where $|m|=2$ and $F=1$, and the four coefficients $\mathcal A$,
$\mathcal B$, $\mathcal C$ and $\mathcal D$ are explicitly
defined by
\begin{eqnarray}
\mathcal
A\equiv24-32\,\beta-3\,{\lambda}^{4}+12\,{\lambda}^{2}\delta
+48\,\delta-80\,\beta\,\delta
\nonumber\\
-6\,\beta\,{\lambda}^{2}\delta+8\,{\beta}^{2}+6\,{\lambda}^{2}
+32\,{\beta}^{2}\delta-6\,{\lambda}^{4}\delta\nonumber\ ,
\end{eqnarray}
\begin{eqnarray}
\mathcal B\equiv -12\,{
\lambda}^{2}-3\,\beta\,{\lambda}^{2}\delta-48+32\,\beta
+48\,{\beta}^{2}\delta
\nonumber\\
-40\,\beta\,\delta-24\,\delta+6\,{\lambda}^{4}-6\,{\lambda}^{2}
\delta+3\,{\lambda}^{4}\delta\nonumber\ ,
\end{eqnarray}
\begin{eqnarray}
\mathcal C\equiv3\,\beta\,{\lambda}^{2}\delta
+40\,\beta\,\delta+64\,\beta-24\,{\beta}^{2}\nonumber
\end{eqnarray}
and
\begin{eqnarray}
\mathcal D\equiv-16\,{\beta}^{2}\delta -16\,{\beta}^{2} \nonumber\ ,
\end{eqnarray}
respectively.

\begin{figure}
\begin{center}
\includegraphics[angle=0,scale=0.45]{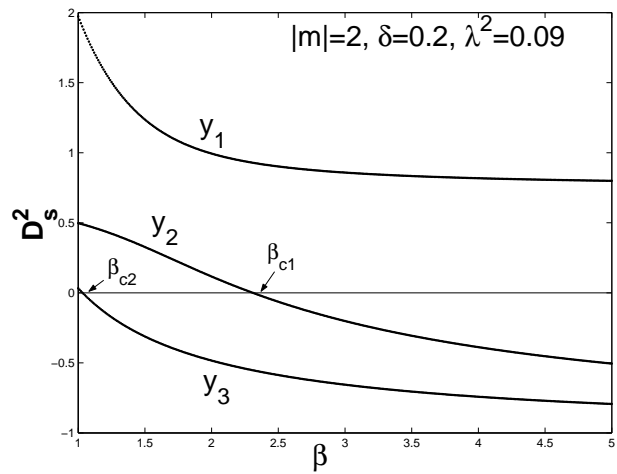}
\caption{\label{f11}Three solution curves of $D_s^2$ versus $\beta$
for the aligned case with $|m|=2$, $\delta=0.2$ and $\lambda^2=0.09$.}
\end{center}
\end{figure}

\begin{figure}
\begin{center}
\includegraphics[angle=0,scale=0.45]{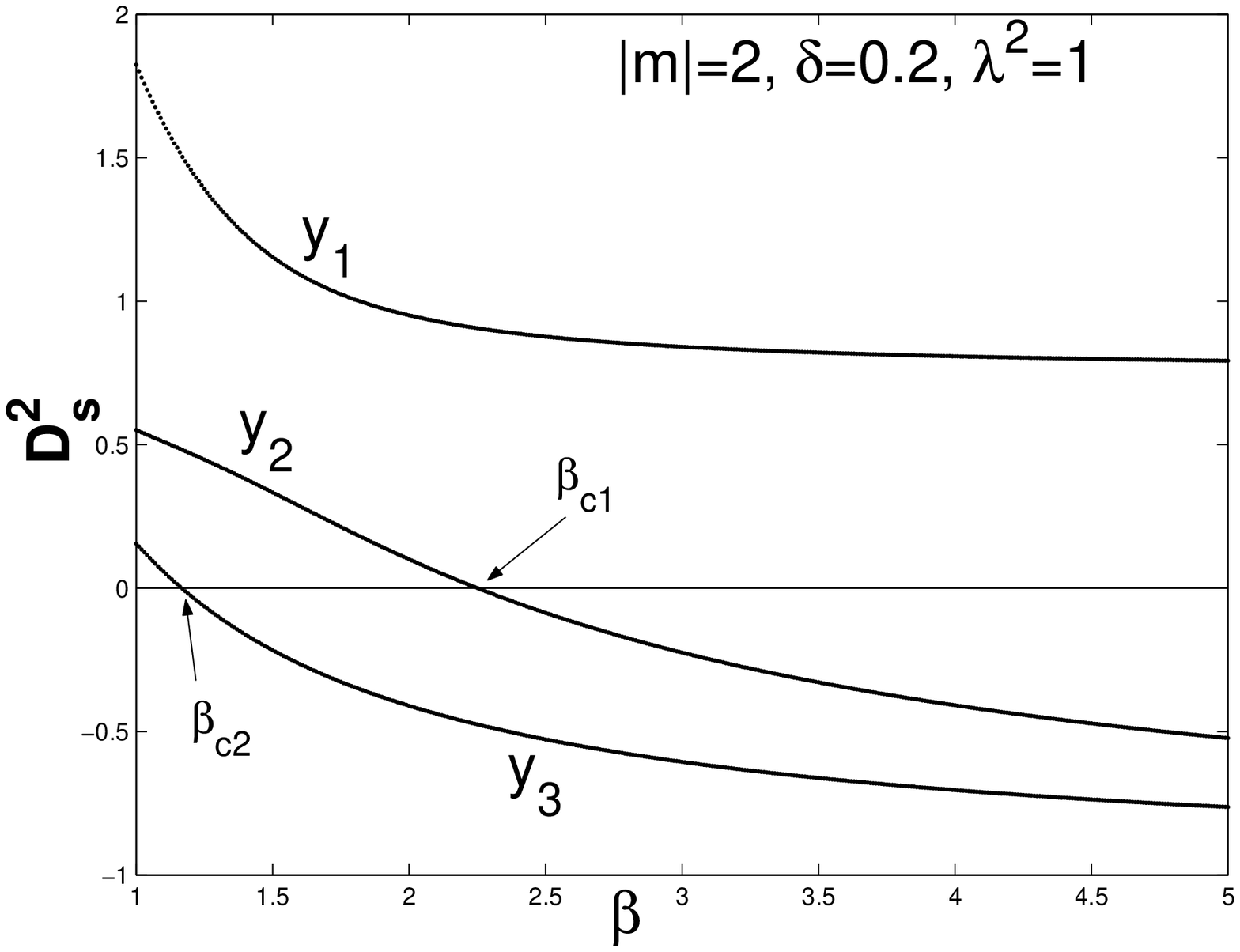}
\caption{\label{f12}Three solution curves of $D_s^2$ versus $\beta$
for the aligned case with $|m|=2$, $\delta=0.2$ and $\lambda^2=1$.}
\end{center}
\end{figure}

\begin{figure}
\begin{center}
\includegraphics[angle=0,scale=0.45]{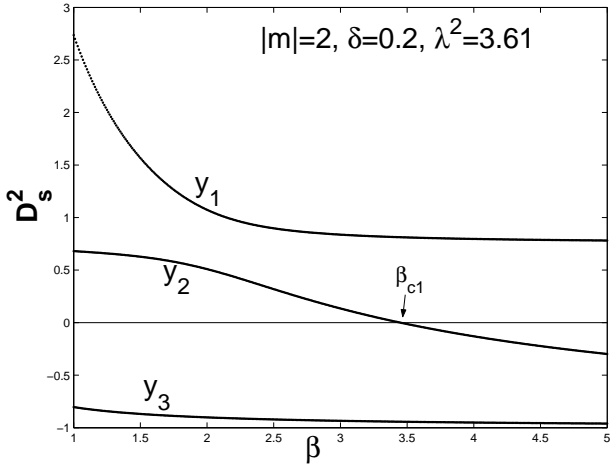}
\caption{\label{f13}Three solution curves of $D_s^2$ versus $\beta$
for the aligned case with $|m|=2$, $\delta=0.2$ and $\lambda^2=3.61$.}
\end{center}
\end{figure}

Formally, three mathematical solutions for $y\equiv D_s^2$ can
be written out analytically from cubic equation (\ref{91}) or
from the more general cubic dispersion relation (\ref{89}) with
$|m|>2$ and $F<1$, although the solution expressions are fairly
involved (see Appendix C for details).
Practically, we explore numerically various
parameter regimes for the three $D_s^2$ solutions.
For specified values of parameters $\delta$ and $\lambda^2$, we
show three mathematical solutions $y\equiv D_s^2$ versus $\beta$
in Figs. \ref{f11}$-$\ref{f13}.

For typical parameters, the three solution branches of $D_s^2$ do
not intersect with each other. For the convenience of discussion,
we use $y_1$, $y_2$ and $y_3$ to denote the upper, middle and lower
solution branches, respectively. Generally speaking, the upper two
solution branches, $y_1$ and $y_2$, are qualitatively similar to
those in a composite unmagnetized SID system (Lou \& Shen 2003),
that is, $y_1$ branch remains always positive, while $y_2$ decreases
monotonically with increasing $\beta$ and becomes negative when
$\beta$ exceeds a certain critical value $\beta_{c1}$. As expected
for small $\lambda^2$ (i.e. weak magnetic field; see Fig. \ref{f11}),
$y_1$ and $y_2$ respectively obey the limits of those in a composite
SIDs without magnetic field [see equations (54) and (55) of Lou \&
Shen (2003)], showing that as $\lambda^2\rightarrow 0$, $y_1$ and
$y_2$ consistently approach the upper and lower branches respectively
of Lou \& Shen (2003). One novel feature is the lowest solution branch
$y_3$ owing to the presence of magnetic field. In most cases, $y_3$ is
negative when $\beta$ varies from $1$ to $+\infty$, but for some
special parameters, $y_3$ may become positive when $\beta$ becomes
smaller than a critical value $\beta_{c2}$. Analytical expressions of
$\beta_{c1}$ and $\beta_{c2}$ for the case of $|m|=2$ can be derived
from a quadratic equation and are given by the following pair
\begin{equation}\label{94}
\begin{split}
\beta_{c1}\equiv \,{\frac1{8+32\,\delta}}
[40\,\delta+16+3\,{\lambda}^{2}\delta
\qquad\qquad\qquad\qquad\\
+\,(64\,{\delta}^{2}+128\,\delta-144\,{\lambda}^{2}{\delta}^{2}
+64-192\,{\lambda}^{2}\delta
\qquad\\
+201\,{\lambda}^{4}{\delta}^{2}+24\,{\lambda}^{4}+
144\,{\lambda}^{4}\delta-48\,{\lambda}^{2})^{1/2}]\quad
\end{split}
\end{equation}
and
\begin{equation}\label{95}
\begin{split}
\beta_{c2}\equiv \,{\frac1{8+32\,\delta}}
[40\,\delta+16+3\,{\lambda}^{2}\delta
\qquad\qquad\qquad\qquad\\
-\,(64\,{\delta}^{2}+128\,\delta-144\,{\lambda}^{2}{\delta}^{2}
+64-192\,{\lambda}^{2}\delta
\qquad\\
+201\,{\lambda}^{4}{\delta}^{2}+24\,{\lambda}^{4}+
144\,{\lambda}^{4}\delta-48\,{\lambda}^{2})^{1/2}]\ ,\quad
\end{split}
\end{equation}
respectively, where $\beta_{c1}$ remains always larger than
$\beta_{c2}$. We note the following. First, there is no
essential mathematical difficulty of obtaining more general
forms of $\beta_{c1}$ and $\beta_{c2}$ for arbitrary $|m|$
values with $0<F<1$. Secondly, by setting $\lambda^2=0$ in
expression (\ref{94}) for $\beta_{c1}$, we obtain
\begin{equation}\label{96}
\beta_{c1}=\frac32\bigg(1+\frac1{4\delta+1}\bigg)\ ,
\end{equation}
consistent with the case of $|m|=2$ in expression (56) for $\beta_c$
by Lou \& Shen (2003). Finally, while $\beta_{c2}$ is real in many
cases [i.e. a positive determinant in both expressions (\ref{94})
and (\ref{95})], $\beta_{c2}$ may be too small to be discernible in
the parameter regime of $\beta\geq 1$. For some special parameters
specified, $\beta_{c2}$ becomes noticeable. For example, in Fig.
\ref{f12} with $\delta=0.2$ and $\lambda^2=1$, we have
$\beta_{c2}=1.1667>1$ according to expression (\ref{95}). In this
case, for $\beta$ smaller than 1.1667 (still quite restrictive),
there exist three positive solutions of $D_s^2$.  For
$\beta_{c2}<\beta<\beta_{c1}$, the upper two branches $y_1$ and
$y_2$ are positive, corresponding to two possible stationary
perturbation modes. When $\beta$ exceeds $\beta_{c1}$, only $y_1$
remains positive, corresponding to one possible stationary
perturbation mode. For the usual case of $\beta_{c2}<1$, there are
at most two possible stationary modes when $\beta\rightarrow1$ (see
Fig. \ref{f13}).

\subsubsection{Phase relationships among perturbation variables}

\begin{figure}
\begin{center}
\includegraphics[scale=0.45]{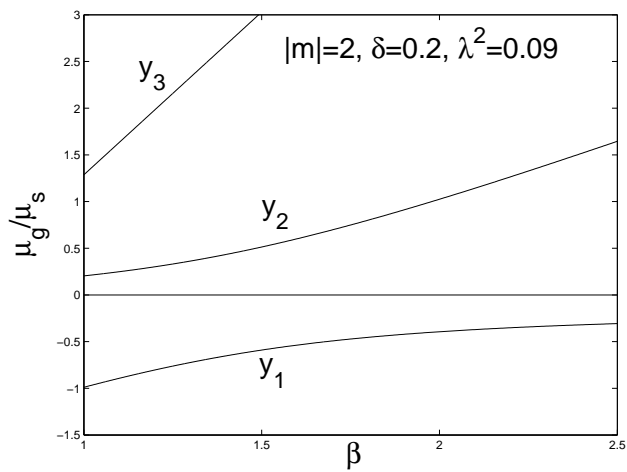}
\caption{\label{f16}Curves of $\mu_g/\mu_s$ versus $\beta$ for the
aligned $|m|=2$ case with $\delta=0.2$ and $\lambda^2=0.09$. }
\end{center}
\end{figure}

\begin{figure}
\begin{center}
\includegraphics[scale=0.45]{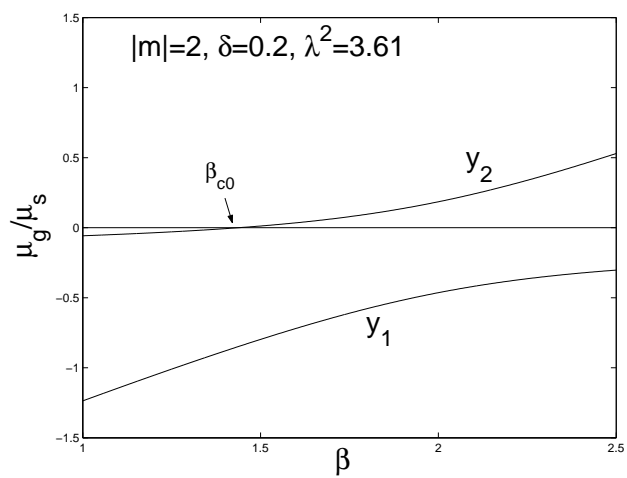}
\caption{\label{f18}Curves of $\mu_g/\mu_s$ versus $\beta$ for the
aligned $|m|=2$ case with $\delta=0.2$ and $\lambda^2=3.61$. The
uppermost branch related to $y_3$ is only discernible for $\beta<1$
and is not shown here. Specifically, this is caused by a negative
$y_3$ when $\beta>1$ (see Fig. \ref{f13}).}
\end{center}
\end{figure}

We now examine phase relationships among the azimuthal magnetic
field and the surface mass density perturbations, because they
may provide clues for magnetized spiral galaxies through optical
and synchrotron radio observations (e.g. Mathewson et al. 1972;
Beck \& Hoernes 1996; Fan \& Lou 1996; Lou \& Fan 1998a, 2002,
2003; Frick et al. 2000, 2001; Lou 2002; Lou et al. 2002). For
the phase relationship between the two surface mass density
perturbations $\mu_g$ and $\mu_s$, a combination of expressions
(\ref{11}), (\ref{20}) and equation (\ref{42}) gives
\begin{equation}\label{97}
\frac{\mu_g}{\mu_s}
=-1-{\frac {\left[y \left({m}^{2}-2\right)-{m}^{2}\right]
\left(1+\delta \right)}{\left|m\right|\left(y+1\right)}}\ 
\end{equation}
where $y\equiv D_s^2$. A substitution of expression (\ref{97})
into equation (\ref{91}) leads to a cubic algebraic equation
of $\mu_g/\mu_s$. We show different curves of $\mu_g/\mu_s$
versus $\beta$ in Figs. \ref{f16} and \ref{f18} by specifying
different values of $\delta$ and $\lambda^2$.

\begin{figure}
\begin{center}
\includegraphics[angle=0,scale=0.45]{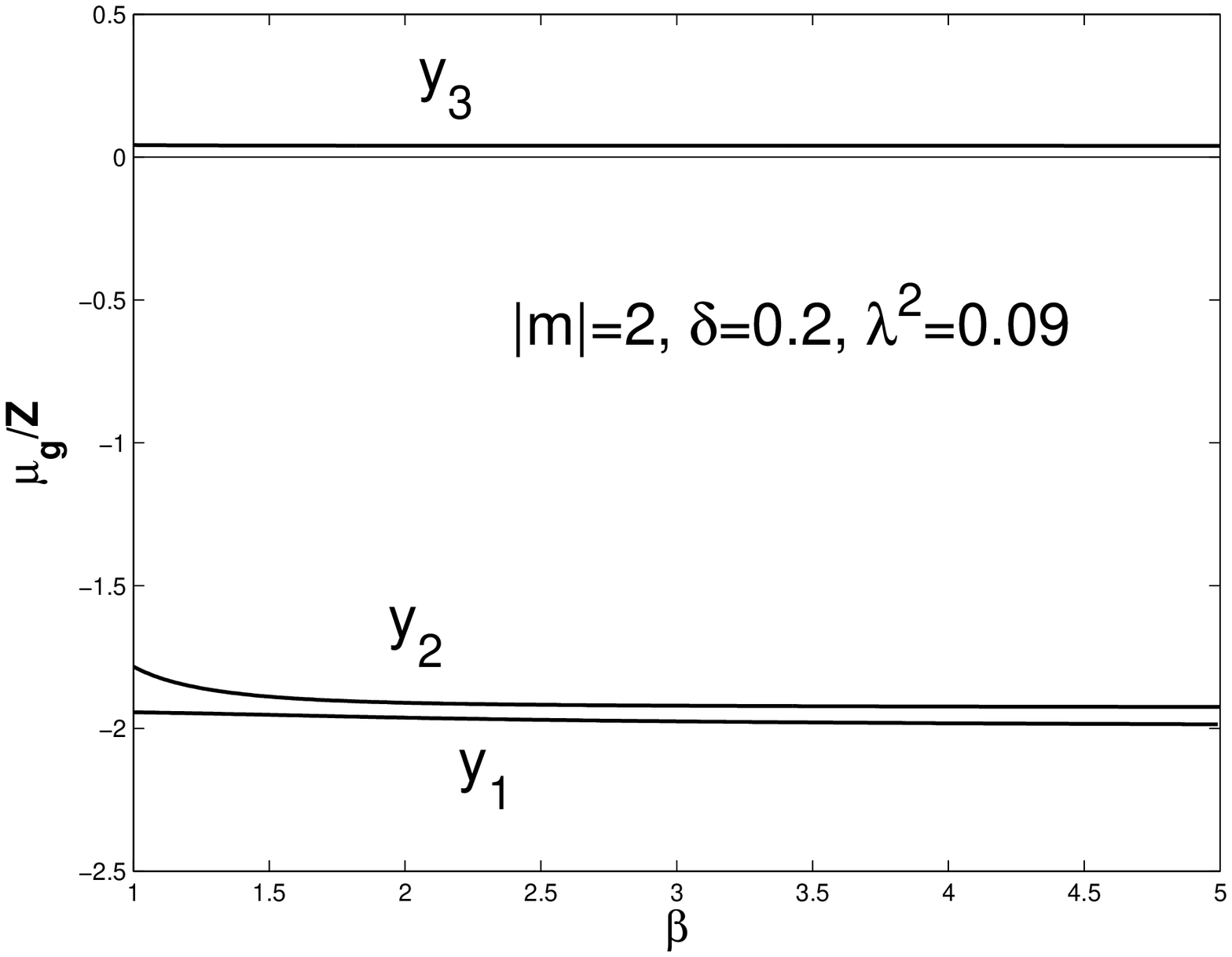}
\caption{\label{f20}Curves of $\mu_g/Z$ versus $\beta$ for the
aligned $|m|=2$ case with $\delta=0.2$ and $\lambda^2=0.09$. }
\end{center}
\end{figure}

\begin{figure}
\begin{center}
\includegraphics[angle=0,scale=0.45]{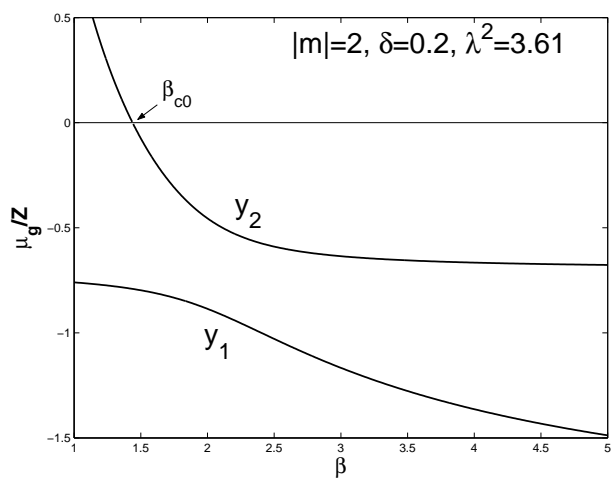}
\caption{\label{f22}Curves of $\mu_g/Z$ versus $\beta$ for the
aligned $|m|=2$ case with $\delta=0.2$ and $\lambda^2=3.61$.
The uppermost branch related to $y_3$ appears somewhat flat
at about $44$ and is not shown here. }
\end{center}
\end{figure}

In parallel, we examine the phase relationship between the
azimuthal magnetic field perturbation $b_{\theta}$ and the
surface mass density perturbation $\Sigma_1^g$ in the gaseous
MSID. By equation (\ref{101}), we introduce a dimensionless
$q$ parameter
\begin{equation}\label{102}
q\equiv-\frac{[m^2D_g^2-\lambda^2(m^2-1/2)]}{2D_g^2-\lambda^2/2}\ ,
\end{equation}
whose sign determines the phase relationship between $Z$ and
$\mu_g$. Using equation (\ref{102}), we may express $D_g^2$
as a function of $q$, namely
\begin{equation}\label{103}
D_g^2={\frac {{\lambda}^{2}\left(q+2\,{m}^{2}-1\right)}
{4\,q+2{m}^{2 }}}\ .
\end{equation}
A combination of equation (\ref{89}) and expression (\ref{90})
gives a cubic algebraic equation of $D_g^2$; and a substitution
of expression (\ref{103}) into the resulting cubic equation of
$D_g^2$ leads to a cubic equation in terms of $q$ (with $|m|=2$).
We present different curves of $q$ versus $\beta$ for different
values of $\delta$ and $\lambda^2$ in Figs. \ref{f20}$-$\ref{f22}.

From expressions (\ref{97}) and (\ref{102}), one can show that both
larger $\mu_g/\mu_s$ and $\mu_g/Z$ correspond to smaller $y$. The
lowest branches of $\mu_g/\mu_s$ or $\mu_g/Z$ are then related to the
uppermost $y_1$; the middle branches of $\mu_g/\mu_s$ or $\mu_g/Z$
are related to the middle $y_2$; and the uppermost branches of
$\mu_g/\mu_s$ or $\mu_g/Z$ are related to the lowest $y_3$ as shown
in Figs. \ref{f16}$-$\ref{f22}.

Several interesting features are noted here for these curves.
First, when $\lambda^2$ is small (see Fig. \ref{f16}), the
lowest and middle branches of $\mu_g/\mu_s$ respectively
follow those in a composite SID system without magnetic fields
[see equations (58) and (59) of Lou \& Shen (2003)], showing
consistently that as $\lambda^2\rightarrow 0$, the middle and
lowest branches of $\mu_g/\mu_s$ correspond to the lower and
upper branches of Lou \& Shen (2003), respectively.

Secondly, $\mu_g/\mu_s$ and $\mu_g/Z$ of each solution branch of $y$
bear definite signs in most cases. For the case of $\delta=0.2$ and
$\lambda^2=0.09$ as shown in Figs. \ref{f11}, \ref{f16} and \ref{f20},
the stationary perturbation mode of $y_1$ branch has a phase
relationship between the surface mass density of the two SIDs (i.e.
$\mu_g/\mu_s$) and a phase relationship between the surface mass
density of the MSID and the azimuthal magnetic field (i.e. $\mu_g/Z$)
being both always out of phase. In most cases, the stationary
perturbation mode of $y_2$ branch has a $\mu_g/\mu_s$ being in phase
(see Fig. \ref{f16}) and a $\mu_g/Z$ being out of phase (see
Fig.\ref{f20}). The stationary perturbation mode of $y_3$ branch has
$\mu_g/\mu_s$ and $\mu_g/Z$ being both always in phase (see Figs.
\ref{f16} and \ref{f20}).

Thirdly, for the stationary perturbation mode of $y_2$ branch, both
$\mu_g/\mu_s$ and $\mu_g/Z$ have a common zero point at $\beta_{c0}$
that increases with increasing $\lambda^2$. For a sufficiently large
$\lambda^2$ (e.g. $\lambda^2=3.61$), $\beta_{c0}$ becomes greater than
$1$. Thus, $\mu_g/\mu_s$ and $\mu_g/Z$ may become zero or even carry
opposite signs (see Figs. \ref{f18} and \ref{f22}). Our analysis show
that the middle $y_2$ branch for $\mu_g/\mu_s$ and $\mu_g/Z$ has a
common root at
\begin{equation}\label{98}
\beta_{c0}
=\,{\frac {(6+3\delta)\lambda^2
+8\,\delta+16}{24(1+\delta)}}\ .
\end{equation}
For a fixed $\delta$, $\beta_{c0}$ increases with increasing
$\lambda^2$. Specifically for $\delta=0.2$ and $\lambda^2=0.09$
in expression (\ref{98}), we have $\beta_{c0}=0.6317$ smaller
than 1. Therefore, the middle $y_2$ branch remains always positive
(see Fig. \ref{f16}). On the other hand, for $\lambda^2=3.61$
in expression (\ref{98}), we have $\beta_{c0}=1.4384$ larger
than 1 (see Fig. \ref{f18}).

It is not surprising that the middle $y_2$ branch of $\mu_g/\mu_s$
and $\mu_g/Z$ has a common root as result of $\mu_g=0$. The special
$\mu_g=0$ case may be a possible situation in our model. By equation
(\ref{68}), it follows that $J_g$ vanishes. As $\mu_s$ and $Z$ do
not vanish, we see that $V$, $\Phi_g$ and consequently $U_g$ do
not vanish by expressions (\ref{49}), (\ref{50}) and (\ref{69}).

Finally, our analysis show that the uppermost branch of $\mu_g/\mu_s$
and $\mu_g/Z$ corresponding to the $y_3$ branch has a root at $\beta=0$.
It follows that $y_3$ remains always positive.

\subsection{Secular Barlike Instabilities}

For the onset of aligned secular barlike instabilities in a
composite MSID system, we here derive the MHD virial theorem
for a composite MSID system (either full or partial) from the
background radial equilibrium conditions (\ref{2}) and (\ref{5})
\begin{equation}\label{104}
-\Sigma_0^sr\Omega_s^2=-\frac{d}{dr}(a_s^2\Sigma^s_0)
-\Sigma_0^s\frac{d\phi_0}{dr}
\end{equation}
and
\begin{equation}\label{105}
-\Sigma_0^gr\Omega_g^2=-\frac{d}{dr}(a_g^2\Sigma^g_0)
-\Sigma_0^g\frac{d\phi_0}{dr}-\Sigma_0^g\frac{C_A^2}{2r}\ .
\end{equation}
Adding equations (\ref{104}) and (\ref{105}), we readily obtain
\begin{equation}\label{106}
(\Sigma_0^s\Omega_s^2+\Sigma_0^g\Omega_g^2)r
=\frac{d}{dr}(a_s^2\Sigma^s_0+a_g^2\Sigma^g_0)
+(\Sigma_0^s+\Sigma_0^g)\frac{d\phi_0}{dr}
+\Sigma_0^g\frac{C_A^2}{2r}.
\end{equation}
Multiplying equation (\ref{106}) by $2\pi r^2dr$ and integrating
from $0$ to a finite radius $R$, that is allowed to approach
infinity eventually, we have the following MHD virial theorem.
\begin{equation}\label{107}
2(\mathcal{T}+\mathcal{U})+\mathcal{W}-\mathcal{M}
=2\pi R^2[a_s^2\Sigma_0^s(R)+a_g^2\Sigma_0^g(R)]
\end{equation}
within radius $R$, where
\begin{equation}\label{122}
\mathcal T\equiv\int_0^R\frac12\Sigma_0^s(r\Omega_s)^2
2\pi rdr+\int_0^R\frac12\Sigma_0^g(r\Omega_g)^22\pi rdr
\end{equation}
is the total rotational kinetic energy of the two SIDs,
\begin{equation}\label{123}
\mathcal
W\equiv-\int_0^Rr(\Sigma_0^s+\Sigma_0^g)\frac{d\phi_0}{dr}2\pi rdr
\end{equation}
is the gravitational energy in the composite MSID system,
\begin{equation}\label{124}
\mathcal U\equiv\int_0^R(a_s^2\Sigma_0^s+a_g^2\Sigma_0^g)2\pi rdr
\end{equation}
is the sum of stellar and gas internal energies and
\begin{equation}\label{125}
\mathcal M\equiv\int_0^R\Sigma_0^g\frac{C_A^2}{2r}2\pi r^2dr
\end{equation}
is the magnetic energy contained in the gaseous MSID component.

For a full composite system with $F=1$, we combine relations
(\ref{11}), (\ref{15}), (\ref{20}) and (\ref{21}) with
expressions (\ref{122})$-$(\ref{125}) to obtain
\begin{equation}\label{108}
\begin{split}
\mathcal T=\int_0^R\frac12\Sigma_0^s(r\Omega_s)^22\pi
rdr+\int_0^R\frac12\Sigma_0^g(r\Omega_g)^22\pi rdr
\\
=a_s^4(D_s^2+1)
\frac{D_s^2+[D_s^2+1+\lambda^2/(2\beta)-1/\beta]\delta}
{2G(1+\delta)}R\ ,
\end{split}
\end{equation}

\begin{eqnarray}\label{109}
\mathcal W=-\int_0^Rr(\Sigma_0^s+\Sigma_0^g)
\frac{d\phi_0}{dr}2\pi rdr
= -\frac{a_s^4(D_s^2+1)^2}{G}R\ ,
\end{eqnarray}

\begin{equation}\label{111}
\mathcal M=\int_0^R\Sigma_0^g\frac{C_A^2}{2r}2\pi r^2dr
=\frac{\lambda^2a_4^2(1+D_s^2)\delta}{2G\beta(1+\delta)}R\ .
\end{equation}
Based on early numerical simulations (Hohl 1971; Miller et al. 1970),
it was known that a thin self-gravitating disc in rotation may rapidly
evolve into bar configurations (e.g. Binney \& Tremaine 1987).
Ostriker \& Peebles (1973) suggested an approximate criterion
$\mathcal T/|\mathcal W|\leq 0.14\pm 0.02$, necessary but not
sufficient, for stability against bar-type instabilities, on the
basis of their $N-$body numerical explorations involving 300 particles.
In the presence of a coplanar azimuthal magnetic field with a radial
scaling of $\propto r^{-1/2}$, Lou (2002) proposed that the ratio
$\mathcal T/|\mathcal W-\mathcal M|$ may play the role of
$\mathcal T/|\mathcal W|$ in an unmagnetized SID to determine
the onset of instability in a single MSID. By the above analogy and a
natural extension, we use $\mathcal T/|\mathcal W-\mathcal M|$ instead
of $\mathcal T/|\mathcal W|$ to examine the onset criterion of
instability in a full composite MSID system in the presence of coplanar
non-axisymmetric aligned MHD perturbations. Using expressions (\ref{108}),
(\ref{109}) and (\ref{111}), we arrive at
\begin{eqnarray}\label{112}
\frac{\mathcal T}{|\mathcal W-\mathcal M|}={\frac
{-2\,\delta+2\,\delta\,\beta+\delta\,{\lambda}^{2}+
\left( 2\,\delta\,\beta+2\,\beta \right)
D_s^2}{2\,\delta\,{\lambda}^{2}+4\,\beta+4 \,\delta\,\beta+ \left(
4\,\beta+4\,\delta\,\beta \right) D_s^2}}\nonumber\\
=\frac12-{\frac {\beta+\delta}{\delta\,{\lambda}^{2}+
2\beta(1+\delta)(D_s^2+1)}}\ ,
\end{eqnarray}
indicating that the value of $\mathcal T/|\mathcal W-\mathcal M|$
falls between $0$ and $1/2$ as in the unmagnetized case (Lou \& Shen
2003) and increases with increasing $D_s^2$. Therefore in a composite
MSID system, the three possible values of $D_s^2$ correspond to three
different values of $\mathcal T/|\mathcal W-\mathcal M|$ ratio; and
larger values of $D_s^2$ correspond to higher
$\mathcal T/|\mathcal W-\mathcal M|$ ratios.

We here examine a few cases to illustrate the utility of
our proposed onset criterion for instability. First, we set $|m|=2$,
$\delta=0.2$, $\beta=1$ and $\lambda^2$=1. From equation (\ref{91}),
one obtains three $D_s^2$ solutions $D_s^2=1.8246$,
$D_s^2=0.5513$ and $D_s^2=0.1553$. Insertions of these three values
of $D_s^2$ into expression (\ref{112}) would give
$\mathcal T/|\mathcal W-\mathcal M|=0.3281$, $\mathcal T/|\mathcal
W-\mathcal M|=0.1941$ and $\mathcal T/|\mathcal W-\mathcal M|=0.0963$,
respectively. As another example, we set $|m|=2$, $\delta=0.2$, $\beta=2$
and $\lambda^2$=1. From equation (\ref{91}), we have $D_s^2=0.9508$,
$D_s^2=0.1001$ and $D_s^2=-0.4103$; the third branch of solution
should be ignored as $D_s^2<0$. The corresponding values of
$\mathcal T/|\mathcal W-\mathcal M|$ for the first two $D_s^2$
solutions are $\mathcal T/|\mathcal W-\mathcal M|=0.2700$ and
$\mathcal T/|\mathcal W-\mathcal M|=0.0986$, respectively.

From these numerical estimates, it is clear that the ratio
$\mathcal T/|\mathcal W-\mathcal M|$ can be much smaller than 0.14
(mainly through the perturbation mode of the $y_2$ branch), analogous
to an unmagnetized composite SID system (Lou \& Shen 2003). Shu
et al. (2000) suggested a correspondence between $\mathcal T/|\mathcal W|$
ratio and the onset of secular bar-like instabilities in a single fluid
SID. It seems natural to extend this suggestion to a single MSID (Lou
2002) and to an unmagnetized composite system of two SIDs (Lou \& Shen
2003). By this analogy, we now propose a further extension of this 
correspondence of ratio $\mathcal T/|\mathcal W-\mathcal M|$ and the 
onset of secular bar-like instabilities in a composite system composed 
of one SID and one MSID. It then appears, as in the case of Lou \& Shen 
(2003), the threshold of $\mathcal T/|\mathcal W-\mathcal M|$ ratio can 
be considerably lowered in a composite SID or MSID system. Qualitatively,
this illustrates that the mutual gravitational coupling tends to make
a disc system more unstable (Jog \& Solomon 1984a,b; Bertin \& Romeo
1988; Romeo 1992; Elmegreen 1995; Jog 1996; Lou \& Fan 1998b; Lou \&
Shen 2003; Shen \& Lou 2003).
\begin{figure}
\begin{center}
\includegraphics[angle=0,scale=0.45]{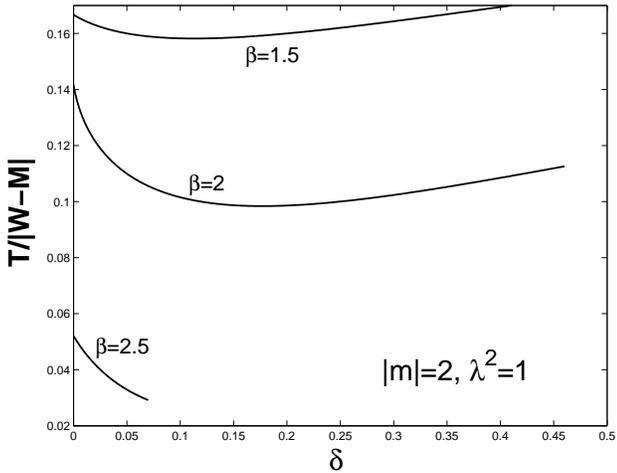}
\caption{\label{f24}Curves of ratio $\mathcal T/|\mathcal W-\mathcal M|$
versus $\delta$ for different values of $\beta=1.5$, $2$, $2.5$ for the
aligned $|m|=2$ case with $\lambda^2=1$.}
\end{center}
\end{figure}

\begin{figure}
\begin{center}
\includegraphics[angle=0,scale=0.45]{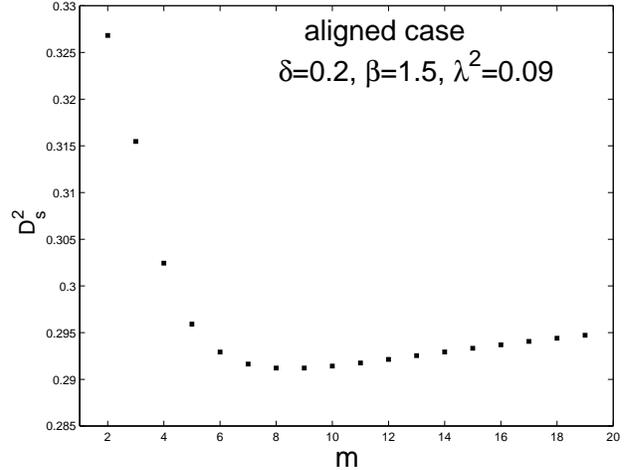}
\caption{\label{f26}The $D_s^2$ solution of the $y_2$ branch versus
$|m|$ with parameters $\delta=0.2$, $\beta=1.5$ and $\lambda^2=0.09$.
The smallest $D_s^2$ occurs at $|m|=9$.}
\end{center}
\end{figure}

\begin{figure}
\begin{center}
\includegraphics[angle=0,scale=0.45]{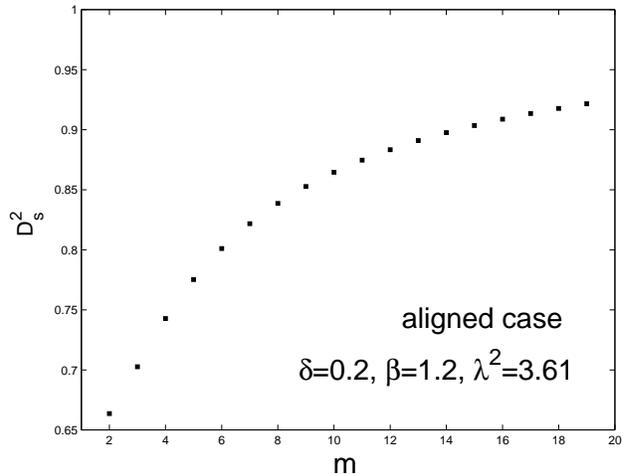}
\caption{\label{f27}The $D_s^2$ solution of the $y_2$ branch versus
$|m|$ with parameters $\delta=0.2$, $\beta=1.2$ and $\lambda^2=3.61$.
The smallest $D_s^2$ occurs at $|m|=2$ in this case.}
\end{center}
\end{figure}

We now explore trends of variations for the ratio
$\mathcal T/|\mathcal W-\mathcal M|$ of the $y_2$ branch. Setting
$|m|=2$, we present curves of $\mathcal T/|\mathcal W-\mathcal M|$
versus $\delta$ when parameters $\beta$ and $\lambda^2$ are specified
(see Fig. \ref{f24}). As there are three $D_s^2$ solution branches
involving three parameters $\delta$, $\beta$ and $\lambda^2$, values
of ratio $\mathcal T/|\mathcal W-\mathcal M|$ can be diversified.
Consequently, instability properties of a composite (M)SIDs system
are more complex than in a single SID. Moreover, possible
perturbation modes of the $D_s^2$ solution of $y_2$ branch can vary
with different values of azimuthal wavenumber $|m|$ as shown in
Figs. \ref{f26} and \ref{f27}.

For parameters $\delta=0.2$, $\beta=1.5$ and $\lambda^2=0.09$,
$D_s^2$ attains its minimum value around $|m|=9$, while for
parameters $\delta=0.2$, $\beta=1.2$ and $\lambda^2=3.61$, the
minimum value of $D_s^2$ occurs at $|m|=2$. In association with
the smallest value of $D_s^2$, the most vulnerable stationary
perturbation configuration varies with chosen parameters. In
other words, a slightly different choice of parameters may lead
to entirely different MHD perturbation configurations. Again, this
example shows the diversity and complexity in a composite MSID
system and indicates that bar-like instabilities (i.e. $m=2$)
may not necessarily be the dominant instability in a composite
MSID system.

\section{UNALIGNED MHD PERTURBATIONS OF LOGARITHMIC SPIRALS}

In this section, we analyze unaligned perturbation configurations
of logarithmic spiral structures in a composite MSID system. 
We first derive the stationary dispersion relation for both full
and partial composite MSID system. We then focus on the full 
case ($F=1$) and address the problem of axisymmetric marginal 
instabilities with $|m|=0$.

\subsection{Stationary Dispersion Relation for Logarithmic Spirals}

To construct stationary MHD perturbation configurations of unaligned
logarithmic spirals in a composite MSID system, we pick the
following potential-density pair (Syer \& Tremaine 1996; Lynden-Bell
\& Lemos 1993; Shu et al. 2000; Lou 2002; Lou \& Fan 2002; Lou \&
Shen 2003), namely
\begin{equation}\label{51}
\mu_s=\sigma_sr^{-3/2+i\alpha}\ ,
\end{equation}
\begin{equation}\label{52}
\mu_g=\sigma_gr^{-3/2+i\alpha}\ ,
\end{equation}
where $\sigma_s$ and $\sigma_g$ are two constant coefficients and
$\alpha$ is the radial scaling parameter related effectively to 
the radial wavenumber, together with
\begin{equation}\label{53}
V=v_sr^{-1/2+i\alpha}+v_gr^{-1/2+i\alpha}\ ,
\end{equation}
where the two constant coefficients $v_s$ and $v_g$ are 
related to $\sigma_s$ and $\sigma_g$ respectively by
\begin{equation}\label{55}
v_s=-2\pi G\mathcal N_m(\alpha)\sigma_s\ ,
\end{equation}
\begin{equation}\label{56}
v_g=-2\pi G\mathcal N_m(\alpha)\sigma_g\ ,
\end{equation}
with $\mathcal N_m(\alpha)\equiv K(\alpha,m)$ being the
Kalnajs function (Kalnajs 1971). Consistently, we may write
\begin{equation}\label{54}
U_g=u_gr^{-1/2+i\alpha}
\end{equation}
and
\begin{equation}\label{74}
U_s=u_sr^{-1/2+i\alpha}\ ,
\end{equation}
where $u_g$ and $u_s$ are two constant coefficients. Using
expressions (\ref{52}), (\ref{53}) and (\ref{54}), equations
(\ref{63}), (\ref{64}), (\ref{66}) and (\ref{67}) lead to
\begin{equation}\label{75}
\begin{split}
\frac{m\Omega_g\sigma_g}{\Sigma_0^g}
+iu_g\bigg(i\alpha+\frac{C_A^2}{2\Omega_g^2r^2}-\frac32\bigg)
\qquad\qquad\\
-\frac{m}{\Omega_gr}\bigg(\frac{a_g^2\sigma_g}{r\Sigma_0^g}
+v_s+v_g\bigg)=0\ ,\quad
\end{split}
\end{equation}
\begin{equation}\label{76}
\begin{split}
\bigg\{m^2\Omega_g^2r^2-\kappa_g^2r^2-C_A^2\bigg[m^2-1
-\bigg(\frac{C_A^2}{2\Omega_g^2r^2}-\frac32\bigg)
\qquad\\
\times\bigg(\frac{C_A^2}{2\Omega_g^2r^2}-2\bigg)\bigg]\bigg\}iu_g
=C_A^2\bigg(i\alpha+2-\frac{C_A^2}{2\Omega_g^2r^2}\bigg)
\qquad\\
\times\bigg[\frac{m\Omega_g\sigma_g}{\Sigma_0^g}
-\frac{m}{\Omega_gr}\bigg(\frac{a_g^2\sigma_g}{r\Sigma_0^g}
+v_s+v_g\bigg)\bigg]
\quad\\
+m\Omega_gr\bigg(i\alpha+\frac32\bigg)
\bigg(\frac{a_g^2\sigma_g}{r\Sigma_0^g}+v_s+v_g\bigg)
-\frac{m\Omega_gC_A^2\sigma_g}{2\Sigma_0^g}\ ,
\end{split}
\end{equation}
\begin{equation}\label{77}
iR=\frac{r^{1/2}B_\theta}{\Omega_gr}iu_gr^{-1+i\alpha}
=\frac{B_\theta iU_g}{\Omega_gr}\ ,
\end{equation}
\begin{equation}\label{78}
Z=-\frac{i\alpha r^{1/2}B_\theta}
{m\Omega_gr}iu_gr^{-1+i\alpha}
=-\frac{i\alpha B_\theta}{m\Omega_gr}iU_g\ .
\end{equation}
A combination of equations (\ref{75}) and (\ref{76})
together with relations (\ref{55}) and (\ref{56}) give
\begin{equation}\label{58}
\begin{split}
\bigg[\bigg(\frac{\mathcal Km}{\Omega_gr}-\mathcal
A\bigg)\bigg(\frac{a_g^2}{r\Sigma_0^g}-2\pi G\mathcal
N_m(\alpha)\bigg)
\quad\qquad\qquad\qquad\\
-\bigg(\mathcal C+\frac{\mathcal Km\Omega_g}
{\Sigma_0^g}\bigg)\bigg]\sigma_g
=\bigg(\frac{\mathcal Km}{\Omega_gr}
-\mathcal A\bigg)2\pi G\mathcal N_m(\alpha)\sigma_s\ ,
\end{split}
\end{equation}
where the three coefficients $\mathcal K$, 
$\mathcal A$ and $\mathcal C$ are defined by
\begin{equation}\label{59}
\mathcal K\equiv\bigg(m-\frac2m\bigg)\Omega_gr
-\frac{(\alpha^2+m^2-1)C_A^2}{m\Omega_gr}\ ,
\end{equation}
\begin{equation}\label{60}
\mathcal A\equiv\frac32\bigg(\frac{C_A^2}
{2\Omega_g^2r^2}-\frac32\bigg)-\alpha^2\ ,
\end{equation}
\begin{equation}\label{61}
\mathcal C\equiv-\frac{C_A^2}{2r\Sigma_0^g}\bigg(\frac{C_A^2}
{2\Omega_g^2r^2}-\frac32\bigg)\ .
\end{equation}
For stationary coplanar perturbations in the stellar disk in
parallel, we set $\omega=0$ in equations (\ref{33})$-$(\ref{35})
and use equations (\ref{51})$-$(\ref{56}) and (\ref{74}) to obtain
\begin{equation}\label{57}
\begin{split}
\bigg[\bigg(m^2+\alpha^2+\frac14\bigg)
\bigg(\frac{a_s^2}{r\Sigma_0^s}-2\pi
G\mathcal N_m(\alpha)\bigg)
\qquad\\
-\frac1{r\Sigma_0^s}(m^2-2)\Omega_s^2r^2\bigg]\sigma_s
\quad\\
=\bigg(m^2+\alpha^2+\frac14\bigg)
2\pi G\mathcal N_m(\alpha)\sigma_g\ .
\end{split}
\end{equation}
Equation (\ref{57}) can also be obtained by setting $C_A=0$ 
in equation (\ref{58}) and exchange subscripts $s$ and $g$.

By equations (\ref{57}) and (\ref{58}), we derive
\begin{equation}\label{62}
\begin{split}
\frac{r}{\Omega_g^2m^2}\bigg[\frac{(m^2-2)}r\Omega_s^2r^2
-\bigg(m^2+\alpha^2+\frac14\bigg)
\qquad\qquad\qquad\\
\times\bigg(\frac{a_s^2}r-2\pi G\mathcal
N_m(\alpha)\Sigma_0^s\bigg)\bigg]
\times\bigg\{m^4\Omega_g^4-m^2\Omega_g^2\bigg[2\Omega_g^2
\quad\qquad\\
+\bigg(\frac{C_A^2+a_g^2}{r^2}-\frac{2\pi G\mathcal
N_m(\alpha)\Sigma_0^g}r\bigg)\bigg(m^2+\alpha^2
+\frac14\bigg)-\frac{2C_A^2}{r^2}\bigg]
\qquad\\
+\frac{m^2C_A^2}{r^2}\bigg[\bigg(\frac{a_g^2}{r^2}-\frac{2\pi
G\mathcal
N_m(\alpha)\Sigma_0^g}r\bigg)\bigg(m^2+\alpha^2-\frac14\bigg)
\qquad\quad\\
-\frac{C_A^2}{4r^2}\bigg]\bigg\} =4\pi^2G^2\mathcal
N_m^2(\alpha)\Sigma_0^g\Sigma_0^s\bigg(m^2+\alpha^2+\frac14\bigg)
\qquad\qquad\\
\times\bigg[\bigg(m^2+\alpha^2+\frac14\bigg)
-\frac{C_A^2}{\Omega_g^2r^2}
\bigg(m^2+\alpha^2-\frac14\bigg)\bigg]\ .\qquad\qquad
\end{split}
\end{equation}
In the absence of the gravitational coupling between the stellar
SID and the gaseous MSID represented by the right-hand side term
of equation (\ref{62}), the left-hand side of dispersion
relation (\ref{62}) would be reduced to two separate dispersion
relation factors. The first dispersion relation would be
\begin{equation}\label{79}
\frac{(m^2-2)}r\Omega_s^2r^2-\bigg(m^2+\alpha^2
+\frac14\bigg)\bigg[\frac{a_s^2}r-2\pi G\mathcal N_m(\alpha)
\Sigma_0^s\bigg]=0\
\end{equation}
for stationary logarithmic spirals in a single stellar SID
without magnetic field. Substituting expressions (\ref{11})
and (\ref{20}) of $\Omega_s$ and $\Sigma_0^s$ with $F=1$
into equation (\ref{79}), we have
\begin{equation}\label{80}
\bigg(m^2+\alpha^2+\frac14\bigg)
\frac {[1-(1+D_s^2)\mathcal N_m(\alpha)]}{D_s^2(m^2-2)}=1\ ,
\end{equation}
which is simply equation (37) of Shu et al.
(2000). The second dispersion relation would be
\begin{equation}\label{81}
\begin{split}
m^4\Omega_g^4-\bigg[2\Omega_g^2
+\bigg(\frac{C_A^2+a^2}{r^2}-\frac{2\pi G
\mathcal N_m(\alpha)\Sigma_0^g}r\bigg)
\qquad\qquad\\
\times\bigg(m^2+\alpha^2+\frac14\bigg)
-\frac{2C_A^2}{r^2}\bigg]m^2\Omega_g^2
+\frac{m^2C_A^2}{r^2}
\qquad\qquad\\
\times\bigg[\bigg(\frac{a^2}{r^2}-\frac{2\pi G
\mathcal N_m(\alpha)\Sigma_0^g}r\bigg)
\bigg(m^2+\alpha^2-\frac14\bigg)-\frac{C_A^2}{4r^2}\bigg]=0
\end{split}
\end{equation}
for stationary logarithmic spirals in a single gaseous MSID
with coplanar magnetic field. Equation (\ref{81}) is simply
equation (3.4.15) of Lou (2002).

Without magnetic field with $C_A=0$, equation (\ref{62}) is
equivalent to the dispersion relation (86) for stationary
logarithmic spiral configurations in a composite SID system
of Lou \& Shen (2003) as it should be.

For numerical computations, there are two useful formulae of
$\mathcal N_m(\alpha)$, namely, the recursion relation
\begin{equation}\label{83}
\mathcal N_{m+1}(\alpha)\mathcal
N_m(\alpha)=[(m+1/2)^2+\alpha^2]^{-1}
\end{equation}
and the asymptotic expansion
\begin{equation}\label{84}
\mathcal N_m(\alpha)\approx(m^2+\alpha^2+1/4)^{-1/2}
\end{equation}
when $m^2+\alpha^2\gg 1$ (Kalnajs 1971; Shu et al. 2000).

For the purpose of examining phase relationship between the
azimuthal magnetic field perturbation $b_{\theta}$ and the
surface mass density perturbation $\Sigma_g^1$ of the gaseous
MSID in the full case of $F=1$, we derive the following
relations. From equations (\ref{75}) and (\ref{57}) with
$F=1$, we readily obtain
\begin{equation}\label{115}
\begin{split}
iu_g=\frac{2\pi
Gm\sigma_g}{[i\alpha+C_A^2/(2\Omega_g^2r^2)-3/2](\Omega_gr)}
\qquad\qquad\\
\times\bigg[\frac{(1+\delta)(1-D_g^2)}{\delta(1+D_g^2-\lambda^2/2)}
-\frac{\mathcal A\ \mathcal N_m(\alpha)}
{(\mathcal A-\mathcal B)}\bigg]\ ,
\end{split}
\end{equation}
where the two coefficients $\mathcal A$ and 
$\mathcal B$ are defined by
\begin{equation}\label{116}
\mathcal A\equiv(m^2+\alpha^2+1/4)-(m^2-2)D_s^2\
\end{equation}
and
\begin{equation}\label{117}
\mathcal B\equiv\frac{\mathcal N_m(\alpha)}
{(1+\delta)}(m^2+\alpha^2+1/4)(1+D_s^2)\ .
\end{equation}
It follows from expressions (\ref{78}) and (\ref{115}) that
\begin{equation}\label{118}
\begin{split}
\frac{Z}{\mu_g}\propto\bigg[\alpha^2
+i\alpha\bigg(\frac{\lambda^2}{2D_g^2}-\frac32\bigg)\bigg]
\quad\qquad\qquad\qquad\\
\times\bigg[\frac{(1+\delta)(1-D_g^2)}
{\delta(1+D_g^2-\lambda^2/2)}
-\frac{\mathcal A\ \mathcal N_m(\alpha)}
{(\mathcal A-\mathcal B)}\bigg]\ ,
\end{split}
\end{equation}
that will be discussed later in our analysis.

\subsection{Marginal Stability for Axisymmetric Disturbances}

\begin{figure}
\begin{center}
\includegraphics[angle=0,scale=0.45]{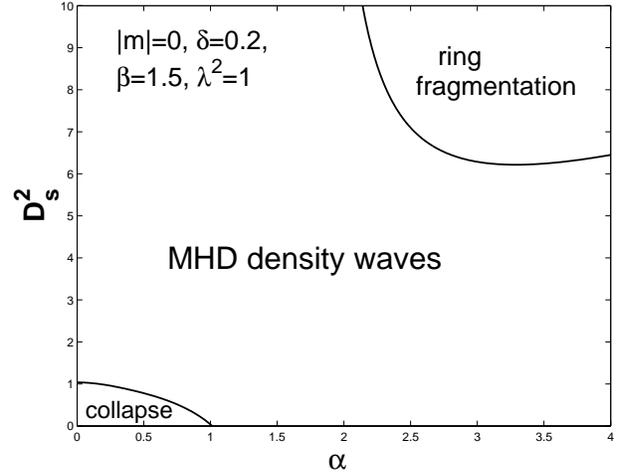}
\caption{\label{f1}The marginal stability curve of $D_s^2$ versus
$\alpha$ with $m=0$, $\delta=0.2$, $\beta=1.5$ and $ \lambda^2=1$.}
\end{center}
\end{figure}

\begin{figure}
\begin{center}
\includegraphics[angle=0,scale=0.45]{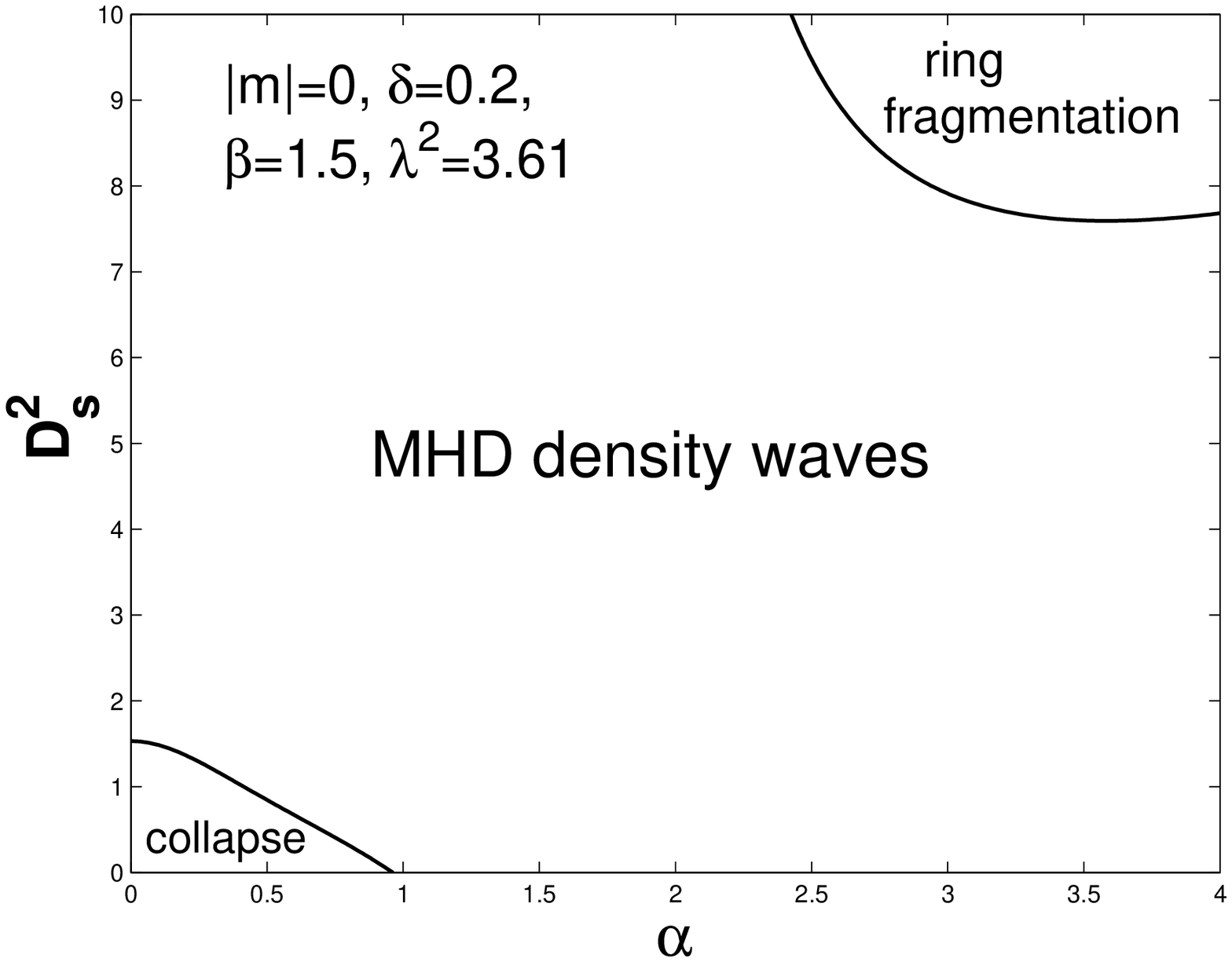}
\caption{\label{f2}The marginal stability curve of $D_s^2$
versus $\alpha$ with $m=0$, $\delta=0.2$, $\beta=1.5$ and
$\lambda^2=3.61$.}
\end{center}
\end{figure}

\begin{figure}
\begin{center}
\includegraphics[angle=0,scale=0.45]{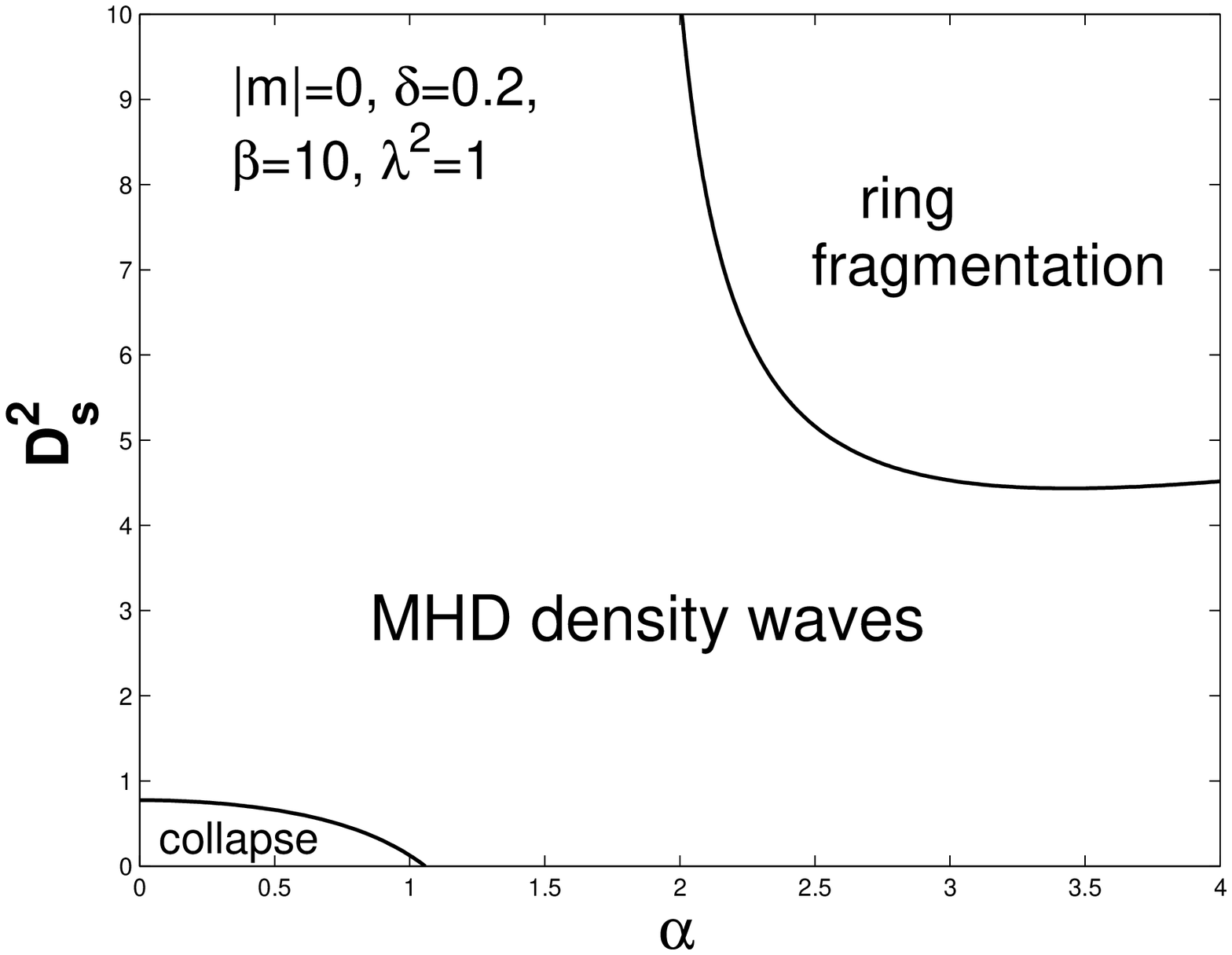}
\caption{\label{f3}The marginal stability curve of $D_s^2$ versus
$\alpha$ with $m=0$, $\delta=0.2$, $\beta=10$ and $ \lambda^2=1$.}
\end{center}
\end{figure}

\begin{figure}
\begin{center}
\includegraphics[angle=0,scale=0.45]{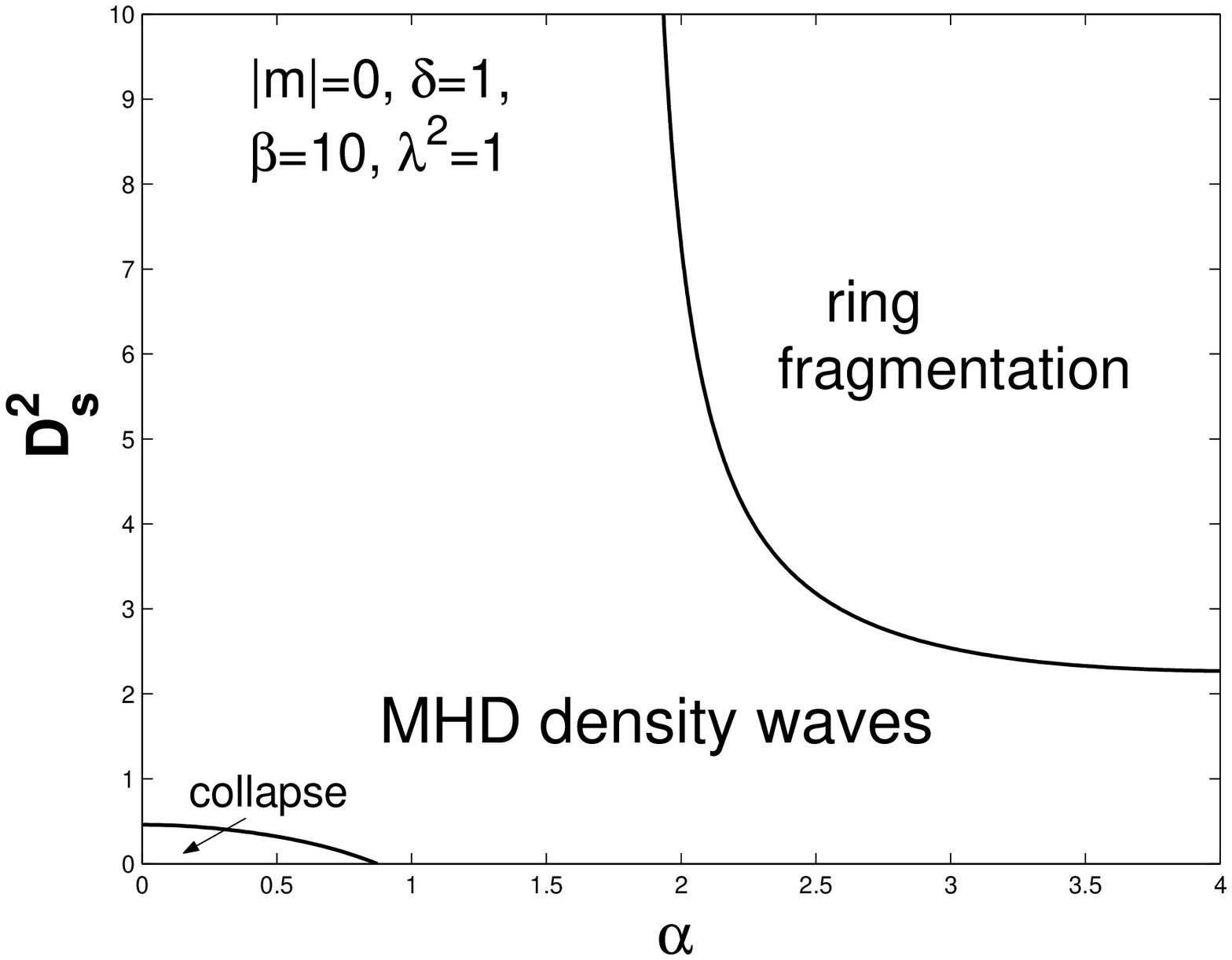}
\caption{\label{f7}The marginal stability curve of $D_s^2$ versus
$\alpha$ with $m=0$, $\delta=1$, $\beta=10$ and $\lambda^2=1$. }
\end{center}
\end{figure}

\begin{figure}
\begin{center}\includegraphics[angle=0,scale=0.45]{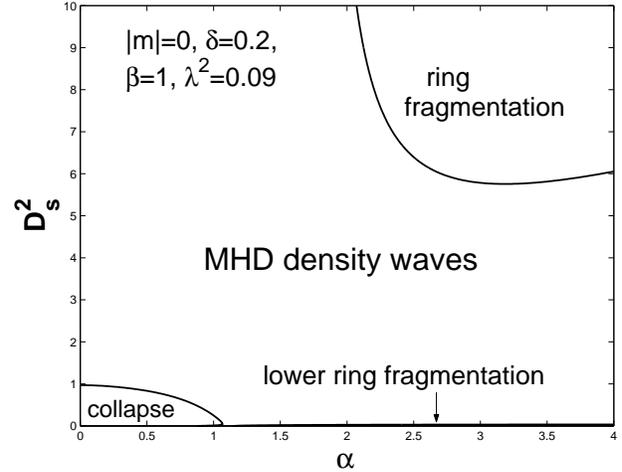}
\caption{\label{f9}The marginal stability curve of $D_s^2$ versus
$\alpha$ with $m=0$, $\delta=0.2$, $\beta=1$ and $\lambda^2=0.09$.
For $\beta\rightarrow 1$, the lower ring fragmentation branch would
gradually rise above the horizontal $\alpha$ axis. }
\end{center}
\end{figure}

While the $|m|=0$ case is somewhat trivial as a rescaling in the
aligned case, it is of considerable interest in the special `spiral'
case with radial oscillations. In this subsection, we analyze this
situation for a full composite MSID system with $F=1$. A substitution
of expressions (\ref{11}), (\ref{15}), (\ref{20}) and (\ref{21})
into equation (\ref{62}) with $\lambda^2\equiv C_A^2/a_g^2$ leads to
\begin{equation}\label{82}
\begin{split}
\bigg\{-2D_s^2-\bigg(\alpha^2+\frac14\bigg)
\bigg[1-\frac{\mathcal N_0(\alpha)(1+D_s^2)}
{1+\delta}\bigg]\bigg\}
\quad\qquad\qquad\\
\times\bigg\{-D_g^2\bigg[2D_g^2
+\bigg(\lambda^2+1
-\frac{\mathcal N_0(\alpha)(1+D_g^2-\lambda^2/2)\delta}
{1+\delta}\bigg)
\qquad\\
\times\bigg(\alpha^2+\frac14\bigg)-2\lambda^2\bigg]
+\lambda^2\bigg[\bigg(1-\frac{\mathcal N_0(\alpha)
(1+D_g^2-\lambda^2/2)\delta}{1+\delta}\bigg)
\hbox{ }\\
\times\bigg(\alpha^2-\frac14\bigg)-\frac{\lambda^2}4\bigg]\bigg\}
-\frac{\mathcal N_0(\alpha)^2D_g^2(1+D_s^2)(1+D_g^2-\lambda^2/2)\delta}
{(1+\delta)^2}
\\
\times\bigg(\alpha^2+\frac 14\bigg)\bigg[\alpha^2+\frac 14
-\lambda^2\bigg(\alpha^2-\frac14\bigg)/D_g^2\bigg]=0\ .\qquad\qquad
\end{split}
\end{equation}
By recursion formula (\ref{83}) and asymptotic expression (\ref{84})
for the Kalnajs functions, we have an approximate expression of
$\mathcal N_0(\alpha)$, namely
\begin{equation}\label{85}
\begin{split}
\mathcal N_0(\alpha)=(\alpha^2+9/4)/(\alpha^2+1/4)
\mathcal N_2(\alpha)
\quad\qquad\qquad\\
=(\alpha^2+9/4)/[(\alpha^2+1/4)(\alpha^2+17/4)^{1/2}]\ .
\end{split}
\end{equation}
A substitution of approximate expression (\ref{85}) for
$\mathcal N_0(\alpha)$ and expression(\ref{87}) for $D_g^2$
into dispersion relation (\ref{82}) yields a cubic algebraic
equation in terms of $D_s^2$; and similarly, a substitution of
approximate expression (\ref{85}) for $\mathcal N_0(\alpha)$
and expression (\ref{90}) for $D_s^2$ yields another cubic
algebraic equation in terms of $D_g^2$.
As noted earlier, we consider the cubic equation
of $D_s^2$ because of the fact that $D_s^2<D_g^2$ when
$\beta\geq 1$. For different values of parameters $\delta$,
$\beta$ and $\lambda^2$, we show curves of $D_s^2$ versus
$\alpha$ in Figs. \ref{f1}$-$\ref{f9}.

\begin{figure}
\begin{center}\includegraphics[angle=0,scale=0.45]{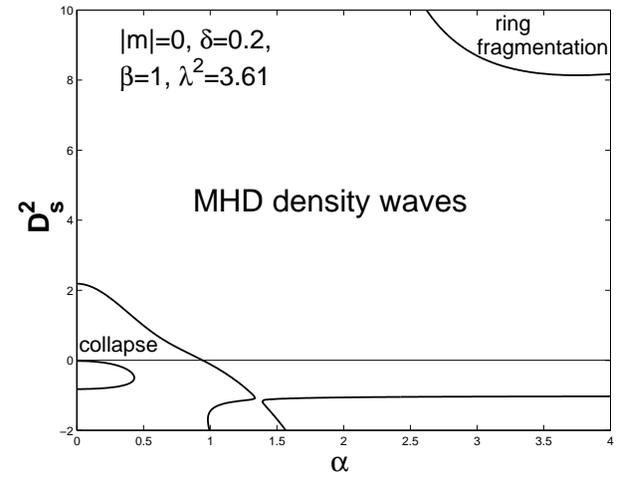}
\caption{\label{f10}The marginal stability curve of $D_s^2$ versus
$\alpha$ with $m=0$, $\delta=0.2$, $\beta=1$ and $
\lambda^2=3.61$. Compared with Fig. \ref{f9}, the increase of
$\lambda$ moves the lower ring fragmentation downward.}
\end{center}
\end{figure}

For the positive portions of solution $D_s^2$ with typical
parameters, it turns out that the basic features of a $D_s^2$
versus $\alpha$ profile are qualitatively similar to those
of a single SID [see fig. 2 of Shu et al. (2000)] and to
those of a composite SID system (see figs. $5-10$ of Lou \&
Shen 2003), except that now trends of variation for $D_s^2$
with three parameters become more complicated. For example, the
lower ring fragmentation curve first discussed in a single
MSID (Lou 2002) becomes negative or indiscernible in most
cases (see Appendix B). For cases of $\beta$ around $1$,
however, the lower ring fragmentation curve of $D_s^2$ may
become positive as shown in Fig. \ref{f9}.

As parameters $\delta$, $\beta$ and $\lambda^2$ vary, we note
several trends of variation in the profiles of $D_s^2$. For example,
when $\lambda^2$ increases for fixed $\delta$ and $\beta$, the
ring fragmentation curves of $D_s^2$ seem to be raised, and the
collapse regime tends to be enlarged by comparing
Figs. \ref{f1} and \ref{f2}. In other words, the presence
of coplanar magnetic field tends to increase the chance of
collapse but reduce the danger of ring fragmentation. This
feature is qualitatively similar to the case of a single
coplanar magnetized SID studied by Lou (2002) (see fig. 1
of Lou 2002). As $\beta$ increases for fixed $\delta$ and
$\lambda^2$, the curves of $D_s^2$ tend to be lowered by
comparing Figs. \ref{f1} and \ref{f3}. When $\delta$ increases
for fixed $\beta$ and $\lambda^2$, the ring fragmentation curves
of $D_s^2$ decrease, and the collapse regime shrinks by comparing
Figs. \ref{f3} and \ref{f7}. Note that as
$\delta$ increases ($\delta\equiv\Sigma_0^g/\Sigma_0^s$), the
influence of gas disc on the stellar disc becomes more important.
Therefore, the gravitational coupling with another disc tends to
suppress the collapse but enhance the danger of ring fragmentation.

One behaviour for the variation of the lower ring 
fragmentation curve appears somewhat surprising.
It seems to be an characteristic feature
involving magnetic field (Lou 2002). As already noted earlier, when
$\beta$ approaches $1$, this curve could rise above the $\alpha$
axis to become a possible perturbation mode (Fig. \ref{f9}).
However, comparing Fig. \ref{f9} with Fig. \ref{f10}, we find that in
typical parameter regimes, the increase of $\lambda$ would force the
lower ring fragmentation curve to go downward. This is completely
different from that of a single MSID (see fig. 1 of Lou 2002).

Through numerical explorations, we realize that the location $\alpha$
of the vertical asymptote for the ring fragmentation branch seems
to be independent of $\delta$, $\beta$ and $\lambda^2$. By letting
the coefficient of the cubic term in the cubic algebraic equation
of $y\equiv D_s^2$ to vanish, the condition that this critical value
$\alpha_c$ must satisfy is once again
\begin{equation}\label{88}
\mathcal N_0(\alpha)(\alpha^2+1/4)=2\ ,
\end{equation}
which is exactly the same as that of Shu et al. (2000), Lou (2002)
and Lou \& Shen (2003) for three related but different SID systems.
From equation (\ref{85}) and asymptotic expansion (\ref{84}), this
critical value $\alpha_c$ is estimated to be 1.759.

Physical interpretations for the marginal instability curves
are clear. As $\alpha$ parameter is a measure for the radial
wavenumber, a smaller $\alpha$ corresponds to a larger radial scale
of perturbations. Thus, the collapse regime indicates a rotationally
modified Jeans instability to which a composite MSID system is
vulnerable when a radial perturbation scale is sufficiently large
(i.e. a sufficiently small $\alpha$). However, when $D_s^2$ becomes
larger, the conservation of angular momentum works against the
inward gravitational force and the Jeans instability is suppressed.

Nevertheless, a composite MSID system is also vulnerable to
Toomre-type instability (Safronov 1960; Toomre 1964) associated
with the ring fragmentation branch (Shu et al. 2000). In various
galactic contexts, there have been numerous studies to identify
an effective $Q$ parameter for a composite disc system (Jog \&
Solomon 1984a,b; Bertin \& Romeo 1988; Kennicutt 1989; Romeo
1992; Elmegreen 1995; Jog 1996; Lou \& Fan 1998b).
In the case of a single SID, Shu et al. (2000) noted that the
minimum of the ring fragmentation curve is closely related to the
Toomre $Q$ parameter (Toomre 1964). In the case of a coplanar
magnetized SID, Lou (2002) and Lou \& Fan (2002) further found
that the minimum of the upper ring fragmentation curve is tightly
associated with the generalized MHD $Q_M$ parameter (Lou \& Fan
1998a). For a composite system of two SIDs, Shen \& Lou (2003)
recently suggested a straightforward $D-$criterion to effectively 
determine the axisymmetric stability against ring fragmentation 
(see Elmegreen 1995 and Jog 1996 for other proposed criteria).
We therefore expect that in the present problem, the minimum of the
upper ring fragmentation curve should be related a magnetic field
modified $D-$criterion to effectively determine the axisymmetric
stability against ring fragmentation in a composite system of one
fluid SID and one MSID.

\subsection{Stationary Logarithmic Spiral Configurations}

\subsubsection{Behaviours of $D_s^2$ solutions}

Parallel to the study of aligned $|m|\geq2$ cases, we 
substitute expressions (\ref{11}), (\ref{15}), (\ref{20}), 
(\ref{21}) and (\ref{87}) into equation (\ref{62}) with 
$\lambda^2\equiv C_A^2/a_g^2$ and obtain a cubic algebraic 
equation of $y\equiv D_s^2$ for unaligned spiral cases. One 
can formally derive three analytic $D_s^2$ solutions (see 
Appendix D) that would be valuable in numerical MHD simulation 
studies of large-scale SID dynamics within the overall MHD 
density-wave scenario.

\begin{figure}
\begin{center}
\includegraphics[angle=0,scale=0.45]{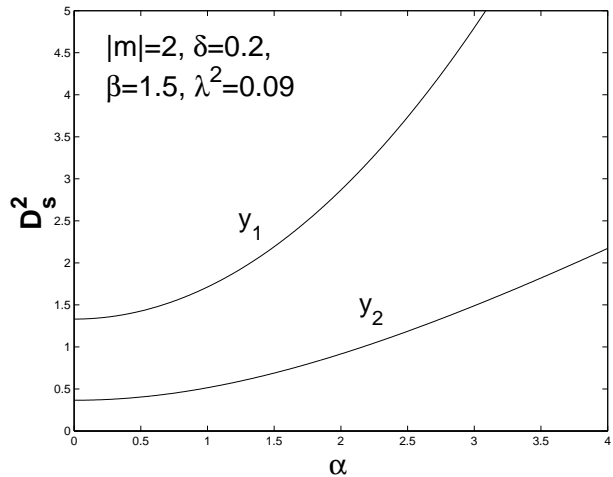}
\caption{\label{f28}Curves of $D_s^2$ versus $\alpha$ for a stationary
unaligned logarithmic spiral with $|m|=2$, $\delta=0.2$, $\beta=1.5$
and $\lambda^2=0.09$.}
\end{center}
\end{figure}

\begin{figure}
\begin{center}
\includegraphics[angle=0,scale=0.45]{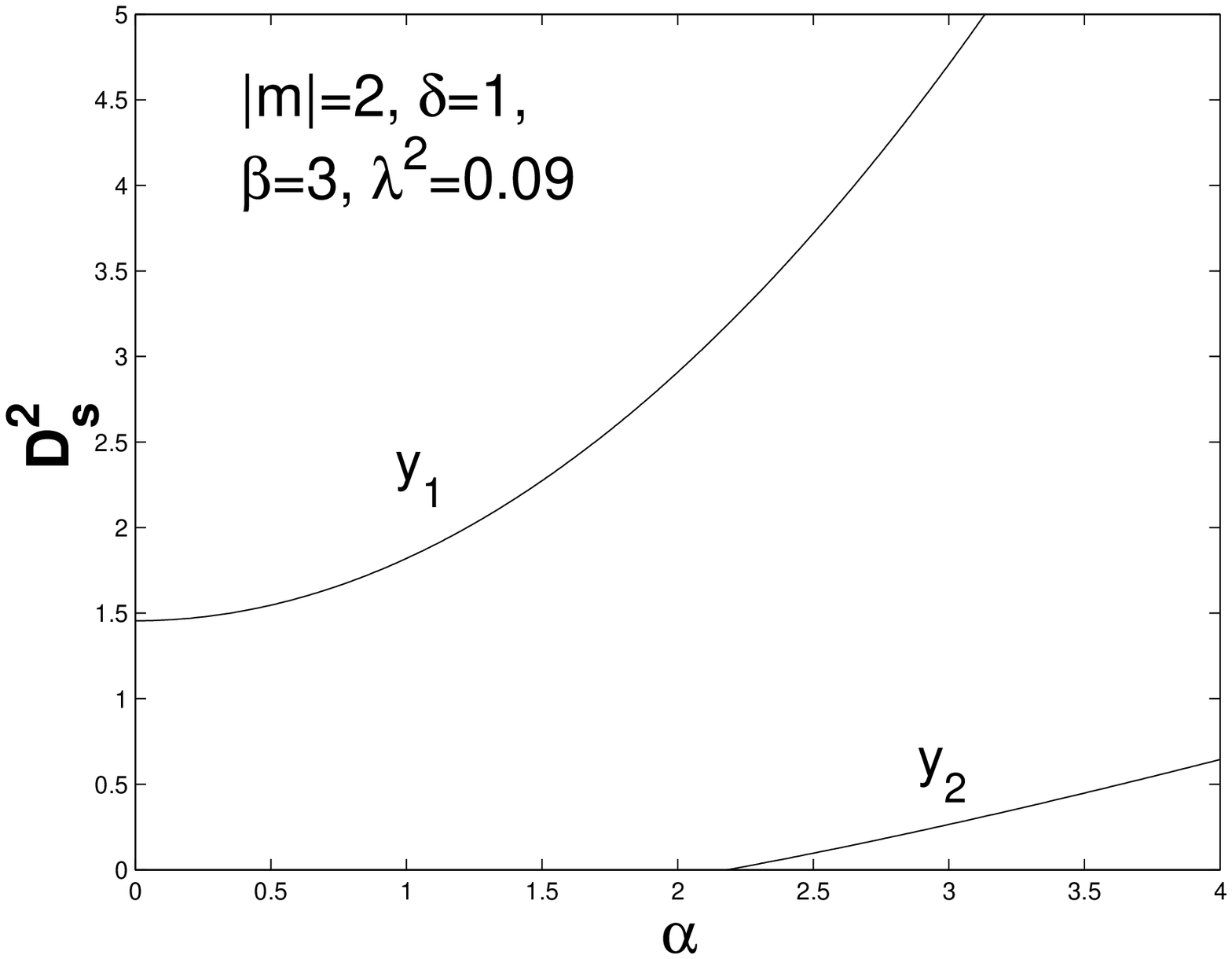}
\caption{\label{f31}Curves of $D_s^2$ versus $\alpha$ for a stationary
unaligned logarithmic spiral with $|m|=2$, $\delta=1$, $\beta=3$ and
$\lambda^2=0.09$.}
\end{center}
\end{figure}

\begin{figure}
\begin{center}
\includegraphics[angle=0,scale=0.45]{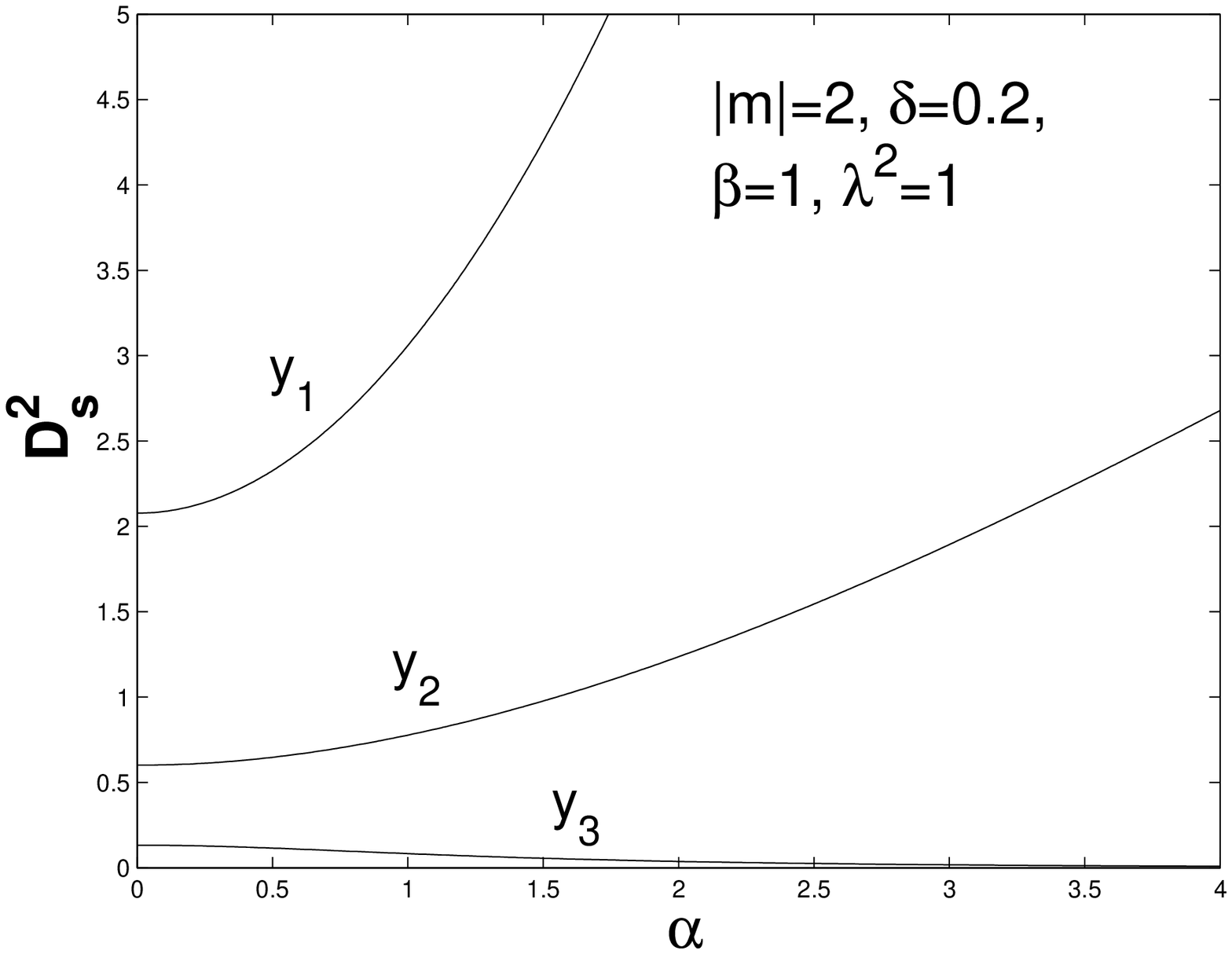}
\caption{\label{f32}Curves of $D_s^2$ versus $\alpha$ for a stationary
unaligned logarithmic spiral with $|m|=2$, $\delta=0.2$, $\beta=1$ and
$\lambda^2=1$. }
\end{center}
\end{figure}

\begin{figure}
\begin{center}
\includegraphics[angle=0,scale=0.45]{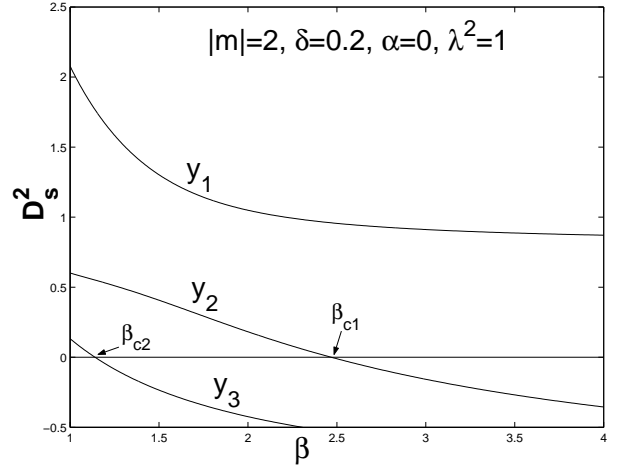}
\caption{\label{f33}Curves of $D_s^2$ versus $\beta$ for a stationary
unaligned logarithmic spiral with $|m|=2$, $\delta=0.2$, $\alpha=0$
and $\lambda^2=1$. }
\end{center}
\end{figure}

To explore parameter regimes, we obtain $D_s^2$ solution curves
numerically for cases of $|m|=1, 2, 3, \cdots$ and so forth. As
expected, there are three solution curves of $D_s^2$ versus $\alpha$
given parameters $|m|$, $\delta$, $\lambda^2$ and $\beta$. In certain
parameter regimes, portions of the lower two $D_s^2$ solution
curves may become negative and hence unphysical. We defer our
discussion of $|m|=1$ case later. In Figs. \ref{f28}$-$\ref{f32},
we show a set of numerical examples with $|m|=2$. Through extensive
numerical examples of stationary logarithmic spiral cases from
$|m|=2$ to $|m|=5$, it is clear that behaviours of $D_s^2$ solutions
and properties of phase relationships among perturbation variables
for $|m|>2$ cases are very similar to those of $|m|=2$ logarithmic
spiral. We therefore carefully examine the $|m|=2$ spiral case for
a relatively simple yet meaningful analysis. In addition, we plot
curves of $D_s^2$ versus $\beta$ by specifying $\alpha$, $\delta$
and $\lambda^2$ in Fig. \ref{f33}.

As in the aligned case, we shall use $y_1$, $y_2$ and $y_3$ to denote
the upper, middle and lower solution branches, respectively. It is
apparent that the three $D_s^2$ solution branches do not intersect with
each other. Similar to the case of axisymmetric disturbances, the lowest
$y_3$ solution branch does not appear above zero in many cases, but
for cases with relatively small $\beta$ value (e.g. $\beta=1$), $y_3$
solution branch can become a possible stationary perturbation mode (see
Figs. \ref{f32} and \ref{f33}). We note that $D_s^2$ always increases
with increasing $\alpha$ for the upper two branches $y_1$ and $y_2$
(see Figs. \ref{f28}$-$\ref{f32}) but decreases with increasing $\beta$
(see Fig. \ref{f33}). The uppermost $y_1$ branch remains always positive,
while the middle $y_2$ branch may cross the horizontal $\alpha$ axis
to become negative in some cases (see Fig. \ref{f31}), similar to the
situation without involving coplanar magnetic field (Lou \& Shen 2003).

In Fig. \ref{f33}, there exist a $\beta_{c1}$ for the middle $y_2$
branch and a $\beta_{c2}$ for the lowest $y_3$ branch for given
$\alpha$, $\delta$ and $\lambda^2$ such that $y_2$ and $y_3$ become
zero and negative with increasing value of $\beta$, respectively.
Parallel to the aligned case, we examine behaviours of the solutions
and obtain the two critical values of $\beta$ analytically. For the
$|m|=2$ case, the analytic forms of $\beta_{c1}$ and $\beta_{c2}$
are given by
\begin{equation}\label{113}
\beta_{c1}=\frac1{2\mathcal A}[-\mathcal B+({\mathcal B}^2
-4\mathcal A\mathcal C)^{1/2}]
\end{equation}
and
\begin{equation}\label{114}
\beta_{c2}=\frac1{2\mathcal A}[-\mathcal B-({\mathcal B}^2
-4\mathcal A\mathcal C)^{1/2}]\ ,
\end{equation}
where definitions of the three coefficients $\mathcal A$, $\mathcal B$
and $\mathcal C$ are given in Appendix A. When $\beta<\beta_{c1}$, the 
middle solution branch $y_2\equiv D_s^2$ is positive, while for 
$\beta>\beta_{c1}$, the middle solution branch $y_2\equiv D_s^2$ becomes 
negative and hence unphysical. For the case of $\delta=0.2$, 
$\lambda^2=0.09$ and $\alpha=0$, we obtain $\beta_{c1}=2.409$ from 
equation (\ref{113}),
consistent with Fig. \ref{f28}; in Fig. \ref{f28}, we have
$\beta=1.5<\beta_{c1}$ and the middle branch $y_2$ is positive
at $\alpha=0$. For $\lambda^2=1.00$, expressions (\ref{113})
and (\ref{114}) give $\beta_{c1}=2.468$ and $\beta_{c2}=1.140$,
respectively, consistent with Fig. \ref{f33}.

\subsubsection{Phase relationships among perturbation variables}

For stationary logarithmic perturbation spirals, we now examine
spatial phase relationships between perturbations of the surface
mass densities of the two SIDs (i.e. $\mu_g$ and $\mu_s$), and
between perturbations of the azimuthal magnetic field and the
surface mass density of the gaseous MSID (i.e. $b_{\theta}$ and
$\Sigma_1^g$). This information might provide necessary and useful
clues for optical and synchrotron radio observational diagnostics
of spiral galaxies.

\begin{figure}
\begin{center}
\includegraphics[angle=0,scale=0.45]
{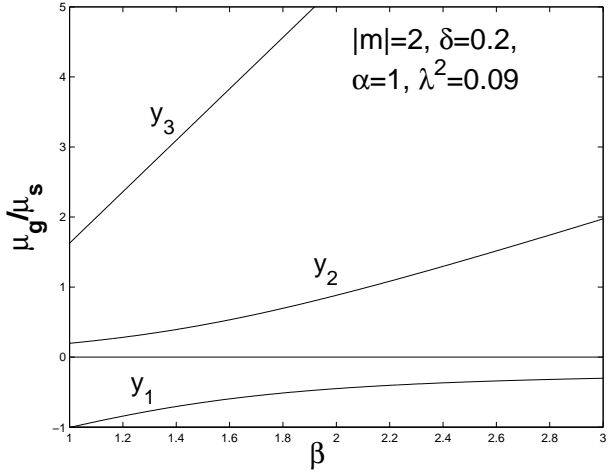}
\caption{\label{f34}Curves of $\mu_g/\mu_s$ versus $\beta$ for
a stationary logarithmic spiral with $|m|=2$, $\delta=0.2$,
$\alpha=1$ and $\lambda^2=0.09$.}
\end{center}
\end{figure}

\begin{figure}
\begin{center}
\includegraphics[angle=0,scale=0.45]
{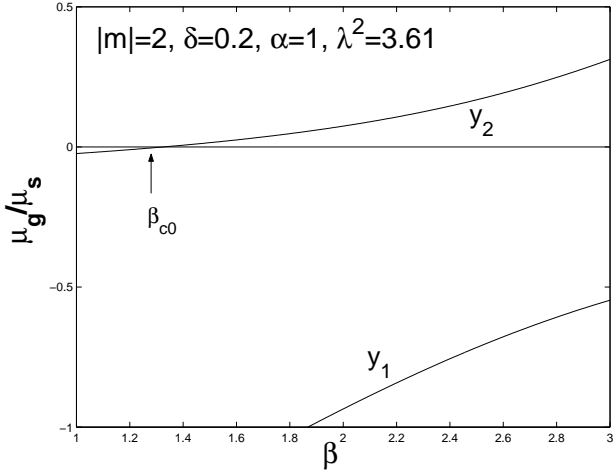}
\caption{\label{f35}Curves of $\mu_g/\mu_s$ versus $\beta$
for a stationary logarithmic spiral $|m|=2$, $\delta=0.2$,
$\alpha=1$ and $\lambda^2=3.61$. The uppermost branch
corresponding to $y_3\equiv D_s^2$ is only discernible for
$\beta<1$ and is not shown here. The branch corresponding
to $y_2\equiv D_s^2$ may cross the horizontal $\beta$ axis. }
\end{center}
\end{figure}

From equation (\ref{57}), one obtains
\begin{equation}\label{119}
\begin{split}
\frac{\mu_g}{\mu_s}=\left( {m}^{2}+{\alpha}^{2}+1/4\right)^{-1}
{\mathcal N_m(\alpha)}^{-1}
\quad\qquad\qquad\\
\times\bigg\lbrace\left( {m}^{2}+{\alpha}^{2}+1/4\right)
\left[ {\frac {(1+\delta)}{(1+y)}}-\mathcal N_m(\alpha)\right]
\quad\\
-{\frac {\left({m}^{2}-2 \right)
\left(1+\delta\right)y}{1+y}}\bigg\rbrace\
\end{split}
\end{equation}
with $y\equiv D_s^2$. Substitution of expression (\ref{119})
into the cubic equation of $D_s^2$ gives a cubic equation of
$\mu_g/\mu_s$, and we plot several curves of $\mu_g/\mu_s$
versus $\beta$ with parameters $\alpha=1$, $\delta=0.2$,
$\lambda^2=0.09$ and $3.61$ in Figs. \ref{f34} and \ref{f35}.

\begin{figure}
\begin{center}
\includegraphics[angle=0,scale=0.45]
{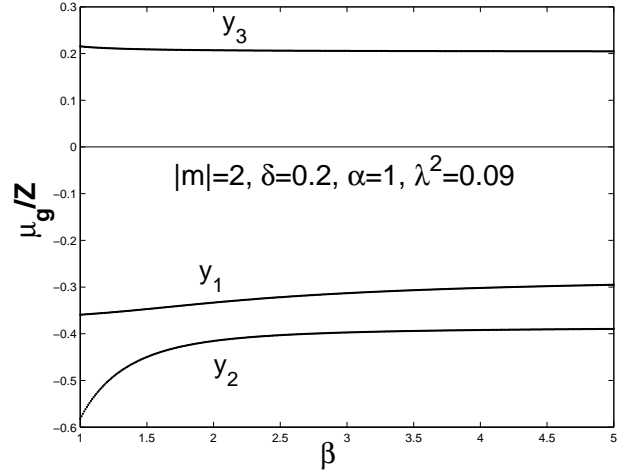}
\caption{\label{f41}Curves of real parts of $\mu_g/Z$ versus $\beta$
for a stationary logarithmic spiral of $|m|=2$, $\delta=0.2$, $\alpha=1$
and $\lambda^2=0.09$. Imaginary parts of $\mu_g/Z$ are not shown here.
While one branch of real part of $\mu_g/Z$ corresponding to $y_1$
remains always negative, this branch does not remain always lower
than the real part of $\mu_g/Z$ branch corresponding to $y_2$. }
\end{center}
\end{figure}

\begin{figure}
\begin{center}
\includegraphics[angle=0,scale=0.45]
{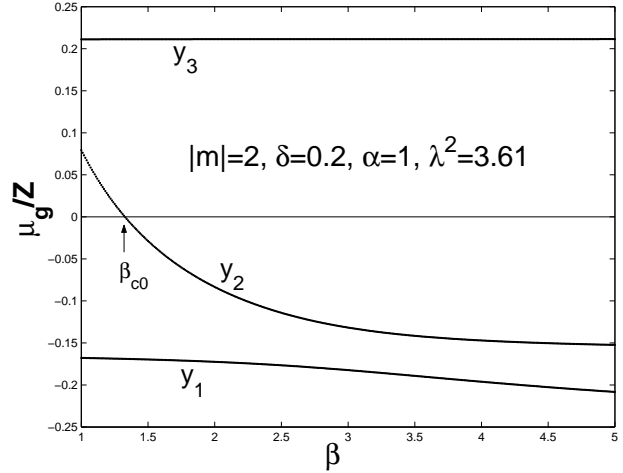}
\caption{\label{f42}Curves of real parts of $\mu_g/Z$ versus $\beta$
for a stationary logarithmic spiral of $|m|=2$, $\delta=0.2$, $\alpha=1$
and $\lambda^2=3.61$. Imaginary parts of $\mu_g/Z$ are not shown here.
The uppermost branch for real part of $\mu_g/Z$ is only discernible for
$\beta<1$ and is not shown here. The branch of real part of $\mu_g/Z$
corresponding to $y_2\equiv D_s^2$ may cross the horizontal $\beta$ axis.}
\end{center}
\end{figure}

For the phase relationship between $\mu_g$ and $Z$, we start
from equation (\ref{118}). For a sufficiently small $\alpha$,
corresponding to a relatively open spiral structure, one may
keep linear terms of $\alpha$ and ignore the $\alpha^2$ term.
It follows that $Z$ is either ahead of or lagging behind $\mu_g$
with a phase difference of about $\sim\pi/2$. On the other hand,
for a sufficiently large $\alpha$ in the tight-winding or WKBJ
regime, one may drop the imaginary part in equation (\ref{118})
(see Lou 2002). When accuracy is not that crucial in a
quantitative analysis, this expression may also be roughly
used in cases of $\alpha\simeq 1$. We can rewrite expression
(\ref{118}) approximately as
\begin{equation}\label{121}
\frac{\mu_g}{Z}\propto\bigg\{\alpha^2
\bigg[\frac{(1+\delta)(1-D_g^2)}
{\delta(1+D_g^2-\lambda^2/2)}
-\frac{\mathcal A \mathcal N_m(\alpha)}
{\mathcal A-\mathcal B}\bigg]\bigg\}^{-1},
\end{equation}
where $\mathcal A$ and $\mathcal B$ are defined by equations
(\ref{116}) and (\ref{117}), respectively. When parameters
$\alpha$, $\beta$, $\delta$ and $\lambda^2$ are specified in
equation (\ref{121}), solutions of $y\equiv D_s^2$ correspond
to values of $\mu_g/Z$. We then present curves of $\mu_g/Z$
versus $\beta$ with parameters $\alpha=1$, $\delta=0.2$,
$\lambda^2=0.09$ and $3.61$ in Figs. \ref{f41} and \ref{f42}. By
equation (\ref{119}), we note again that a smaller $\mu_g/\mu_s$
corresponds to a larger $y\equiv D_s^2$, reminiscent of the same
variation trend of correspondence in the aligned case described
in Section 3. Such a relatively simple correspondence does not
hold for $\mu_g/Z$ that should be analyzed with extra care.
Through extensive numerical explorations, we note that the
highest branch of $\mu_g/Z$ always relates to the lowest $y_3$
branch, while the $y_1$ branch of $\mu_g/Z$ (being always negative)
does not remain always less than the $y_2$ branch of $\mu_g/Z$ as
shown in Figs. \ref{f41} and \ref{f42}. These phase relationships
among branches of $\mu_g/\mu_s$ and $\mu_g/Z$ and the corresponding
branches of $y\equiv D_s^2$ are shown explicitly in
Figs. \ref{f34}$-$\ref{f42}.

The unaligned spiral cases here are fairly similar to the aligned
cases of Section 3, except that the imaginary part of $\mu_g/Z$ does
not vanish in general. For the stationary mode of $y_1$ branch, the
phase relationship between the surface mass density of the two SIDs
(i.e. $\mu_g/\mu_s$) and the phase relationship between the surface
mass density of the MSID and the azimuthal magnetic field
(i.e. $\mu_g/Z$) are always out of phase (see Figs.
\ref{f34}$-$\ref{f42}).\footnote{To be more precise, $\mu_g/Z$ being
`out of phase' means that the real part of $\mu_g/Z$ is negative. As
$\alpha$ ranges from sufficiently small to relatively large values,
the phase difference between $\mu_g$ and $Z$ ranges from $\pi/2$ to
$\pi$, either ahead or lagging behind. Likewise, $\mu_g/Z$ being
`in phase' implies that the phase difference between $\mu_g$ and $Z$
ranges from $0$ to $\pi/2$ as $\alpha$ varies from relatively large
to sufficiently small values.}

For the unaligned perturbation mode of $y_2$ branch in most cases,
$\mu_g/\mu_s$ is in phase, while the real part of $\mu_g/Z$ is always
negative. Taking into account of the imaginary part of $\mu_g/Z$, the
phase difference between $\mu_g$ and $Z$ is $\pi/2\sim\pi$ (see Figs.
\ref{f34} and \ref{f41}). This $y_2$ branch has a zero point at
$\beta_{c0}$ (where $\mu_g=0$) that increases with increasing
$\lambda^2$. When $\lambda^2$ becomes sufficiently large (e.g.
$\lambda^2=3.61$) with $\beta>1$, the value of $\beta_{c0}$ can
become larger than $1$. It is then possible for $\mu_g/\mu_s$ and
$\mu_g/Z$ to become zero at $\beta=\beta_{c0}$ and reverse their
respective signs for $1<\beta<\beta_{c0}$ as shown in Figs. \ref{f35}
and \ref{f42}. Taking into account of the imaginary part of $\mu_g/Z$,
this implies that the phase difference of $\mu_g$ and $Z$ switch from
between $\pi/2\sim\pi$ to between $0\sim\pi/2$ as $\beta$ varies from
$\beta>\beta_{c0}$ to $\beta<\beta_{c0}$. Meanwhile during the same
$\beta$ transition, the phase difference between $\mu_s$ and $\mu_g$
switch from in phase to out of phase. The special situation of
$\mu_g=0$ at $\beta=\beta_{c0}$, occurring for stationary
perturbation mode of $y_2$ branch, is very similar to that of the
aligned case studied earlier. With $\mu_g=J_g=0$ but $\mu_s\neq 0$,
$Z\neq 0$, $\Phi_g\neq 0$ and $U_g\neq 0$, this may correspond to a
unique solution.

For the stationary perturbation mode of $y_3$ branch, both $\mu_g/\mu_s$
and the real part of $\mu_g/Z$ remain positive and therefore $\mu_g$ and
$\mu_s$ are always in phase (see Figs. \ref{f34}$-$\ref{f42}); taking
into account of the imaginary part of $\mu_g/Z$, the phase difference
between $\mu_g$ and $Z$ ranges from $0$ to $\pi/2$.

The analytic expression of $\beta_{c0}$
for $|m|=2$ and $\alpha=1$ is given by
\begin{equation}\label{120}
\begin{split}
\beta_{c0}=\frac{1}{609(1+\delta)}
[{\lambda}^{2}(17\sqrt{21}+68\delta+68)
\quad\qquad \\
+168\,\delta\,+42\,\sqrt {21}+168]\ ,
\end{split}
\end{equation}
that increases with increasing $\lambda^2$. For $\delta=0.2$ and 
$\lambda^2=0.09$ in equation (\ref{120}), one has $\beta_{c0}=0.5589$ 
that is not shown explicitly in Figs. \ref{f34} and \ref{f41} (see 
Fig. E1 in Appendix E).
For $\delta=0.2$ and $\lambda^2=3.61$, one has $\beta_{c0}=1.3271$, 
consistent with Figs. \ref{f35} and \ref{f42}.

\subsubsection{Solution behaviours of unaligned $|m|=1$ case}

\begin{figure}
\begin{center}
\includegraphics[angle=0,scale=0.45]{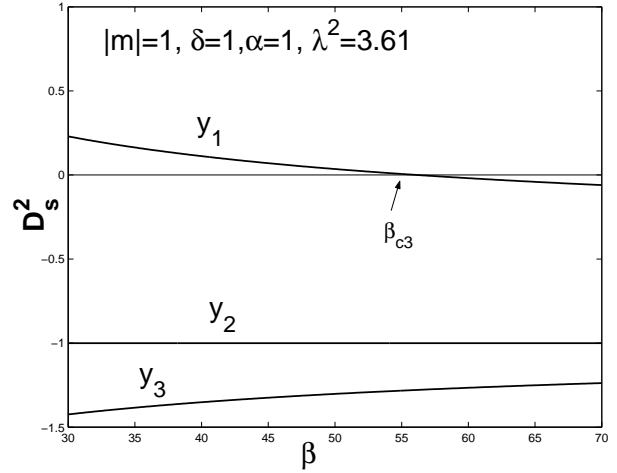}
\caption{\label{f37}Curves of $y\equiv D_s^2$ versus $\beta$ for
stationary unaligned perturbations of $|m|=1$ with $\delta=1$,
$\alpha=1$ and $\lambda^2=3.61$. Solution branch of $y_1\equiv D_s^2$
may go across the horizontal $\beta$ axis and has a root at about $56$.}
\end{center}
\end{figure}

\begin{figure}
\begin{center}
\includegraphics[angle=0,scale=0.45]{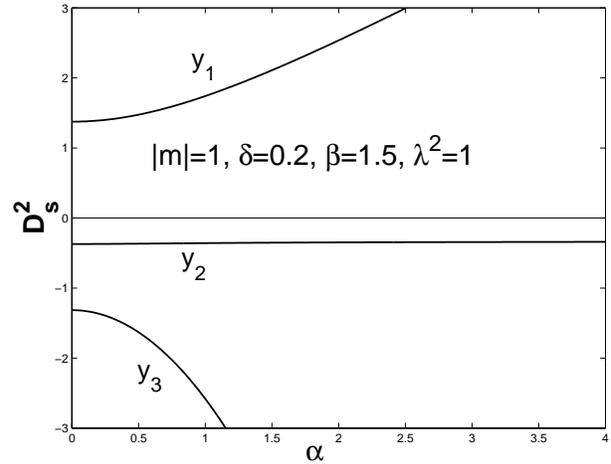}
\caption{\label{f38}Curves of $y\equiv D_s^2$ versus $\alpha$ for
stationary unaligned perturbations of $|m|=1$ with $\delta=0.2$,
$\beta=1.5$ and $\lambda^2=1$.}
\end{center}
\end{figure}

We now turn to the somewhat special case of $|m|=1$. Mathematically,
the qualitative difference between cases of $|m|=1$ and $|m|\geq 2$
is perhaps due to the fact that $\mathcal N_m(\alpha)<1$ for
$|m|\geq 2$ but otherwise for $|m|=1$ (Lou \& Shen 2003). We now
examine behaviours of $y\equiv D_s^2$ of the $|m|=1$ case shown in 
Figs. \ref{f37} and \ref{f38}.

It turns out that solution curves of $|m|=1$ case indeed differ from
those of $|m|=2$ case. For example, the middle and lower branches of
$y\equiv D_s^2$ remain always negative and hence unphysical (see Figs.
\ref{f37} and \ref{f38}). Portions of the upper branch of $y_1\equiv
D_s^2$ may become zero or even negative for a sufficiently large
$\beta$ (Fig. \ref{f37}). That is, when $\beta$ exceeds a critical
value $\beta_{c3}$, $y_1\equiv D_s^2$ would become negative. The
analytical form of this critical $\beta_{c3}$ is given expression
(\ref{126}).
These properties are similar to those in the case of a composite
unmagnetized SID system studied by Lou \& Shen (2003). The three
solution branches of $y\equiv D_s^2$ do not intersect with each other.
The increase of $\beta$ moves both $y_1$ and $y_2$ branches downward
(see Fig. \ref{f37}) and slightly lifts the lowest $y_3$ branch.

One can derive an expression for the critical point of $\beta$
where $y_1=0$. For $\alpha=1$, the analytical form of
this $\beta_{c3}$ is
\begin{equation}\label{126}
\begin{split}
\beta_{c3}\equiv\frac1{(12\delta-8)}
[1+21\delta+5\lambda^2+20\lambda^2\delta
\qquad\qquad\\
+(81+81\delta^2+430\lambda^2\delta
+162\delta+300\lambda^2\delta^2
\quad\\
+130\lambda^2+140\lambda^4\delta
+580\lambda^4\delta^2-15\lambda^4)^{1/2}]\ .
\end{split}
\end{equation}
For $\delta=1$ and $\lambda^2=3.61$, expression (\ref{126})
gives $\beta_{c3}=56.143$, consistent with the upper curve in
Fig. \ref{f37}. There is no essential mathematical difficulty
of obtaining a more general analytical form of $\beta_{c3}$
for arbitrary values of $\alpha$, although the expression may
appear somewhat complicated.

\begin{figure}
\begin{center}
\includegraphics[angle=0,scale=0.45]{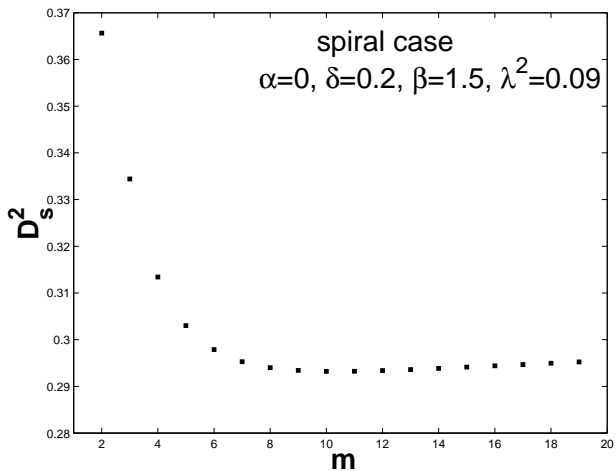}
\caption{\label{f39}Solutions of $D_s^2$ for the $y_2$ branch versus
$|m|$ with parameters $\alpha=0$, $\delta=0.2$, $\beta=1.5$ and
$\lambda^2=0.09$. The smallest $D_s^2$ appears at $|m|=10$.}
\end{center}
\end{figure}

\begin{figure}
\begin{center}
\includegraphics[angle=0,scale=0.45]{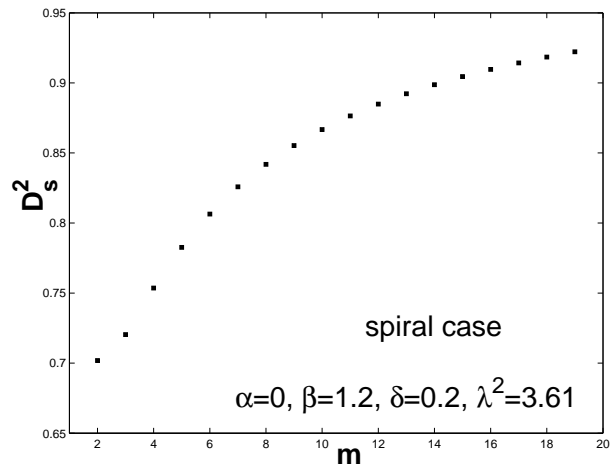}
\caption{\label{f40}Solutions of $D_s^2$ for the $y_2$ branch
versus $|m|$ with parameters $\alpha=0$, $\delta=0.2$, $\beta=1.2$
and $\lambda^2=3.61$. The smallest $D_s^2$ appears at $|m|=2$.}
\end{center}
\end{figure}

Physically, one may exchange the role of $|m|$ and $\alpha$
as they both contribute to the total dimensionless wavenumber.
Given a fixed $\alpha$ value, a composite MSID system may
support different values of $|m|$ (see fig. 3 of Shu et al. 2000
and figs. 14 and 15 of Lou \& Shen 2003). This is shown here in
Figs. \ref{f39} and \ref{f40}. Analogous to the aligned case, the
smallest value of $y_2\equiv D_s^2$ occurs at an azimuthal
periodicity of $|m|=10$ with parameters $\alpha=0$, $\delta=0.2$,
$\beta=1.5$ and $\lambda^2=0.09$. As a comparison, for the same
$\delta=0.2$ and $\alpha=0$ yet a smaller $\beta=1.2$ and a stronger
magnetic field $\lambda^2=3.61$, the smallest value of $y_2\equiv
D_s^2$ then occurs at $|m|=2$. Again, we see that the stability of
a composite (M)SID system appears to be more complicated than those 
of a single (M)SID system by the involvement of three parameters. 
As parameters $\delta$, $\beta$ and $\lambda^2$ vary, vulnerable
configurations with the smallest $D_s^2$ value change accordingly.

\section{DISCUSSION AND SUMMARY}

In this final section we outline some potential applications of our
analysis in the context of a nearby spiral galaxy NGC 6946, and then
summarize our theoretical results and suggestions.

\subsection{Galactic Applications}

By the hydrodynamic density-wave scenario for a disc galaxy,
large-scale high gas density spiral arms are vulnerable to
molecular cloud and star formation activities on smaller scales
and are therefore luminous in optical bands. As synchrotron radio
observations also reveal large-scale spiral structures in
association with optical spiral arms (e.g. Mathewson et al.
1972; Sofue et al. 1986; Beck et al. 1996), it is necessary to
incorporate galactic magnetic field embedded in the interstellar
medium (ISM) and an relativistic cosmic-ray gas (Lou \& Fan 2003)
in the MHD density-wave scenario for a deeper physical
understanding (Fan \& Lou 1996; Lou \& Fan 1998a).

The interlacing of optical and magnetic spiral arms in the nearby
spiral galaxy NGC 6946 was first revealed by Beck \& Hoernes (1996)
in the inner disc portion with an almost rigid rotation. Similar
features of displaced optical and radio arms were also reported
in portions of spiral galaxies IC 342 (e.g. Krause et al. 1989)
and M83 (NGC 5236; e.g. Sukumar \& Allen 1989; Neininger 1992;
Neininger et al. 1993).
%
These observational results led Fan \& Lou (1996) to introduce
theoretical concepts of slow MHD density waves (SMDWs) and fast MHD
density waves (FMDWs) in magnetized spiral galaxies (Lou \& Fan 1998a;
Lou, Yuan \& Fan 2001) and to show that perturbation enhancements of
azimuthal magnetic field and gas mass density of SMDW are phase shifted
by $\gsim\pi/2$ relative to each other. However, several wavelet
analysis on multi-wavelength data of NGC 6946 (e.g. Frick et al. 2000,
2001) further revealed large-scale spiral arms extended well into the
outer disc portion with a more or less flat rotation curve (e.g.
Tacconi \& Young 1989; Sofue 1996; Ferguson et al. 1998), that seems
to challenge earlier arguments that SMDWs are largely confined within
the inner disc portion of almost rigid rotation.

For coplanar MHD perturbation structures of stationary logarithmic
spirals in an MSID model with a flat rotation curve, Lou (2002) has
shown the existence of SMDWs. For stationary SMDWs of Lou (2002),
logarithmic spiral enhancements of gas surface mass density and azimuthal
magnetic field are phase shifted by about $\pi/2\sim\pi$. As effects of
long-range self-gravity, magnetic field and disc differential rotation
are included in the model analysis, SMDW patterns in a single MSID
model can indeed support extended manifestations of interlaced optical
and magnetic spiral arms that persist well into the outer disc portion
with a flat rotation curve (Lou \& Fan 2002). Moreover, Lou (2002)
further suggested that significant phase shifts between $0\sim\pi/2$
related to stationary logarithmic spiral structures of FMDWs within a
single MSID.

Given idealizations of our composite MSID model, the three possible
stationary logarithmic spiral patterns might be conceptually relevant
to magnetized spiral galaxies. Specifically, for the upper two solution
branches of $y_1$ and $y_2$, the phase difference of the stationary
logarithmic spiral enhancements between azimuthal magnetic field and
surface mass gas density is between $\pi/2\sim\pi$,
supporting the idea that
interlaced optical and magnetic spiral arms may persist through the
inner disc portion to the outer portion (with a phase difference
$\geq\pi/2$). For a sufficiently large $\alpha$ in the tight-winding
or WKBJ regime, this phase difference approaches $\sim\pi$, while for
a sufficiently small $\alpha$ in the open regime, this phase
difference approaches $\pi/2$. Note that in our modelling, the phase
difference of $0\sim\pi/2$ occurs mainly through the $y_3$ solution
branch that is usually unphysical with $D_s^2$ being negative.

As sketched earlier (e.g. Lou \& Fan 1998a, 2000a, b; Lou et al. 2002),
a typical disc galaxy involves a stellar disc, a magnetized gas disc
and a massive dark matter halo. Our theoretical model of a composite
MSID system, idealized and simplified in many aspects, does contain
these basic elements. With necessary qualifications in mind, it is
then of considerable interest to reveal possible types of large-scale
structures and their characteristic features in the stationary MHD
density-wave scenario and relate them to galactic diagnostics and
observations.

For the $y_1$ solution with largest value of $D_s^2$, stationary
logarithmic spiral density arms in gaseous MSID and in stellar SID
are spatially out of phase; in particular, spiral gas density and
magnetic field arms are siginificantly phase shifted relative to
each other with a phase difference between $\pi/2 \sim\pi$.

For the $y_2$ solution with middle value of $D_s^2$, stationary
logarithmic spiral density arms in gaseous MSID and in stellar SID
are spatially in phase; meanwhile, spiral gas density and magnetic
field arms are siginificantly phase shifted relative to each other
with a phase difference between $\pi/2 \sim\pi$ for $\beta>1$ and
$\beta>\beta_{c0}$. At the qualitative level, this case appears
comparable to interlaced optical and synchrotron radio arms of
NGC 6946 (Beck \& Hoernes 1996; Fan \& Lou 1996; Frick et al. 2000,
2001; Lou \& Fan 2002). 
On the other hand, for $1<\beta<\beta_{c0}$, stationary logarithmic
spiral density arms in gaseous MSID and in stellar SID are spatially
out of phase; meanwhile, spiral gas density and magnetic field arms
are phase shifted relative to each other with a phase difference
between $0\sim \pi/2$.

By the above analogy of unaligned spiral configurations, our results
for aligned stationary MHD perturbations with $|m|=2$ may bear possible
diagnostics for optical and synchrotron radio observations of galactic
bar structures and their spatial phase relationships. It is
straightforward to compare Figs. \ref{f16}$-$\ref{f22} of the aligned
cases with Figs. \ref{f34}$-$\ref{f42} and infer spatial phase
relationships of stellar bars, gas bars and synchroton radio bars
involving magnetic fields. For example, by Figs. \ref{f16} and \ref{f20},
the $y_1$ solution describes a configuration in which the stellar density
bar and the gas density bar are phase shifted relative to each other with
a $\pi$ phase difference, while the gas density bar and the synchrotron
radio bar (i.e. magnetic bar) are phase shifted relative to each other
with a $\pi$ phase difference.
The $y_2$ solution describes a configuration in which the stellar density
bar and the gas density bar coincide, while the gas density bar and the
synchrotron radio bar (i.e. magnetic bar) are phase shifted relative to
each other with a $\pi$ phase difference.
By Figs. \ref{f18} and \ref{f22}, the $y_1$ solution describes a
qualitatively similar bar configuration as the $y_1$ configuration of
Figs. \ref{f16} and \ref{f20}. For $\beta>\beta_{c0}$ [see definition
(\ref{98}) for $\beta_{c0}$], the $y_2$ solution describes a
qualitatively similar bar
configuration as the $y_2$ configuration of Figs. \ref{f16} and \ref{f20}.
For $1<\beta<\beta_{c0}$, the $y_2$ solution describes a configuration in
which the stellar density bar and the gas density bar are phase shifted
relative to each other with a $\pi$ phase difference, while the gas
density bar and the synchrotron radio bar (i.e. magnetic bar) coincide.
In context of observations, the prominent bar structure of the nearby 
barred spiral galaxy M83 (e.g. Sukumar \& Allen 1989; Neininger 1992; 
Neininger et al. 1993) is worth being analyzed in more details in both 
optical and synchrotron radio bands.

\subsection{Summary}

In terms of theoretical development of MSID model and in reference to
previous theoretical analyses of Shu et al. (2000) on zero-frequency
(i.e. stationary) aligned and unaligned perturbation configurations of
an isopedically magnetized SID, of Lou (2002) on stationary aligned and
unaligned perturbation configurations of a coplanarly magnetized SID,
and of Lou \& Shen (2003) on stationary aligned and unaligned
perturbation structures in a composite system of two-fluid SIDs, we
have constructed analytically in this paper stationary aligned and 
unaligned perturbation configurations in either full or partial 
composite system consisting of a stellar SID and a gaseous MSID. While 
this composite model of one SID and one MSID is highly idealized in 
many aspects, it does contain several necessary and more realistic
elements that are pertinent to structures and dynamics of disc
galaxies. Galactic structures may not be stationary in general, yet
the stationary global MHD perturbation structures might represent a
subclass, provide insights for a disc galaxy, and serve as benchmarks
for numerical MHD simulations. Given our model specifications, we have
reached the following conclusions and suggestions.

For aligned MHD perturbation structures, we derived stationary
dispersion relation (\ref{46}) for both full and partial composite
system of one SID and one MSID. Mathematically, there are three
possible $D_s^2$ values for the square of dimensionless rotation
parameter $D_s$ of the stellar SID (or equivalently, $D_g^2$).
Physically, only those solutions with $D_s^2>0$ and $D_g^2>0$
may be conceptually applicable.

In the aligned case of barred configurations with $|m|=2$ and $F=1$
for a full composite MSID system, the uppermost branch remains always
physically possible, with spatial phase relationships $\mu_g/\mu_s<0$
and $\mu_g/Z<0$ (where $Z$ is the amplitude of azimuthal magnetic
field perturbation); that is, both $\mu_g$ and $\mu_s$ pair and
$\mu_g$ and $Z$ pair are out of phase. These qualitative features
for possible bars may be detectable observationally. When parameter
$\beta$ is smaller than a critical value $\beta_{c1}$ defined by
(\ref{94}), the middle solution branch of $y\equiv D_s^2$ is positive
and thus physically possible. Corresponding phase relationships of
$\mu_g/\mu_s$ and of $\mu_g/Z$ are in phase and out of phase,
respectively. But when $\beta<\beta_{c0}$ defined by expression
(\ref{98}), $\mu_g/\mu_s$ and of $\mu_g/Z$ can become zero or even
change to opposite sign (see Figs. \ref{f18} and \ref{f22}). This
special $\mu_g=0$ case might be an additional possibility. The
lowest branch of $y\equiv D_s^2$ is physically possible only when
$\beta<\beta_{c2}$ defined by expression (\ref{95}). In many cases
$\beta_{c2}$ is smaller than $1$, this branch is thus invalid.
However in some cases (see Fig. \ref{f12}), this branch can rise up
to become positive and thus physically possible. The corresponding
phase relationships of $\mu_g/\mu_s$ and $\mu_g/Z$ are both always
positive and thus in phase.

For aligned perturbations, the case of $|m|=0$ can be made to
effectively rescale the axisymmetric background. Eccentric $|m|=1$
displacements may occur for arbitrary $D_s^2$ values in a full
composite MSID system. In contrast, for a partial composite MSID
system, $D_s^2$ can no longer be arbitrary for $|m|=1$ (Lou 2002;
Lou \& Shen 2003).

For coplanar MHD perturbation structures of unaligned logarithmic
spirals, we obtained the stationary dispersion relation (\ref{62})
for both full and partial composite MSID system. Mathematically,
there are three possible sets of solutions for $D_s^2$ or
equivalently $D_g^2$.

For stationary logarithmic spiral configurations with $|m|=2$
and $F=1$ in a full composite MSID system, the uppermost $D_s^2$
solution branch is always physically possible, with corresponding
$\mu_g/\mu_s$ being always out of phase and with the phase
difference between $\mu_g$ and $Z$ in the range $\pi/2\sim\pi$.
The middle branch of $D_s^2$ solution is physically possible
when $\beta<\beta_{c1}$ defined by expression (\ref{113}). There
also exists a special $\beta_{c0}$ defined by expression (\ref{120});
when $\beta$ decreases below $\beta_{c0}$, the corresponding
$\mu_g/\mu_s$ changes from being in phase to being out of phase
(see Fig. \ref{f35}), and the phase difference between $\mu_g$ and
$Z$ changes from being within $\pi/2\sim\pi$ to being within
$0\sim\pi/2$ (see Fig. \ref{f42}). The special $\mu_g=0$ case may
occur as well. The lowest branch of $D_s^2$ solution can be
physically possible when $\beta<\beta_{c2}$ defined by expression
(\ref{114}). Similar to the aligned case, while this lowest $D_s^2$
solution branch is often invalid as $\beta_{c2}$ is usually less
than $1$, it may become physically possible for $\beta$ within
interval $[1,\beta_{c2}]$ (see Figs. \ref{f32} and \ref{f33}) for
cases with $\beta_{c2}>1$. Correspondingly, $\mu_g/\mu_s$ remains
always in phase and the phase difference between $\mu_g$ and $Z$
falls in the range $0\sim\pi/2$.

For the marginal case of $|m|=0$ with radial propagations, the
collapse and ring fragmentation regimes exist, while the lower
ring fragmentation curve, which is a novel feature in a single
MSID (Lou 2002), does not often appear to be positive. For cases
with $\beta\simeq 1$, this lower ring fragmentation branch may
rise to become positive and will descend with increasing $\lambda^2$
as shown in Figs. \ref{f9} and \ref{f10}. This qualitative trend
differs from that of a single MSID studied by Lou (2002). We
further examined trends of variation as parameters $\delta$,
$\beta$ and $\lambda^2$ vary, and concluded that the mutual
gravitational coupling between the two SIDs appears to reduce
collapse regime but to increase the danger of ring fragmentation.
Physical interpretations on collapse and ring fragmentation
regimes are provided (Shu et al. 2000; Lou 2002; Lou \& Shen
2003; Shen \& Lou 2003). Consistent with prior analyses, we
realize that the ring fragmentation curve has a fixed vertical
asymptote at $\mathcal N_0(\alpha)(\alpha^2+1/4)=2$ with
$\alpha\sim 1.759$ approximately; this condition for the vertical
asymptote remains exactly the same as that of Shu et al. (2000),
Lou (2002) and Lou \& Shen (2003). In reference to these previous
analyses, it is highly suggestive that the effective $Q$ parameter in
a composite MSID system should be closely related to the minimum
of the ring fragmentation curve (Shen \& Lou 2003).

The unaligned spiral case of $|m|=1$ is somewhat special. Being
negative, the two lower solution branches of $D_s^2$ are both
unphysical. The uppermost $D_s^2$ solution branch is physically
possible when $\beta<\beta_{c3}$. The analytic expression of
$\beta_{c3}$ with $\alpha=1$ is given by equation (\ref{126}).

We have also derived the MHD virial theorem in a composite MSID
system, following the suggestion of Lou (2002) that ratio
$\mathcal T/|\mathcal W-\mathcal M|$ may play the role of ratio
$\mathcal T/|\mathcal W|$ in determining the onset criterion for
bar-like instability when coplanar magnetic field is involved.
Similar to a composite system of two fluid SIDs of Lou \& Shen
(2003), we found that solution $D_s^2$ of the middle $y_2$ branch
may correspond to $\mathcal T/|\mathcal W-\mathcal M|$ ratios
being considerably lower than the oft-quoted value of $\sim 0.14$
(Ostriker \& Peebles 1973; Binney \& Tremaine 1987)). This
indicates that a composite (M)SID system tends to be less stable.
This is just one perspective of showing that the stability in a
composite MSID system is far more complicated than that of a
single MSID. In particular, additional solution branches allowed
by more dynamic freedoms as well as more parameters $\delta$,
$\beta$ and $\lambda^2$ do introduce extra dimensions for the
stability problems. For example, as these dimensionless
parameters vary, most vulnerable configurations related to the
smallest $D_s^2$ change.

While our model for a composite system of MSIDs is highly idealized,
theoretical results obtained here provide a conceptual basis and
useful clues for optical and synchrotron radio observational
diagnostics of spiral galaxies of stars and gaseous ISM with magnetic
field coupled by mutual gravity on large scales. In particular, our
analytical solutions for stationary MHD perturbation configurations
in a composite MSID system as well as analytical expressions for
critical points are important and valuable for benchmarking numerical
codes designed for large-scale dynamics of magnetized gaseous discs.

\section*{Acknowledgments}
This research was supported in part by the ASCI Center for
Astrophysical Thermonuclear Flashes at the University of Chicago
under Department of Energy contract B341495, by the Special Funds
for Major State Basic Science Research Projects of China, by the
Tsinghua Center for Astrophysics, by the Collaborative Research
Fund from the National Natural Science Foundation of China (NSFC)
for Young Outstanding Overseas Chinese Scholars (NSFC 10028306) at
the National Astronomical Observatory, Chinese Academy of Sciences,
by NSFC grant 10373009 at the Tsinghua University,
and by the Yangtze Endowment from the Ministry of Education through
the Tsinghua University. Affiliated institutions of Y.Q.L. share
the contribution.

\begin{appendix}
\section{}
For the spiral case of $|m|=2$, a substitution of expressions
(\ref{11}), (\ref{15}), (\ref{20}), (\ref{21}) and (\ref{87})
into equation (\ref{62}) would lead to a cubic algebraic
equation in terms of $y\equiv D_s^2$. By setting the constant
term of this cubic equation equal to zero, we derive conditions
for $y=0$, namely
\begin{equation}
\mathcal A\beta^2+\mathcal B\beta+\mathcal C=0\ ,
\end{equation}
where the three coefficients $\mathcal A$,
$\mathcal B$ and $\mathcal C$ are defined by
\begin{eqnarray}
\mathcal A\equiv-578\,{\it \mathcal
N_2(\alpha)}\,\delta-32\,{\alpha}^{4}{\it \mathcal
N_2(\alpha)}\,\delta-272\,{\it \mathcal
N_2(\alpha)}\,\delta\,{\alpha}^{2}\nonumber\\ +272\,{\it \mathcal
N_2(\alpha)}+64\,{\alpha}^{2}{\it \mathcal N_2(\alpha)}
-272-64\,{\alpha}^{2}\nonumber\\-64\,\delta\,{\alpha}^{2}
-272\,\delta\ ,
\end{eqnarray}
\begin{equation}
\begin{split}
\mathcal B\equiv-400\,{\alpha}^{2}{\it\mathcal
N_2(\alpha)}+272\,{\it\mathcal
N_2(\alpha)}\,\delta\,{\alpha}^{2}+32\,{\alpha}^{4}
\qquad\qquad\qquad\qquad\\
+32\,{\alpha}^{4}{\it\mathcal
N_2(\alpha)}\,\delta+32\,{\alpha}^{4}{ \lambda}^{2}+578\,
{\it\mathcal N_2(\alpha)}\,\delta-34\,{\it\mathcal
N_2(\alpha)}\,{\lambda}^{2}
\qquad\\
-1122\,{\it\mathcal
N_2(\alpha)}+400\,{\alpha}^{2}+1122+144\,\delta\,{\lambda}^{2}{
\alpha}^{2}+32\,{\alpha}^{4}\delta\,{\lambda}^{2}
\qquad\\
+34\,\delta\,{\lambda }^{2}-144\,{\it\mathcal
N_2(\alpha)}\,{\lambda}^{2}{\alpha}^{2}+34\,{\lambda}^{2}
-32\,{\alpha}^{4}{\it\mathcal N_2(\alpha)}+1122\,\delta
\qquad\\
+16\,{\alpha}^{4}{\it\mathcal N_2(\alpha)}\,
\delta\,{\lambda}^{2}+400\,\delta\,{\alpha}^{2}
+32\,{\alpha}^{4}\delta
+144\,{\lambda}^{2}{\alpha}^{2}
\qquad\\
+120\,{\it\mathcal N_2(\alpha)}\,\delta\,{\lambda}^{2}{\alpha}^{2}+221\,
{\it\mathcal N_2(\alpha)}\,\delta\,{\lambda}^{2}-32\,{\alpha}^{4}
{\it\mathcal N_2(\alpha)}\,{\lambda}^{2}
\qquad\quad\\
\end{split}
\end{equation}
and
\begin{equation}
\begin{split}
\mathcal C
\equiv-850-264\,{\lambda}^{2}{\alpha}^{2}-264\,\delta\,{\lambda}^{2}{\alpha}
^{2}-32\,{\alpha}^{4}+96\,{\lambda}^{4}\delta\,{\alpha}^{2}
\quad\\
+96\,{\lambda}^{4}{\alpha}^{2}+16\,{\alpha}^{4}{\lambda}^{4}\delta
-48\,{\alpha}^{4}\delta\,{\lambda}^{2}
-850\,\delta-32\,{\alpha}^{4}\delta
\qquad\\
+16\,{\lambda}^{4}{\alpha}^{4}-336\,{\alpha}^{2}-255\,{\lambda}^{2}-255\,
\delta\,{\lambda}^{2}
\qquad\\
-336\,\delta\,{\alpha}^{2}+119\,{\lambda}^{4}
\delta+119\,{\lambda}^{4}-48\,{\alpha}^{4}{\lambda}^{2}
\qquad\\
+850\,{\it\mathcal N_2(\alpha) }+264\,{\it\mathcal
N_2(\alpha)}\,{\lambda}^{2}{\alpha}^{2}-96\,{\lambda}^{4}
{\it\mathcal N_2(\alpha)}\,{\alpha}^{2}
\qquad\\
+255\,{\it \mathcal
N_2(\alpha)}\,{\lambda}^{2}-119\,{\lambda}^{4}{\it\mathcal
N_2(\alpha)}+336\,{\alpha}^{2}{\it\mathcal N_2(\alpha)}
\qquad\\
+32\,{\alpha}^{4}{\it\mathcal
N_2(\alpha)}-16\,{ \alpha}^{4}{\lambda}^{4}{\it\mathcal
N_2(\alpha)}+48\,{\alpha}^{4}{\it\mathcal N_2(\alpha)}\,
{\lambda}^{2}\ .\qquad
\end{split}
\end{equation}

\section{}
\begin{figure}
\begin{center}
\includegraphics[angle=0,scale=0.45]{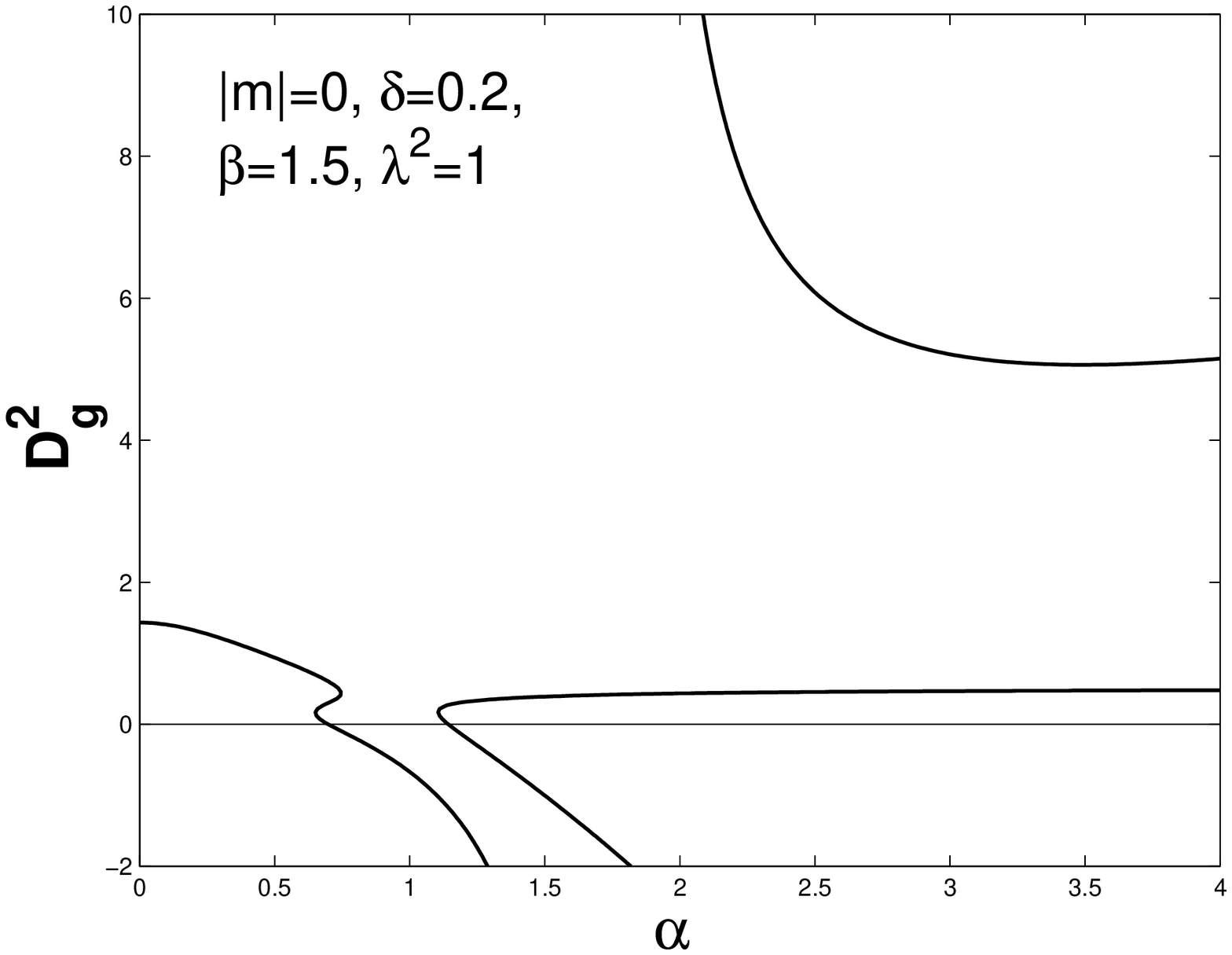}
\caption{\label{f4}The marginal stability curve of $D_g^2$
versus $\alpha$ with parameters $|m|=0$, $\delta=0.2$,
$\beta=1.5$ and $\lambda^2=1$. Fig. \ref{f5} presents
corresponding curves of $D_s^2$ versus $\alpha$. }
\end{center}
\end{figure}

\begin{figure}
\begin{center}
\includegraphics[angle=0,scale=0.45]{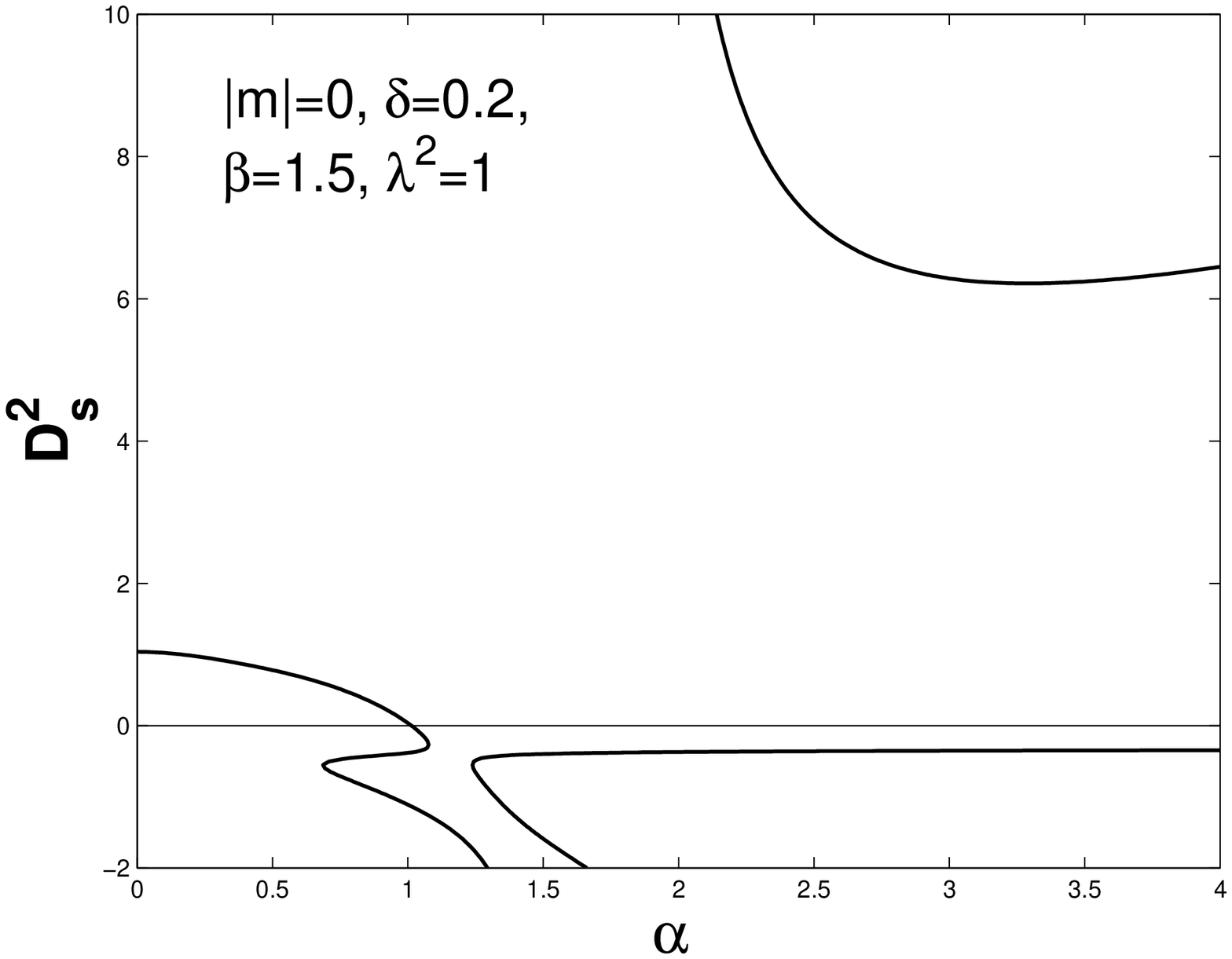}
\caption{\label{f5}The marginal stability curve of $D_s^2$ versus
$\alpha$ with parameters $|m|=0$, $\delta=0.2$, $\beta=1.5$,
$\lambda^2=1$. Fig. \ref{f4} presents corresponding curves of
$D_g^2$ versus $\alpha$.}
\end{center}
\end{figure}

In the presence of gravitational coupling between one SID and
one MSID, we expect three sets of independent mathematical
solutions in general. However, for cases of marginal stabilities
for spiral configurations with $|m|=0$, the solution curves
qualitatively resemble those determined for a single SID (Shu
et al. 2000) in most cases, and the two extra solutions turn out
to be negative. Moreover, the lower ring fragmentation curve in
a single MSID (see fig. 1 of Lou 2002) remains often indiscernible
(see Figs. \ref{f1}$-$\ref{f9}). Here, we provide a complete
structure for solution curves in Figs. \ref{f4} and \ref{f5}.

We note that the basic positive $D_g^2$ versus $\alpha$ profile is
qualitatively similar to fig. 1 of Lou (2002) (see Fig. \ref{f4}).
However, the $D_s^2$ curve corresponding to the lower ring
fragmentation curve of $D_g^2$ is negative in most cases, indicating
that this branch is physically invalid (see Figs. \ref{f4} and
\ref{f5}). On the other hand, when $\beta$ approaches $1$, this
branch of $D_s^2$ can be raised above zero and thus becomes
physically valid in this limit (see Fig. \ref{f9}).

\section{}

In this Appendix C, we provide the complete form of the
cubic equation of $y\equiv D_s^2$ for the aligned case of
coplanar MHD perturations in a composite MSID system for
interested readers or potential users in numerical MHD
simulations. From dispersion relation (\ref{46}) and
relation (\ref{87}), we readily derive  
\begin{equation}
\mathcal Ay^3+\mathcal By^2+\mathcal Cy+\mathcal D=0\ ,
\end{equation}
where the four coefficients $\mathcal A$, $\mathcal B$,
$\mathcal C$ and $\mathcal D$ are defined by
\begin{equation}
\begin{split}
\mathcal A\equiv-16\, \left| m \right| {\beta}^{2}-4\,m^{4}
{\beta}^{2}(1+\delta)
+16\,{m}^{2} \left| m \right| {\beta}^{2}
\qquad\\
-4\,{m}^{4} \left| m \right|
{\beta} ^{2}-16\, \left| m \right|
\delta\,{\beta}^{2}
+8\,{m}^{2}\delta\,{\beta}^{2}
\quad\qquad\\
-4\,{m}^{4} 
\left| m \right| \delta\,{\beta}^{2}+8\,
m^{2}{\beta}^{2}
+16\,{m}^{2} \left| m
 \right| \delta\,{\beta}^{2}\ ,\qquad
\end{split}
\end{equation}

\begin{equation}
\begin{split}
\mathcal B\equiv-4\,{m}^{4}\delta\,{\beta}^{2}-16\, 
m^2\beta-4\,{m}^{4} \left| m \right|
{\beta}^{2}
+2\,{m}^{4}\delta\,{ \lambda}^{2}\beta
\\
-32\, \left| m \right| {\beta}^{2}+32\, \left| m
 \right| \delta\,\beta
-4\,{m}^{4} \left| m \right| \delta\,{\beta}^{2}
\\
+32\, \left| m \right| \beta-40\,{m}^{2} \left| m \right| \delta\,
\beta
+4\,{m}^{4}\delta\,\beta+16\,{m}^{2}\delta\,{\beta}^{2}
\\
+24\, m^2{\beta}^{2}+24\,{m}^{2} \left|
m \right| {\beta}^{2}-32\, \left| m \right|
\delta\,{\beta}^{2}
\\
+12\,{m }^{4} \left| m \right|
\delta\,\beta-8\,{m}^{2}\delta\,\beta-12\, m^4{\beta}^{2}
\\
+4\,\delta\, {\lambda}^{2}\beta+12\,m^4\beta
+24\,{m}^{2} \left| m \right| \delta\,{\beta}^{2}
\\
-6\,{m}^{2}
\delta\,{\lambda}^{2}\beta+12\,{m}^{4} \left| m \right|
\beta-40\,{m}^ {2} \left| m \right| \beta \ ,
\end{split}
\end{equation}

\begin{equation}
\begin{split}
\mathcal C\equiv24\,m^2\left|m\right|
\delta+2\,{\lambda}^2m^2-8\,m^4 
+24\,m^2\left|m\right|+m^4{\lambda}^4
\\
-{\lambda}^{4} m^2
-16\, \left| m \right| \delta+{m}^{4}{\lambda}^{4}
\left| m \right| -32\, m^2\beta +24 \, m^2{\beta}^{2}
\\
-8\,{m}^{4}
\left| m \right| -16\, \left| m \right| 
+6\,{m}^{2}{\lambda}^{2}\left| m \right| \delta
-2\,m^4{\lambda}^2\left|m\right|\delta
\\
-3\, \left| m \right| {\lambda}^{4}{m}^{2}\delta
+6\,{m}^{2}{ \lambda}^{2} \left| m \right|\ 
-3\,{m}^{2}{\lambda}^{4} \left| m\right| 
\\
-2\,{m}^{4}{\lambda}^2\left|m\right| 
-2\,{m}^{4}{\lambda }^2 
-4\, \left|m\right|{\lambda}^2
-4\, \left|m\right|\delta\,{\lambda}^2
\\
-4\,m^2
\delta\,{\lambda}^{2}\beta -8\,{m}^{2}\delta\,\beta
+4\,m^4\delta\, \beta^2
+8\,m^2\delta\,{\beta}^2
\\
-8\,m^4 \left|m \right| 
\delta +8\, m^2-24\,m^2\left|m\right|\beta
+24\,m^4\beta 
\\
-12\,m^4{\beta}^2
+4\,m^4\left|m\right|{\beta}^2
-24\,m^2\left|m\right|\delta\, \beta
+m^4{\lambda}^4\left|m\right|\delta
\\
+4\,m^4\left|m\right|\delta\,{\beta}^2
+4\,\delta\,{\lambda}^{2}\beta
-16\,\left| m \right|{\beta}^{2}
+32\,\left|m\right|\beta
\\
+2\,\lambda^4\left|m\right|
-16\, \left|m\right|\delta\,{\beta}^2
+32\,\left|m\right|\delta\,\beta
+2\,{\lambda}^4\left|m\right|\delta\ 
\end{split}
\end{equation}
and

\begin{equation}
\begin{split}
\mathcal D\equiv-8\,{m}^2 \left| m \right|
\delta+2\,{\lambda}^{2}m^2-8\, m^4 
-8\,{m}^{2} \left| m \right| +{m}^{4}{\lambda}^{4} 
\qquad\\
-{\lambda}^{4}m^2
-{m}^{4}{\lambda}^{4} \left| m \right| -16\, m^2\beta
+8\, m^2 {\beta}^{2} +8\,{m}^{4} \left| m \right| 
\qquad\\
-2\,{m}^{2}{
\lambda}^{2} \left| m \right| \delta+2\,{m}^{4}{\lambda}^{2}
\left| m \right| \delta
+ \left| m \right| {\lambda}^{4}{m}^{2}\delta
-2\,{m}^{2}{\lambda}^{2} \left| m \right| 
\\
+{m}^{2}{\lambda}^{4} \left| m
 \right| 
+2\,{m}^{4}{\lambda}^{2} \left| m \right| 
-2\,{m}^{4}{\lambda }^{2} 
-4\,{m}^{4}\delta\,\beta
+2\,{m}^{2}\delta\,{\lambda}^{2}\beta
\\
-2\,{m}^{4}\delta\,{\lambda}^{2}
\beta+4\,{m}^{4}\delta\,{\beta}^{2}+8\,{m}^{4} \left| m \right|
\delta +8\, m^2 
-8\,{m}^{2} \left| m
 \right| {\beta}^{2}
\\
+16\,{m}^{2} \left| m \right| \beta
+12\,{m}^{4}\beta-12\,{m}^{4} \left| m
 \right| \beta-4\,{m}^{4}{\beta }^{2}
\\
+4\,{m}^{4}\left| m \right|
{\beta}^{2}-12\,{m}^{4}\left| m
 \right|\delta\,\beta+16\,{m}^{2}\left|m\right|\delta\,\beta
\\
-{m}^{4}{\lambda}^{4} \left| m \right|\delta-8\,{m}^{2}
\left| m\right|\delta\,{\beta}^{2}+4\,{m}^{4}\left| m\right|
\delta\,{\beta}^{2}\ .
\end{split}
\end{equation}
One can solve cubic equation (C1) for possible aligned MHD
perturbation configurations in a composite MSID system by 
specifying parameters $m$, $\delta$, $\beta$ and $\lambda$.

\section{}

In this Appendix D, we provide the complete form of the cubic 
equation of $y\equiv D_s^2$ for the unaligned spiral case of 
coplanar MHD perturbations in a composite MSID system for
interested readers or potential users in numerical MHD
simulations. From dispersion relation (\ref{46}) and relation
(\ref{87}), we readily derive 
\begin{equation}
\mathcal Ay^3+\mathcal By^2+\mathcal Cy+\mathcal D=0\ ,
\end{equation}
where the four coefficients $\mathcal A$, $\mathcal B$,
$\mathcal C$ and $\mathcal D$ are defined by
\begin{equation}
\begin{split}
\mathcal A\equiv32\,{m}^{2}\mathcal
N_m(\alpha){\alpha}^{2}{\beta}^{2}+32\,{m}^{2}\mathcal
N_m(\alpha)\delta\,{\alpha}^{2}{ \beta}^{2}
\qquad\qquad\\
-16\,\mathcal N_m(\alpha){\beta}^{2}
+128\,\delta\,{\beta}^{2}-128\,{m}^{2}{
\beta}^{2}+32\,{m}^{4}{\beta}^{2}+128\,{\beta}^{2}
\quad\\
-16\,\mathcal
N_m(\alpha)\delta\,{\beta }^{2}-64\,\mathcal
N_m(\alpha){\alpha}^{2}{\beta}^{2}-56\,\mathcal
N_m(\alpha){m}^{2}{\beta}^{2}
\qquad\\
+32\,{m}^{4} 
\mathcal N_m(\alpha){\beta}^{2}-64\,
\mathcal N_m(\alpha)\delta\,{\alpha}^{2}{\beta}^{2}
+32\,{m}^{4}\mathcal N_m(\alpha)\delta\, {\beta}^{2}
\qquad\\
-56\,\mathcal N_m(\alpha)\delta\,{m}^{2}{\beta}^{2}
+32\,{m}^{4}\delta\,{\beta}^2-128\,{m}^{2}\delta\,{\beta}^{2}\ ,\qquad
\end{split}
\end{equation}

\begin{equation}
\begin{split}
\mathcal B\equiv-16\,{m}^{4}\mathcal
N_m(\alpha)\delta\,{\lambda}^{2}\beta-16\,{m}^{2}\mathcal
N_m(\alpha)\delta\,{\lambda}^2{\alpha}^{2}\beta
\\
-32\,{m}^{2}\mathcal
N_m(\alpha){\lambda}^{2}{\alpha}^{2}\beta
\\
+30\,\mathcal
N_m(\alpha) \beta-128\,{m}^{2}\mathcal
N_m(\alpha){\alpha}^{2}\beta+64\,\mathcal
N_m(\alpha)\delta\,{\alpha}^{2}\beta
\\
-32
\,{m}^{4}\mathcal N_m(\alpha)\delta\,\beta-32\,{m}^2\mathcal
N_m(\alpha)\delta\,{\alpha}^2\beta-32\,{m}^2{\alpha}^{2}\delta\,\beta
\\
+56\,\mathcal
N_m(\alpha)\delta\,{m}^2\beta+64\,{\alpha}^2\beta-240\,\delta\,\beta
\\
-96\,{m}^{4}\beta+96\,\mathcal
N_m(\alpha){m}^{2}\beta+112\,\mathcal N_m(\alpha){
\alpha}^{2}\beta
\\
+64\,{\alpha}^{2}\delta\,\beta-32\,{m}^{2}{\alpha}^{2}
\beta-96\,{m}^{4}\mathcal N_m(\alpha)\beta
\\
+16\,\mathcal
N_m(\alpha)\delta\,\beta+16\,{\lambda}^{2}\beta-32\,
{m}^{2}\delta\,{\lambda}^{2}{\alpha}^{2}\beta
\\
+64\,\delta\,{\lambda}^{2}{\alpha}^{2}\beta-24\,\mathcal
N_m(\alpha)\delta\,{\lambda}^{2}\beta-32\,{m}^{2}{\lambda
}^{2}{\alpha}^{2}\beta
\\
-8\,\mathcal
N_m(\alpha){\lambda}^{2}{m}^{2}\beta-16\,\mathcal
N_m(\alpha){\lambda}^{2
}{\alpha}^{2}\beta-32\,{\alpha}^{4}\mathcal
N_m(\alpha){\lambda}^{2}\beta
\\
-2\,\mathcal
N_m(\alpha){\lambda}^{
2}\beta+64\,{\lambda}^{2}{\alpha}^{2}\beta-8\,{\lambda}^{2}{m}^{2}
\beta
\\
-32\,{\alpha}^{4}\mathcal N_m(\alpha)\beta
+32\,\mathcal N_m(\alpha)\delta\,{\alpha}^2\beta\,{\lambda}^2
+44\,\mathcal N_m(\alpha)\delta\,m^2\beta\,{\lambda}^2
\\
-144\,\mathcal N_m(\alpha)\delta\,{\alpha}^2{\beta}^2
-128\,\mathcal N_m(\alpha)\delta\,m^2{\beta}^2
-32\,m^2{\alpha}^2\delta\,{\beta}^{2}
\\
+96\,m^2\mathcal N_m(\alpha){\alpha}^2{\beta}^2
+32\,m^4{\beta}^2-48\,\mathcal N_m(\alpha)\beta^2
+64\,{\alpha}^2{\beta}^2
\\
+272\,\delta\,{\beta}^{2}
-168\,\mathcal N_m(\alpha)m^2{\beta}^2
-192\,\mathcal N_m(\alpha){\alpha}^2{\beta}^2
\\
+64\, \alpha^2\delta\,{\beta}^2
-32\,m^2{\alpha}^2{\beta}^2
+96\,m^4\mathcal N_m(\alpha){\beta}^2
\\
-34\,\mathcal N_m(\alpha)\delta\,{\beta}^2
+32\,m^4\mathcal N_m(\alpha)\delta\, \beta^2
-32\,{\alpha}^{4}\mathcal N_m(\alpha)\delta\,{\beta}^2
\\
-200\,m^2{\beta}^2-240\,\beta+272\,{\beta}^2
\\
-96\,m^4\delta\,\beta-8\,m^2
\delta\,{\lambda}^2\beta+312\,m^2\delta\,\beta
\\
+32\,m^4\delta
\,{\beta}^2-200\,m^2\delta\,{\beta}^2+312\,m^2\beta
+16\, \delta\,{\lambda}^2\beta \ ,
\end{split}
\end{equation}

\begin{equation}
\begin{split}
\mathcal C\equiv112-96\,\delta\,{\lambda}^{2}{\alpha}^{2}-\mathcal
N_m(\alpha){\lambda}^{2}+{\lambda}^{4} \mathcal N_m(\alpha)
\\
+16\,{m}^{4}{\lambda}^{2}-36\,{m}^{2}\delta\,{\lambda}^{2}+20\,{
\lambda}^{4}{m}^{2}\delta
\\
+48\,{m}^{2}{\lambda}^{2}{\alpha}^{2}+8\,
\delta\,{\lambda}^{2}-8\,{m}^{4}{\lambda}^{4}\delta
\\
+16\,{m}^{4}\delta
\,{\lambda}^{2}+20\,{m}^{2}{\lambda}^{4}+16\,{m}^{4}\mathcal
N_m(\alpha){\lambda}^{2}
\\
+8\, \mathcal
N_m(\alpha){\lambda}^2{\alpha}^2-64\,{\alpha}^2
-96\,{\lambda}^2{\alpha}^2
\\
-40\,{m}^2\mathcal
N_m(\alpha)-64\,{\alpha}^2\delta-48\,{\alpha}^2
\mathcal N_m(\alpha)
\\
+32\,{\alpha}^2{\lambda}^4+2\,{\lambda}^4
\mathcal N_m(\alpha)m^2-8\,{\lambda}^4\delta
\\
+16\, m^2\mathcal N_m(\alpha)\delta\,
{\lambda}^2{\alpha}^2\beta
-16\,m^2{\lambda}^4\delta\,{\alpha}^2
-24\,{m}^2{\lambda}^4\mathcal N_m(\alpha){\alpha}^2
\\
+48\,m^2\delta\,{\lambda}^2{\alpha}^2
+32\,m^2{\alpha}^2+32\,m^2 {\alpha}^2\delta
\\
-8\,m^4{\lambda}^4+64\,m^2\mathcal
N_m(\alpha){\lambda}^2{\alpha}^2-64\,m^2
\mathcal N_m(\alpha){\lambda}^2{\alpha}^2\beta
\\
+16\,{\alpha}^4\mathcal N_m(\alpha)\delta\,{\lambda}^2\beta+60\,\mathcal
N_m(\alpha)\beta-256\,m^2\mathcal N_m(\alpha){\alpha}^2\beta
\\
+80\,\mathcal
N_m(\alpha)\delta\,{\alpha}^2\beta+32\,{\alpha}^4\mathcal
N_m(\alpha)\delta\,\beta+32\,m ^2\mathcal
N_m(\alpha)\delta\,{\alpha}^2\beta
\\
+96\,m^2{\alpha}^2\delta\,\beta+72
\,\mathcal
N_m(\alpha)\delta\,m^2\beta-48\,{\alpha}^2\beta
\\
-270\,\delta\,\beta+192\,
\mathcal N_m(\alpha)m^2\beta+224\,\mathcal
N_m(\alpha){\alpha}^2\beta
\\
-48\,{\alpha}^2\delta\,\beta+96\,m^2{\alpha}^2\beta-192\,m^4
\mathcal N_m(\alpha)\beta
\\
+18\,\mathcal N_m(\alpha)\delta\,\beta+32\,{\alpha}^2
{\lambda}^4\delta+18\,{\lambda}^{2}\beta
\nonumber
\end{split}
\end{equation}
\begin{equation}
\begin{split}
+80\,\delta\,{\lambda}^{2}{\alpha}^2\beta
-27\,\mathcal N_m(\alpha)\delta\,{\lambda}^2\beta
+32\,{\alpha}^4\delta\,{\lambda}^2\beta
\\
-16\,\mathcal
N_m(\alpha){\lambda}^{2}{m}^{2}\beta- 32\,\mathcal
N_m(\alpha){\lambda}^{2}{\alpha}^{2}\beta-64\,{\alpha}^{4}\mathcal
N_m(\alpha){\lambda}^{2} \beta
\\
-4\,\mathcal
N_m(\alpha){\lambda}^{2}\beta+80\,{\lambda}^{2}{\alpha}^{2}\beta
+32\,{\alpha}^{4}{\lambda}^{2}\beta
\\
+112\,\delta+32\,{\alpha}^{4}\beta+32\,{
\alpha}^{4}\delta\,\beta
\\
-64\,{\alpha}^{4}\mathcal
N_m(\alpha)\beta+24\,\mathcal N_m(\alpha)\delta\,{\alpha}^
{2}\beta\,{\lambda}^{2}-96\,{m}^{2}\mathcal
N_m(\alpha)\delta\,{\alpha}^{2}{\beta}^{2}
\\
+36\,\mathcal N_m(\alpha)\delta\,{m}^{2}\beta\,{\lambda}^{2}
-96\,\mathcal N_m(\alpha)\delta\,{\alpha}^{2}{\beta }^{2}-88\,
\mathcal N_m(\alpha)\delta\,{m}^{2}{\beta}^{2}
\\
-64\,m^2{\alpha}^2\delta\,\beta^2+96\,m^2
\mathcal N_m(\alpha){\alpha}^2{\beta}^2-32\,m^4{\beta}^2
\\
-48\,\mathcal N_m(\alpha){\beta}^{2}
+128\,{\alpha}^{2}{\beta}^{2}+160\,\delta\,{\beta}^{2}
\\
-168\,\mathcal N_m(\alpha)m^2{\beta}^2-192\,
\mathcal N_m(\alpha){\alpha}^2{\beta}^2
+128\,{\alpha}^2\delta\,{\beta}^2
\\
-64\,{m}^2{\alpha}^2{\beta}^2
+96\,m^4\mathcal N_m(\alpha) {\beta}^2
-20\, \mathcal N_m(\alpha)\delta\,{\beta}^2
\\
-184\,m^2\delta-32\,m^4
\mathcal N_m(\alpha)\delta\,{\beta}^2-64\,\alpha^4
\mathcal N_m(\alpha)\delta\,{\beta}^2
\\
-16\,m^2{\beta}^2-184\,m^2+8\,{\lambda}^2-270\,\beta
\\
+48\,\alpha^4
\mathcal N_m(\alpha){\lambda}^{2}+160\,{\beta}^{2}
-8\,{m}^{4}{\lambda}^{4}\mathcal N_m(\alpha)
-36\,{\lambda}^{2}{m}^{2}
\\
+64\,m^4-8\,{\lambda}^4+64\,m^4\delta
\\
-16\,{\alpha}^{4}{\lambda}^4\mathcal N_m(\alpha)
+96\,m^2\mathcal N_m(\alpha){\alpha}^{2}
-16\,{\lambda}^4m^2{\alpha}^2
\\
+216\,{m}^{2}\delta\,\beta-32\,m^4\delta\,{\beta}^{2}
-16\,m^2\delta\,{\beta}^2
\\
+216\,m^2\beta+32\,{\alpha}^4\mathcal N_m(\alpha)
+64 \,m^4\mathcal N_m(\alpha)
\\
+18\,\delta\,{\lambda}^2\beta-14\,\mathcal N_m(\alpha) \ ,
\end{split}
\end{equation}

\begin{equation}
\begin{split}
\mathcal D\equiv14-8\,\delta\,{\lambda}^{2}{\alpha}^{2}-\mathcal
N_m(\alpha){\lambda}^{2}+{\lambda}^{4}\mathcal
N_m(\alpha)
\\
-16\,m^4{\lambda}^2-2\,{\lambda}^4m^2\delta-64\,m^2
{\lambda}^2{\alpha}^{2}
\\
+\delta\,{\lambda}^{2}+8\,{m}^{4}{\lambda}^{4}
\delta-16\,{m}^{4}\delta\,{\lambda}^{2}
\\
-2\,m^2{\lambda}^4-48\,{\alpha}^4\delta\,{\lambda}^{2}+16\,{m}^{4}
\mathcal N_m(\alpha){\lambda}^{2}
\\
+16\,{m}^{4} \mathcal N_m(\alpha)\delta\,{\lambda}^{2}\beta+8\,
\mathcal N_m(\alpha){\lambda}^{2}{\alpha}^{2}+48\,{\alpha}^2
\\
-8\,{\lambda}^{2}{\alpha}^{2}-40\,{m}^{2}\mathcal
N_m(\alpha)+48\,{\alpha}^{2}\delta
\\
-48\,{\alpha}^{2}\mathcal N_m(\alpha)+2\,{\lambda}^{4}\mathcal
N_m(\alpha){m}^{2}-{\lambda}^{4}\delta
\\
+16\,{\alpha}^{4}{\lambda}^4\delta+32\,m^2
\mathcal N_m(\alpha)\delta\,{\lambda}^2{\alpha}^2
\beta +24\,{m}^{2}{\lambda}^{4}\delta\,{\alpha}^{2}
\\
-24\,m^2{\lambda}^4\mathcal N_m(\alpha){\alpha}^2
-32\,{\alpha}^{4}-64\,m^2\delta\,{\lambda}^2{\alpha}^2
\\
+16\,{\lambda}^4{\alpha}^4-96\,m^2{\alpha}^2
-48\,{\alpha}^{4}{\lambda}^{2}
\\
-96\,m^2{\alpha}^2\delta-32\,{\alpha}^4\delta
+8\,{m}^{4}{\lambda}^{4}
+64\,m^2\mathcal N_m(\alpha){\lambda}^2{\alpha}^2
\\
-32\,m^2\mathcal N_m(\alpha)
{\lambda}^2{\alpha}^2\beta+16\,{\alpha}^4
\mathcal N_m(\alpha)\delta\,{\lambda}^2\beta
\\
+30\,\mathcal N_m(\alpha)\beta-128\,m^2
\mathcal N_m(\alpha){\alpha}^2\beta
+16\,\mathcal N_m(\alpha)\delta\,{\alpha}^2\beta
\\
+32\,m^4\mathcal N_m(\alpha)\delta\, \beta
+32\,{\alpha}^4\mathcal N_m(\alpha)\delta\,\beta
\\
+64\,m^2\mathcal N_m(\alpha)\delta\,{\alpha}^2\beta
+128\,{m}^{2}{\alpha}^2\delta\,\beta
+16\,\mathcal N_m(\alpha)\delta\,m^2\beta
\\
-112\,{\alpha}^{2}\beta
-30\,\delta\,\beta+96\,m^4\beta+96\,\mathcal
N_m(\alpha){m}^{2} \beta
\\
+112\,\mathcal N_m(\alpha){\alpha}^2\beta-112\,{\alpha}^2
\delta\,\beta+128\,m^2{\alpha}^2\beta
\\
-96\,m^4\mathcal N_m(\alpha)\beta
+2\,\mathcal N_m(\alpha)\delta\,\beta+2\,{\lambda}^2\beta
\\
+32\,m^2\delta\,{\lambda}^2{\alpha}^2\beta+16\,\delta
\,{\lambda}^2{\alpha}^2\beta-3\,\mathcal N_m(\alpha)
\delta\,{\lambda}^2\beta
\\
+32\,m^2{\lambda}^2{\alpha}^2\beta
+32\,{\alpha}^4\delta\,{\lambda}^2\beta
-8\,\mathcal N_m(\alpha){\lambda}^2m^2\beta
\\
-16\,\mathcal N_m(\alpha){\lambda}^{2}{\alpha}^{2}\beta
-32\,{\alpha}^{4}\mathcal N_m(\alpha){\lambda}^{2}\beta
-2\,\mathcal N_m(\alpha){\lambda}^{2}\beta
\\
+16\,\lambda^2{\alpha}^2\beta
+32\,{\alpha}^4{\lambda}^2\beta+8\,{\lambda}^{2}m^2\beta
+14\,\delta +32\,{\alpha}^{4}\beta
\\
+32\,{\alpha}^4\delta\,\beta
-32\,{\alpha}^{4}\mathcal N_m(\alpha)\beta
-8\,\mathcal N_m(\alpha)\delta\,{\alpha}^2\beta \,{\lambda}^2
\\
-64\,m^2\mathcal N_m(\alpha)\delta\,{\alpha}^2{\beta}^2
-8\,\mathcal N_m(\alpha)\delta \,m^2\beta\,{\lambda}^2
\\
-16\,\mathcal N_m(\alpha)\delta\,{\alpha}^2{\beta}^2
-16 \,\mathcal N_m(\alpha)\delta\,m^2{\beta}^2
-32\,m^2{\alpha}^2\delta\,{\beta}^2
\\
+32\,m^2\mathcal N_m(\alpha){\alpha}^2{\beta}^2
-32\,m^4{\beta}^2-16\,\mathcal N_m(\alpha){\beta}^2
\nonumber
\end{split}
\end{equation}
\begin{equation}
\begin{split}
+64\,{\alpha}^2{\beta}^2+16\,\delta\,{\beta}^2
-56\,\mathcal N_m(\alpha){m }^2{\beta}^2
\\
-64\,\mathcal N_m(\alpha){\alpha}^2{\beta}^2+64\,{\alpha}^2\delta\,
{\beta}^{2}-32\,{m}^{2}{\alpha}^{2}{\beta}^{2}
\\
+32\,m^4\mathcal N_m(\alpha){\beta}^{2} 
-2\,\mathcal N_m(\alpha)\delta\,{\beta}^2
+40\,{m}^{2}\delta
\\
-32\,m^4\mathcal N_m(\alpha)\delta\,{\beta}^2
-32\,{\alpha}^4\mathcal N_m(\alpha)\delta\,{\beta}^2
+56\,{m}^{2}{\beta}^2
\\
+40\, m^2+{\lambda}^2-30\,\beta
+48\,{\alpha}^4\mathcal N_m(\alpha){\lambda}^{2}
\\
+16\, \beta^2-8\,m^4{\lambda}^4\mathcal N_m(\alpha)
-64\,m^4-\lambda^4
\\
-64\,m^4\delta-16\,{\alpha}^4{\lambda}^4
\mathcal N_m(\alpha)+96\,m^2
\mathcal N_m(\alpha){\alpha}^{2}
\\
+24\,{\lambda}^4m^2{\alpha}^2+96\,m^4\delta\,\beta+8\,m^2
\delta\,{\lambda}^2\beta
-96\,m^2\delta\,\beta
\\
-32\,m^4\delta\,{\beta}^{2}
+56\,m^2\delta\,{\beta}^{2}-96\,m^2\beta
+32\,{\alpha}^4\mathcal N_m(\alpha)
\\
+64\,{m}^{4}\mathcal N_m(\alpha)+2\,\delta\,{\lambda}^2\beta
-14\,\mathcal N_m(\alpha) \ .
\end{split}
\end{equation}

\section{}

In section 4.3.2, we indicated that there exists a root 
$\beta_{c0}$ for the $y_2$ solution branch of the $\mu_g/Z$,
and in the case of Figure \ref{f41}, the value of this
$\beta_{c0}$ is about 0.5589 by formula (\ref{120}). However, 
one might worry that from Figure \ref{f41} in which $\beta$ 
ranges from 1 to 5, the trend of variation of $y_2$ may not
lead to such a $\beta_{c0}$ where $y_2$ vanishes. Here, we 
provide the $\mu_g/Z$ versus $\beta$ curve in which $\beta$ 
ranges from 0.2 to 1.5 (all other parameters are exactly the
same as those of Figure \ref{f41} and only the $y_2$ branch 
is shown here) to support our statement in the main text. In 
Figure E1, it is clear that this $y_2$ branch does indeed go 
across the horizontal $\beta$ axis at about 0.5589 as expected
from formula (\ref{120}).
\begin{figure}
\begin{center}
\includegraphics[angle=0,scale=0.45]{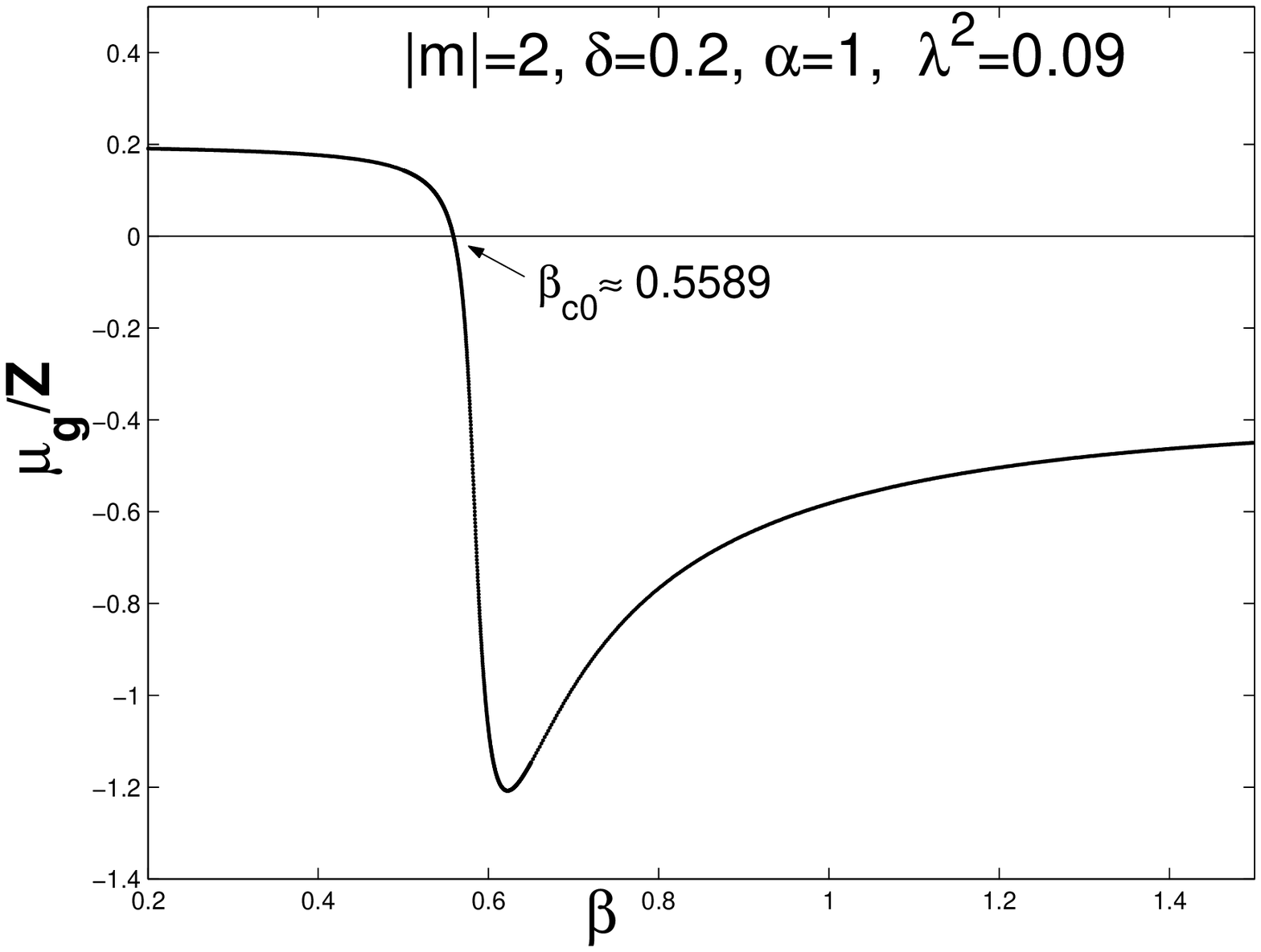}
\caption{The phase relationship between the perturbation of gas 
density $\mu_g$ and that of azimuthal magnetic field $Z$, with 
$|m|=2$, $\delta=0.2$, $\alpha=1$ and $\lambda^2=0.09$. This 
branch goes across the horizontal $\beta$ axis consistent with 
formula (\ref{120}).  }
\end{center}
\end{figure}

\end{appendix}

\end{document}